\documentclass[fleqn,usenatbib]{mnras}

\usepackage{newtxtext,newtxmath}

\usepackage{enumitem}
\usepackage[T1]{fontenc}
\usepackage{xcolor}
\usepackage{textcomp}
\usepackage{fancyvrb}
\usepackage{lineno}
\usepackage{comment}
\DeclareRobustCommand{\VAN}[3]{#2}
\let\VANthebibliography\thebibliography
\def\thebibliography{\DeclareRobustCommand{\VAN}[3]{##3}\VANthebibliography}

\usepackage{graphicx}	
\usepackage{amsmath}	
\usepackage{multirow}

\title[Galactic transient sources with CTAO]{Galactic transient sources with the Cherenkov Telescope Array Observatory}

\author[K.~Abe et al]{\parbox{\textwidth}{\small\centering%
  K.~Abe$^{\ref{AFFIL::JapanUTokai}}$
  S.~Abe$^{\ref{AFFIL::JapanUTokyoICRR}}$
  J.~Abhir$^{\ref{AFFIL::SwitzerlandETHZurich}}$
  A.~Abhishek$^{\ref{AFFIL::ItalyUSienaandINFN}}$
  F.~Acero$^{\ref{AFFIL::FranceCEAIRFUDAp},\ref{AFFIL::SpainFSLACIRLCNRSIAC}}$
  A.~Acharyya$^{\ref{AFFIL::USAUAlabamaTuscaloosa}}$
  R.~Adam$^{\ref{AFFIL::FranceOCotedAzur},\ref{AFFIL::FranceLLREcolePolytechnique}}$
  A.~Aguasca-Cabot$^{\ref{AFFIL::SpainICCUB}}$
  I.~Agudo$^{\ref{AFFIL::SpainIAACSIC}}$
  A.~Aguirre-Santaella$^{\ref{AFFIL::UnitedKingdomICCUDurham}}$
  J.~Alfaro$^{\ref{AFFIL::ChileUPontificiaCatolicadeChile}}$
  R.~Alfaro$^{\ref{AFFIL::MexicoUNAMMexico}}$
  N.~Alvarez-Crespo$^{\ref{AFFIL::SpainUCMAltasEnergias}}$
  R.~Alves~Batista$^{\ref{AFFIL::SpainIFTUAMCSIC}}$
  J.-P.~Amans$^{\ref{AFFIL::FranceObservatoiredeParis}}$
  E.~Amato$^{\ref{AFFIL::ItalyOArcetri}}$
  G.~Ambrosi$^{\ref{AFFIL::ItalyUPerugiaandINFN}}$
  F.~Ambrosino$^{\ref{AFFIL::ItalyORoma}}$
  E.~O.~Ang\"uner$^{\ref{AFFIL::TurkeyTubitak}}$
  L.~A.~Antonelli$^{\ref{AFFIL::ItalyORoma}}$
  C.~Aramo$^{\ref{AFFIL::ItalyINFNNapoli}}$
  C.~Arcaro$^{\ref{AFFIL::ItalyINFNPadova}}$
  T.~T.~H.~Arnesen$^{\ref{AFFIL::SpainIAC}}$
  K.~Asano$^{\ref{AFFIL::JapanUTokyoICRR}}$
  Y.~Ascasibar$^{\ref{AFFIL::SpainIFTUAMCSIC}}$
  J.~Aschersleben$^{\ref{AFFIL::NetherlandsUGroningen}}$
  H.~Ashkar$^{\ref{AFFIL::FranceLLREcolePolytechnique}}$
  L.~Augusto~Stuani$^{\ref{AFFIL::BrazilIFSCUSaoPaulo}}$
  D.~Baack$^{\ref{AFFIL::GermanyUDortmundTU}}$
  M.~Backes$^{\ref{AFFIL::NamibiaUNamibia},\ref{AFFIL::SouthAfricaNWU}}$
  C.~Balazs$^{\ref{AFFIL::AustraliaUMonash}}$
  M.~Balbo$^{\ref{AFFIL::SwitzerlandUGenevaDPNC}}$
  A.~Baquero~Larriva$^{\ref{AFFIL::SpainUCMAltasEnergias},\ref{AFFIL::EcuadorUAzuay}}$
  V.~Barbosa~Martins$^{\ref{AFFIL::GermanyDESY}}$
  U.~Barres~de~Almeida$^{\ref{AFFIL::BrazilCBPF},\ref{AFFIL::BrazilIAGUSaoPaulo}}$
  J.~A.~Barrio$^{\ref{AFFIL::SpainUCMAltasEnergias}}$
  L.~Barrios-Jim\'enez$^{\ref{AFFIL::SpainIAC}}$
  I.~Batkovi\'c$^{\ref{AFFIL::ItalyUPadovaandINFN}}$
  R.~Batzofin$^{\ref{AFFIL::GermanyUPotsdam}}$
  J.~Baxter$^{\ref{AFFIL::JapanUTokyoICRR}}$
  J.~Becerra~Gonz\'alez$^{\ref{AFFIL::SpainIAC}}$
  J.~Becker~Tjus$^{\ref{AFFIL::GermanyUBochum}}$
  R.~Belmont$^{\ref{AFFIL::FranceCEAIRFUDApUParisCiteaffiliatedpersonnel}}$
  W.~Benbow$^{\ref{AFFIL::USACfAHarvardSmithsonian}}$
  J.~Bernete$^{\ref{AFFIL::SpainCIEMAT}}$
  K.~Bernl\"ohr$^{\ref{AFFIL::GermanyMPIK}}$
  A.~Berti$^{\ref{AFFIL::GermanyMPP}}$
  B.~Bertucci$^{\ref{AFFIL::ItalyUPerugiaandINFN}}$
  V.~Beshley$^{\ref{AFFIL::UkraineIAPMMLviv}}$
  P.~Bhattacharjee$^{\ref{AFFIL::FranceLAPPUSavoieMontBlanc}}$
  S.~Bhattacharyya$^{\ref{AFFIL::SloveniaUNovaGoricaCAC}}$
  C.~Bigongiari$^{\ref{AFFIL::ItalyORoma}}$
  E.~Bissaldi$^{\ref{AFFIL::ItalyPolitecnicoBari},\ref{AFFIL::ItalyINFNBari}}$
  O.~Blanch$^{\ref{AFFIL::SpainIFAEBIST}}$
  J.~Blazek$^{\ref{AFFIL::CzechRepublicFZU}}$
  F.~Bocchino$^{\ref{AFFIL::ItalyOPalermo}}$
  C.~Boisson$^{\ref{AFFIL::FranceObservatoiredeParis}}$
  J.~Bolmont$^{\ref{AFFIL::FranceLPNHEUSorbonne}}$
  G.~Bonnoli$^{\ref{AFFIL::ItalyOBrera},\ref{AFFIL::ItalyINFNPisa}}$
  A.~Bonollo$^{\ref{AFFIL::ItalyIUSSPaviaINAF},\ref{AFFIL::ItalyUTrento}}$
  P.~Bordas$^{\ref{AFFIL::SpainICCUB}}$
  Z.~Bosnjak$^{\ref{AFFIL::CroatiaUZagreb}}$
  E.~Bottacini$^{\ref{AFFIL::ItalyUPadovaandINFN}}$
  M.~B\"ottcher$^{\ref{AFFIL::SouthAfricaNWU}}$
  F.~Bradascio$^{\ref{AFFIL::FranceCEAIRFUDPhP}}$
  E.~Bronzini$^{\ref{AFFIL::ItalyOASBologna}}$
  A.~M.~Brown$^{\ref{AFFIL::UnitedKingdomUDurham}}$
  G.~Brunelli$^{\ref{AFFIL::ItalyOASBologna}}$
  A.~Bulgarelli$^{\ref{AFFIL::ItalyOASBologna}}$
  I.~Burelli$^{\ref{AFFIL::ItalyUUdineandINFNTrieste}}$
  C.~Burger-Scheidlin$^{\ref{AFFIL::IrelandDIAS}}$
  L.~Burmistrov$^{\ref{AFFIL::SwitzerlandUGenevaDPNC}}$
  M.~Burton$^{\ref{AFFIL::UnitedKingdomArmaghObservatoryandPlanetarium},\ref{AFFIL::AustraliaUNewSouthWales}}$
  M.~Buscemi$^{\ref{AFFIL::ItalyINFNCatania}}$
  J.~Cailleux$^{\ref{AFFIL::FranceObservatoiredeParis}}$
  A.~Campoy-Ordaz$^{\ref{AFFIL::SpainUABandCERESIEEC}}$
  B.~K.~Cantlay$^{\ref{AFFIL::ThailandUKasetsart},\ref{AFFIL::ThailandNARIT}}$
  G.~Capasso$^{\ref{AFFIL::ItalyOCapodimonte}}$
  A.~Caproni$^{\ref{AFFIL::BrazilUCidadeSPaulo}}$
  R.~Capuzzo-Dolcetta$^{\ref{AFFIL::ItalyORoma},\ref{AFFIL::ItalyURomaSapienza}}$
  P.~Caraveo$^{\ref{AFFIL::ItalyIASFMilano}}$
  S.~Caroff$^{\ref{AFFIL::FranceLAPPUSavoieMontBlanc}}$
  A.~Carosi$^{\ref{AFFIL::ItalyORoma}}$
  R.~Carosi$^{\ref{AFFIL::ItalyINFNPisa}}$
  E.~Carquin$^{\ref{AFFIL::ChileUTecnicaFedericoSantaMaria}}$
  M.-S.~Carrasco$^{\ref{AFFIL::FranceCPPMUAixMarseille}}$
  E.~Cascone$^{\ref{AFFIL::ItalyOCapodimonte}}$
  F.~Cassol$^{\ref{AFFIL::FranceCPPMUAixMarseille}}$
  L.~Castaldini$^{\ref{AFFIL::ItalyOASBologna}}$
  N.~Castrejon$^{\ref{AFFIL::SpainUAlcala}}$
  A.~J.~Castro-Tirado$^{\ref{AFFIL::SpainIAACSIC}}$
  D.~Cerasole$^{\ref{AFFIL::ItalyUandINFNBari}}$
  G.~Ceribella$^{\ref{AFFIL::GermanyMPP}}$
  M.~Cerruti$^{\ref{AFFIL::FranceAPCUParisCite}}$
  P.~M.~Chadwick$^{\ref{AFFIL::UnitedKingdomUDurham}}$
  S.~Chaty$^{\ref{AFFIL::FranceAPCUParisCite}}$
  A.~W.~Chen$^{\ref{AFFIL::SouthAfricaUWitwatersrand}}$
  M.~Chernyakova$^{\ref{AFFIL::IrelandDCU}}$
  A.~Chiavassa$^{\ref{AFFIL::ItalyINFNTorino},\ref{AFFIL::ItalyUTorino}}$
  J.~Chudoba$^{\ref{AFFIL::CzechRepublicFZU}}$
  L.~Chytka$^{\ref{AFFIL::CzechRepublicFZU}}$
  G.~M.~Cicciari$^{\ref{AFFIL::ItalyUPalermo},\ref{AFFIL::ItalyINFNCatania}}$
  A.~Cifuentes$^{\ref{AFFIL::SpainCIEMAT}}$
  C.~H.~Coimbra~Araujo$^{\ref{AFFIL::BrazilUFPR}}$
  M.~Colapietro$^{\ref{AFFIL::ItalyOCapodimonte}}$
  V.~Conforti$^{\ref{AFFIL::ItalyOASBologna}}$
  J.~L.~Contreras$^{\ref{AFFIL::SpainUCMAltasEnergias}}$
  J.~Cortina$^{\ref{AFFIL::SpainCIEMAT}}$
  A.~Costa$^{\ref{AFFIL::ItalyOCatania}}$
  H.~Costantini$^{\ref{AFFIL::FranceCPPMUAixMarseille}}$
  G.~Cotter$^{\ref{AFFIL::UnitedKingdomUOxford}}$
  P.~Cristofari$^{\ref{AFFIL::FranceObservatoiredeParis}}$
  O.~Cuevas$^{\ref{AFFIL::ChileUdeValparaiso}}$
  Z.~Curtis-Ginsberg$^{\ref{AFFIL::USAUWisconsin}}$
  A.~D'A{\`\i}$^{\ref{AFFIL::ItalyIASFPalermo}}$
  G.~D'Amico$^{\ref{AFFIL::NorwayUBergen}}$
  F.~D'Ammando$^{\ref{AFFIL::ItalyRadioastronomiaINAF}}$
  S.~Dai$^{\ref{AFFIL::AustraliaUWesternSydney}}$
  F.~Dazzi$^{\ref{AFFIL::ItalyINAF}}$
  M.~de~Bony~de~Lavergne$^{\ref{AFFIL::FranceCEAIRFUDAp}}$
  V.~De~Caprio$^{\ref{AFFIL::ItalyOCapodimonte}}$
  G.~De~Cesare$^{\ref{AFFIL::ItalyOASBologna}}$
  F.~De~Frondat~Laadim$^{\ref{AFFIL::FranceObservatoiredeParis}}$
  E.~M.~de~Gouveia~Dal~Pino$^{\ref{AFFIL::BrazilIAGUSaoPaulo}}$
  B.~De~Lotto$^{\ref{AFFIL::ItalyUUdineandINFNTrieste}}$
  M.~De~Lucia$^{\ref{AFFIL::ItalyINFNNapoli}}$
  D.~de~Martino$^{\ref{AFFIL::ItalyOCapodimonte}}$
  R.~de~Menezes$^{\ref{AFFIL::ItalyINFNTorino},\ref{AFFIL::ItalyUTorino}}$
  M.~de~Naurois$^{\ref{AFFIL::FranceLLREcolePolytechnique}}$
  E.~de~Ona~Wilhelmi$^{\ref{AFFIL::GermanyDESY}}$
  V.~de~Souza$^{\ref{AFFIL::BrazilIFSCUSaoPaulo}}$
  L.~del~Peral$^{\ref{AFFIL::SpainUAlcala}}$
  M.~V.~del~Valle$^{\ref{AFFIL::BrazilIAGUSaoPaulo}}$
  C.~Delgado$^{\ref{AFFIL::SpainCIEMAT}}$
  M.~Dell'aiera$^{\ref{AFFIL::FranceLAPPUSavoieMontBlanc}}$
  M.~Della~Valle$^{\ref{AFFIL::ItalyOCapodimonte},\ref{AFFIL::ItalyINFNNapoli}}$
  D.~della~Volpe$^{\ref{AFFIL::SwitzerlandUGenevaDPNC}}$
  D.~Depaoli$^{\ref{AFFIL::GermanyMPIK}}$
  J.~Devin$^{\ref{AFFIL::FranceLUPMUMontpellier}}$
  T.~Di~Girolamo$^{\ref{AFFIL::ItalyUNapoli},\ref{AFFIL::ItalyINFNNapoli}}$
  A.~Di~Piano$^{\ref{AFFIL::ItalyOASBologna},\ref{AFFIL::ItalyUModena}}$
  F.~Di~Pierro$^{\ref{AFFIL::ItalyINFNTorino}}$
  R.~Di~Tria$^{\ref{AFFIL::ItalyUandINFNBari}}$
  L.~Di~Venere$^{\ref{AFFIL::ItalyINFNBari}}$
  S.~Diebold$^{\ref{AFFIL::GermanyIAAT}}$
  C.~Dignam$^{\ref{AFFIL::IrelandDCU}}$
  R.~Dima$^{\ref{AFFIL::ItalyUPadovaandINFN}}$
  A.~Dinesh$^{\ref{AFFIL::SpainUCMAltasEnergias}}$
  J.~Djuvsland$^{\ref{AFFIL::NorwayUBergen}}$
  R.~M.~Dominik$^{\ref{AFFIL::GermanyUDortmundTU}}$
  D.~Dominis~Prester$^{\ref{AFFIL::CroatiaURijeka}}$
  A.~Donini$^{\ref{AFFIL::ItalyORoma}}$
  D.~Dorner$^{\ref{AFFIL::GermanyUWurzburg},\ref{AFFIL::SwitzerlandETHZurich}}$
  J.~D\"orner$^{\ref{AFFIL::GermanyUBochum}}$
  M.~Doro$^{\ref{AFFIL::ItalyUPadovaandINFN}}$
  C.~Dubos$^{\ref{AFFIL::FranceIJCLab}}$
  L.~Ducci$^{\ref{AFFIL::GermanyIAAT}}$
  V.~V.~Dwarkadas$^{\ref{AFFIL::USAUChicagoDAA}}$
  J.~Ebr$^{\ref{AFFIL::CzechRepublicFZU}}$
  C.~Eckner$^{\ref{AFFIL::SloveniaUNovaGoricaCAC},\ref{AFFIL::FranceLAPTh}}$
  K.~Egberts$^{\ref{AFFIL::GermanyUPotsdam}}$
  S.~Einecke$^{\ref{AFFIL::AustraliaUAdelaide}}$
  D.~Els\"asser$^{\ref{AFFIL::GermanyUDortmundTU}}$
  G.~Emery$^{\ref{AFFIL::FranceCPPMUAixMarseille}}$
  M.~Errando$^{\ref{AFFIL::USAWashingtonU}}$
  C.~Escanuela$^{\ref{AFFIL::GermanyMPIK}}$
  P.~Escarate$^{\ref{AFFIL::ChileEscIngElec}}$
  M.~Escobar~Godoy$^{\ref{AFFIL::USASCIPP}}$
  J.~Escudero~Pedrosa$^{\ref{AFFIL::SpainIAACSIC}}$
  P.~Esposito$^{\ref{AFFIL::ItalyIUSSPaviaINAF},\ref{AFFIL::ItalyIASFMilano}}$
  D.~Falceta-Goncalves$^{\ref{AFFIL::BrazilEACHUSaoPaulo}}$
  S.~Fegan$^{\ref{AFFIL::FranceLLREcolePolytechnique}}$
  Q.~Feng$^{\ref{AFFIL::USAUUtah}}$
  G.~Ferrand$^{\ref{AFFIL::CanadaUManitoba},\ref{AFFIL::JapanRIKEN}}$
  F.~Ferrarotto$^{\ref{AFFIL::ItalyINFNRomaLaSapienza}}$
  E.~Fiandrini$^{\ref{AFFIL::ItalyUPerugiaandINFN}}$
  A.~Fiasson$^{\ref{AFFIL::FranceLAPPUSavoieMontBlanc}}$
  M.~Filipovic$^{\ref{AFFIL::AustraliaUWesternSydney}}$
  V.~Fioretti$^{\ref{AFFIL::ItalyOASBologna}}$
  M.~Fiori$^{\ref{AFFIL::ItalyOPadova}}$
  H.~Flores$^{\ref{AFFIL::FranceObservatoiredeParis}}$
  L.~Foffano$^{\ref{AFFIL::ItalyIAPS}}$
  L.~Font~Guiteras$^{\ref{AFFIL::SpainUABandCERESIEEC}}$
  G.~Fontaine$^{\ref{AFFIL::FranceLLREcolePolytechnique}}$
  A.~Franckowiak$^{\ref{AFFIL::GermanyUBochumPhysAst}}$
  S.~Fr\"ose$^{\ref{AFFIL::GermanyUDortmundTU}}$
  Y.~Fukazawa$^{\ref{AFFIL::JapanUHiroshima}}$
  Y.~Fukui$^{\ref{AFFIL::JapanUNagoya}}$
  S.~Funk$^{\ref{AFFIL::GermanyUErlangenECAP}}$
  A.~Furniss$^{\ref{AFFIL::USASCIPP}}$
  G.~Galanti$^{\ref{AFFIL::ItalyIASFMilano}}$
  G.~Galaz$^{\ref{AFFIL::ChileUPontificiaCatolicadeChile}}$
  C.~Galelli$^{\ref{AFFIL::FranceObservatoiredeParis}}$
  Y.~A.~Gallant$^{\ref{AFFIL::FranceLUPMUMontpellier}}$
  S.~Gallozzi$^{\ref{AFFIL::ItalyORoma}}$
  V.~Gammaldi$^{\ref{AFFIL::SpainUniversidadSanPabloCEU},\ref{AFFIL::SpainIFTUAMCSIC}}$
  M.~Garczarczyk$^{\ref{AFFIL::GermanyDESY}}$
  C.~Gasbarra$^{\ref{AFFIL::ItalyINFNRomaTorVergata}}$
  D.~Gasparrini$^{\ref{AFFIL::ItalyINFNRomaTorVergata}}$
  M.~Gaug$^{\ref{AFFIL::SpainUABandCERESIEEC}}$
  A.~Ghalumyan$^{\ref{AFFIL::ArmeniaNSLAlikhanyan}}$
  F.~Gianotti$^{\ref{AFFIL::ItalyOASBologna}}$
  M.~Giarrusso$^{\ref{AFFIL::ItalyINFNCatania}}$
  J.~G.~Giesbrecht~Formiga~Paiva$^{\ref{AFFIL::BrazilCBPF}}$
  N.~Giglietto$^{\ref{AFFIL::ItalyPolitecnicoBari},\ref{AFFIL::ItalyINFNBari}}$
  F.~Giordano$^{\ref{AFFIL::ItalyUandINFNBari}}$
  M.~Giroletti$^{\ref{AFFIL::ItalyRadioastronomiaINAF}}$
  R.~Giuffrida$^{\ref{AFFIL::ItalyOPalermo}}$
  A.~Giuliani$^{\ref{AFFIL::ItalyIASFMilano}}$
  J.-F.~Glicenstein$^{\ref{AFFIL::FranceCEAIRFUDPhP}}$
  J.~Glombitza$^{\ref{AFFIL::GermanyUErlangenECAP}}$
  P.~Goldoni$^{\ref{AFFIL::FranceAPCUParisCiteCEAaffiliatedpersonnel}}$
  J.~M.~Gonz\'alez$^{\ref{AFFIL::ChileUAndresBello}}$
  M.~M.~Gonz\'alez$^{\ref{AFFIL::MexicoUNAMMexico}}$
  J.~Goulart~Coelho$^{\ref{AFFIL::BrazilUFES}}$
  T.~Gradetzke$^{\ref{AFFIL::GermanyUDortmundTU}}$
  J.~Granot$^{\ref{AFFIL::IsraelOpenUniversityofIsrael},\ref{AFFIL::USAGWUWashingtonDC}}$
  R.~Grau$^{\ref{AFFIL::SpainIFAEBIST}}$
  L.~Gr\'eaux$^{\ref{AFFIL::FranceIJCLab}}$
  D.~Green$^{\ref{AFFIL::GermanyMPP}}$
  J.~G.~Green$^{\ref{AFFIL::GermanyMPP}}$
  T.~Greenshaw$^{\ref{AFFIL::UnitedKingdomULiverpool}}$
  I.~Grenier$^{\ref{AFFIL::FranceCEAIRFUDApUParisCiteaffiliatedpersonnel}}$
  G.~Grolleron$^{\ref{AFFIL::FranceLPNHEUSorbonne}}$
  J.~Grube$^{\ref{AFFIL::UnitedKingdomKingsCollege}}$
  O.~Gueta$^{\ref{AFFIL::ItalyCTAOBologna}}$
  J.~Hackfeld$^{\ref{AFFIL::GermanyUBochum},\ref{AFFIL::GermanyUDortmundTU}}$
  D.~Hadasch$^{\ref{AFFIL::JapanUTokyoICRR}}$
  P.~Hamal$^{\ref{AFFIL::CzechRepublicFZU}}$
  W.~Hanlon$^{\ref{AFFIL::USACfAHarvardSmithsonian}}$
  S.~Hara$^{\ref{AFFIL::JapanUYamanashiGakuin}}$
  V.~M.~Harvey$^{\ref{AFFIL::AustraliaUAdelaide}}$
  T.~Hassan$^{\ref{AFFIL::SpainCIEMAT}}$
  K.~Hayashi$^{\ref{AFFIL::JapanNITSendaiNatori},\ref{AFFIL::JapanUTokyoICRR}}$
  B.~He{\ss}$^{\ref{AFFIL::GermanyIAAT}}$
  L.~Heckmann$^{\ref{AFFIL::GermanyMPP},\ref{AFFIL::AustriaUInnsbruck}}$
  M.~Heller$^{\ref{AFFIL::SwitzerlandUGenevaDPNC}}$
  J.~Hinton$^{\ref{AFFIL::GermanyMPIK}}$
  N.~Hiroshima$^{\ref{AFFIL::JapanUTokyoICRR},\ref{AFFIL::JapanUYokohamaNational}}$
  B.~Hnatyk$^{\ref{AFFIL::UkraineAstObsofUKyiv}}$
  R.~Hnatyk$^{\ref{AFFIL::UkraineAstObsofUKyiv}}$
  W.~Hofmann$^{\ref{AFFIL::GermanyMPIK}}$
  D.~Horan$^{\ref{AFFIL::FranceLLREcolePolytechnique}}$
  P.~Horvath$^{\ref{AFFIL::CzechRepublicUOlomouc}}$
  T.~Hovatta$^{\ref{AFFIL::FinlandFINCA},\ref{AFFIL::FinlandUAalto}}$
  M.~Hrabovsky$^{\ref{AFFIL::CzechRepublicUOlomouc}}$
  D.~Hrupec$^{\ref{AFFIL::CroatiaUOsijek}}$
  M.~Iarlori$^{\ref{AFFIL::ItalyCETEMPSandUandINFNAquila}}$
  T.~Inada$^{\ref{AFFIL::JapanUTokyoICRR}}$
  F.~Incardona$^{\ref{AFFIL::ItalyOCatania}}$
  S.~Inoue$^{\ref{AFFIL::JapanUChiba},\ref{AFFIL::JapanUTokyoICRR}}$
  Y.~Inoue$^{\ref{AFFIL::JapanUOsaka},\ref{AFFIL::JapanRIKEN}}$
  F.~Iocco$^{\ref{AFFIL::ItalyUNapoli},\ref{AFFIL::ItalyINFNNapoli}}$
  A.~Iuliano$^{\ref{AFFIL::ItalyINFNNapoli}}$
  M.~Jamrozy$^{\ref{AFFIL::PolandUJagiellonian}}$
  P.~Janecek$^{\ref{AFFIL::CzechRepublicFZU}}$
  F.~Jankowsky$^{\ref{AFFIL::GermanyLSW}}$
  P.~Jean$^{\ref{AFFIL::FranceIRAPUToulouse}}$
  I.~Jim\'enez~Mart{\'\i}nez$^{\ref{AFFIL::GermanyMPP}}$
  J.~Jimenez~Quiles$^{\ref{AFFIL::SpainIFAEBIST}}$
  W.~Jin$^{\ref{AFFIL::USAUCLA}}$
  C.~Juramy-Gilles$^{\ref{AFFIL::FranceLPNHEUSorbonne}}$
  J.~Jurysek$^{\ref{AFFIL::CzechRepublicFZU}}$
  O.~Kalekin$^{\ref{AFFIL::GermanyUErlangenECAP}}$
  D.~Kantzas$^{\ref{AFFIL::FranceLAPTh}}$
  V.~Karas$^{\ref{AFFIL::CzechRepublicASU}}$
  H.~Katagiri$^{\ref{AFFIL::JapanUIbaraki}}$
  J.~Kataoka$^{\ref{AFFIL::JapanUWaseda}}$
  S.~Kaufmann$^{\ref{AFFIL::UnitedKingdomUDurham}}$
  D.~Kerszberg$^{\ref{AFFIL::SpainIFAEBIST}}$
  B.~Kh\'elifi$^{\ref{AFFIL::FranceAPCUParisCite}}$
  D.~B.~Kieda$^{\ref{AFFIL::USAUUtah}}$
  R.~Kissmann$^{\ref{AFFIL::AustriaUInnsbruck}}$
  T.~Kleiner$^{\ref{AFFIL::GermanyDESY}}$
  G.~Kluge$^{\ref{AFFIL::NorwayUOslo}}$
  W.~Klu\'zniak$^{\ref{AFFIL::PolandNicolausCopernicusAstronomicalCenter}}$
  Y.~Kobayashi$^{\ref{AFFIL::JapanUChiba},\ref{AFFIL::JapanUTokyoICRR}}$
  K.~Kohri$^{\ref{AFFIL::JapanNAOJ},\ref{AFFIL::JapanKEK}}$
  N.~Komin$^{\ref{AFFIL::SouthAfricaUWitwatersrand}}$
  A.~Kong$^{\ref{AFFIL::JapanUTokyoICRR}}$
  P.~Kornecki$^{\ref{AFFIL::FranceObservatoiredeParis}}$
  K.~Kosack$^{\ref{AFFIL::FranceCEAIRFUDAp}}$
  D.~Kostunin$^{\ref{AFFIL::GermanyDESY}}$
  G.~Kowal$^{\ref{AFFIL::BrazilEACHUSaoPaulo}}$
  H.~Kubo$^{\ref{AFFIL::JapanUTokyoICRR}}$
  J.~Kushida$^{\ref{AFFIL::JapanUTokai}}$
  A.~La~Barbera$^{\ref{AFFIL::ItalyIASFPalermo}}$
  N.~La~Palombara$^{\ref{AFFIL::ItalyIASFMilano}}$
  M.~L\'ainez$^{\ref{AFFIL::SpainUCMAltasEnergias}}$
  A.~Lamastra$^{\ref{AFFIL::ItalyORoma}}$
  J.~Lapington$^{\ref{AFFIL::UnitedKingdomULeicester}}$
  S.~Lazarevi\'c$^{\ref{AFFIL::AustraliaUWesternSydney}}$
  J.~Lazendic-Galloway$^{\ref{AFFIL::AustraliaUMonash}}$
  M.~Lemoine-Goumard$^{\ref{AFFIL::FranceLP2IUBordeaux}}$
  J.-P.~Lenain$^{\ref{AFFIL::FranceLPNHEUSorbonne}}$
  F.~Leone$^{\ref{AFFIL::ItalyUCatania}}$
  E.~Leonora$^{\ref{AFFIL::ItalyINFNCatania}}$
  G.~Leto$^{\ref{AFFIL::ItalyOCatania}}$
  E.~Lindfors$^{\ref{AFFIL::FinlandUTurku}}$
  I.~Liodakis$^{\ref{AFFIL::FinlandFINCA}}$
  S.~Lloyd$^{\ref{AFFIL::UnitedKingdomUDurham}}$
  S.~Lombardi$^{\ref{AFFIL::ItalyORoma},\ref{AFFIL::ASISpaceScienceDataCenter}}$
  F.~Longo$^{\ref{AFFIL::ItalyUandINFNTrieste}}$
  R.~L\'opez-Coto$^{\ref{AFFIL::SpainIAACSIC}}$
  M.~L\'opez-Moya$^{\ref{AFFIL::SpainUCMAltasEnergias}}$
  A.~L\'opez-Oramas$^{\ref{AFFIL::SpainIAC}}$\ref{CONTACTAOUTHOR::1}
  S.~Loporchio$^{\ref{AFFIL::ItalyPolitecnicoBari},\ref{AFFIL::ItalyINFNBari}}$
  J.~Lozano~Bahilo$^{\ref{AFFIL::SpainUAlcala}}$
  F.~Lucarelli$^{\ref{AFFIL::ItalyORoma}}$
  P.~L.~Luque-Escamilla$^{\ref{AFFIL::SpainUJaen}}$
  E.~Lyard$^{\ref{AFFIL::SwitzerlandUGenevaISDC}}$
  O.~Macias$^{\ref{AFFIL::NetherlandsUAmsterdam}}$
  J.~Mackey$^{\ref{AFFIL::IrelandDIAS}}$
  G.~Maier$^{\ref{AFFIL::GermanyDESY}}$
  P.~Majumdar$^{\ref{AFFIL::IndiaSahaInstitute}}$
  M.~Makariev$^{\ref{AFFIL::BulgariaINRNEBAS}}$
  M.~Mallamaci$^{\ref{AFFIL::ItalyUPalermo},\ref{AFFIL::ItalyINFNCatania}}$
  D.~Mandat$^{\ref{AFFIL::CzechRepublicFZU}}$
  M.~Manganaro$^{\ref{AFFIL::CroatiaURijeka}}$
  G.~Manic\`o$^{\ref{AFFIL::ItalyINFNCatania},\ref{AFFIL::ItalyUCatania}}$
  P.~Marinos$^{\ref{AFFIL::USAStanford}}$
  M.~Mariotti$^{\ref{AFFIL::ItalyUPadovaandINFN}}$
  S.~Markoff$^{\ref{AFFIL::NetherlandsUAmsterdam}}$
  I.~M\'arquez$^{\ref{AFFIL::SpainIAACSIC}}$
  P.~Marquez$^{\ref{AFFIL::SpainIFAEBIST}}$
  G.~Marsella$^{\ref{AFFIL::ItalyUPalermo},\ref{AFFIL::ItalyINFNCatania}}$
  G.~A.~Mart{\'\i}nez$^{\ref{AFFIL::SpainCIEMAT}}$
  M.~Mart{\'\i}nez$^{\ref{AFFIL::SpainIFAEBIST}}$
  O.~Martinez$^{\ref{AFFIL::SpainUCMElectronica},\ref{AFFIL::SpainUPCMadrid}}$
  C.~Marty$^{\ref{AFFIL::FranceIRAPUToulouse}}$
  A.~Mas-Aguilar$^{\ref{AFFIL::SpainUCMAltasEnergias}}$
  M.~Mastropietro$^{\ref{AFFIL::ItalyORoma}}$
  G.~Maurin$^{\ref{AFFIL::FranceLAPPUSavoieMontBlanc}}$
  D.~Mazin$^{\ref{AFFIL::JapanUTokyoICRR},\ref{AFFIL::GermanyMPP}}$
  S.~McKeague$^{\ref{AFFIL::IrelandDCU}}$
  D.~Melkumyan$^{\ref{AFFIL::GermanyDESY}}$
  S.~Menchiari$^{\ref{AFFIL::SpainIAACSIC},\ref{AFFIL::ItalyOArcetri}}$
  S.~Mereghetti$^{\ref{AFFIL::ItalyIASFMilano}}$
  E.~Mestre$^{\ref{AFFIL::SpainICECSIC}}$
  J.-L.~Meunier$^{\ref{AFFIL::FranceLPNHEUSorbonne}}$
  D.~M.-A.~Meyer$^{\ref{AFFIL::SpainICECSIC}}$
  D.~Miceli$^{\ref{AFFIL::ItalyINFNPadova}}$
  M.~Miceli$^{\ref{AFFIL::ItalyUPalermo},\ref{AFFIL::ItalyOPalermo}}$
  M.~Michailidis$^{\ref{AFFIL::GermanyIAAT}}$
  J.~Micha{\l}owski$^{\ref{AFFIL::PolandIFJ}}$
  T.~Miener$^{\ref{AFFIL::SwitzerlandUGenevaDPNC}}$
  J.~M.~Miranda$^{\ref{AFFIL::SpainUCMElectronica},\ref{AFFIL::SpainIPARCOSInstitute}}$
  A.~Mitchell$^{\ref{AFFIL::GermanyUErlangenECAP}}$
  M.~Mizote$^{\ref{AFFIL::JapanUKonan}}$
  T.~Mizuno$^{\ref{AFFIL::JapanHASC}}$
  R.~Moderski$^{\ref{AFFIL::PolandNicolausCopernicusAstronomicalCenter}}$
  L.~Mohrmann$^{\ref{AFFIL::GermanyMPIK}}$
  M.~Molero$^{\ref{AFFIL::SpainIAC}}$
  C.~Molfese$^{\ref{AFFIL::ItalyINAF}}$
  E.~Molina$^{\ref{AFFIL::SpainIAC}}$
  T.~Montaruli$^{\ref{AFFIL::SwitzerlandUGenevaDPNC}}$
  A.~Moralejo$^{\ref{AFFIL::SpainIFAEBIST}}$
  D.~Morcuende$^{\ref{AFFIL::SpainIAACSIC}}$
  K.~Morik$^{\ref{AFFIL::GermanyUDortmundTU},\ref{AFFIL::LamarrInstituteGermany}}$
  A.~Morselli$^{\ref{AFFIL::ItalyINFNRomaTorVergata}}$
  E.~Moulin$^{\ref{AFFIL::FranceCEAIRFUDPhP}}$
  V.~Moya~Zamanillo$^{\ref{AFFIL::SpainUCMAltasEnergias}}$
  K.~Munari$^{\ref{AFFIL::ItalyOCatania}}$
  T.~Murach$^{\ref{AFFIL::GermanyDESY}}$
  A.~Muraczewski$^{\ref{AFFIL::PolandNicolausCopernicusAstronomicalCenter}}$
  H.~Muraishi$^{\ref{AFFIL::JapanUKitasato}}$
  S.~Nagataki$^{\ref{AFFIL::JapanRIKEN}}$
  T.~Nakamori$^{\ref{AFFIL::JapanUYamagata}}$
  L.~Nava$^{\ref{AFFIL::ItalyOBrera}}$
  A.~Nayak$^{\ref{AFFIL::UnitedKingdomUDurham}}$
  R.~Nemmen$^{\ref{AFFIL::BrazilIAGUSaoPaulo},\ref{AFFIL::USAStanford}}$
  J.~P.~Neto$^{\ref{AFFIL::BrazilURioGrandedoNortePhys},\ref{AFFIL::BrazilURioGrandedoNorteIIP}}$
  L.~Nickel$^{\ref{AFFIL::GermanyUDortmundTU}}$
  J.~Niemiec$^{\ref{AFFIL::PolandIFJ}}$
  D.~Nieto$^{\ref{AFFIL::SpainUCMAltasEnergias}}$
  M.~Nievas~Rosillo$^{\ref{AFFIL::SpainIAC}}$
  M.~Niko{\l}ajuk$^{\ref{AFFIL::PolandUBiaystok}}$
  L.~Nikoli\'c$^{\ref{AFFIL::ItalyUSienaandINFN}}$
  K.~Nishijima$^{\ref{AFFIL::JapanUTokai}}$
  K.~Noda$^{\ref{AFFIL::JapanUChiba},\ref{AFFIL::JapanUTokyoICRR}}$
  D.~Nosek$^{\ref{AFFIL::CzechRepublicUPrague}}$
  V.~Novotny$^{\ref{AFFIL::CzechRepublicUPrague}}$
  S.~Nozaki$^{\ref{AFFIL::GermanyMPP}}$
  M.~Ohishi$^{\ref{AFFIL::JapanUTokyoICRR}}$
  Y.~Ohtani$^{\ref{AFFIL::JapanUTokyoICRR}}$
  A.~Okumura$^{\ref{AFFIL::JapanUNagoyaISEE},\ref{AFFIL::JapanUNagoyaKMI}}$
  J.-F.~Olive$^{\ref{AFFIL::FranceIRAPUToulouse}}$
  R.~A.~Ong$^{\ref{AFFIL::USAUCLA}}$
  R.~Orito$^{\ref{AFFIL::JapanUTokushima}}$
  M.~Orlandini$^{\ref{AFFIL::ItalyOASBologna}}$
  E.~Orlando$^{\ref{AFFIL::ItalyUandINFNTrieste}}$
  S.~Orlando$^{\ref{AFFIL::ItalyOPalermo}}$
  M.~Ostrowski$^{\ref{AFFIL::PolandUJagiellonian}}$
  J.~Otero-Santos$^{\ref{AFFIL::SpainIAACSIC}}$
  I.~Oya$^{\ref{AFFIL::GermanyCTAOHeidelberg}}$
  I.~Pagano$^{\ref{AFFIL::ItalyOCatania}}$
  A.~Pagliaro$^{\ref{AFFIL::ItalyIASFPalermo}}$
  M.~Palatiello$^{\ref{AFFIL::ItalyORoma}}$
  G.~Panebianco$^{\ref{AFFIL::ItalyOASBologna}}$
  D.~Paneque$^{\ref{AFFIL::GermanyMPP}}$
  F.~R.~Pantaleo$^{\ref{AFFIL::ItalyINFNBari},\ref{AFFIL::ItalyPolitecnicoBari}}$
  A.~Papitto$^{\ref{AFFIL::ItalyORoma}}$
  J.~M.~Paredes$^{\ref{AFFIL::SpainICCUB}}$
  N.~Parmiggiani$^{\ref{AFFIL::ItalyOASBologna}}$
  B.~Patricelli$^{\ref{AFFIL::ItalyORoma},\ref{AFFIL::ItalyUPisa}}$
  A.~Pe'er$^{\ref{AFFIL::GermanyMPP}}$
  M.~Pech$^{\ref{AFFIL::CzechRepublicFZU}}$
  M.~Pecimotika$^{\ref{AFFIL::CroatiaURijeka},\ref{AFFIL::CroatiaIRB}}$
  U.~Pensec$^{\ref{AFFIL::FranceLPNHEUSorbonne},\ref{AFFIL::FranceObservatoiredeParis}}$
  M.~Peresano$^{\ref{AFFIL::GermanyMPP}}$
  J.~P\'erez-Romero$^{\ref{AFFIL::SloveniaUNovaGoricaCAC}}$
  M.~Persic$^{\ref{AFFIL::ItalyOPadova},\ref{AFFIL::ItalyOandINFNTrieste}}$
  P.-O.~Petrucci$^{\ref{AFFIL::FranceIPAGUGrenobleAlpes}}$
  O.~Petruk$^{\ref{AFFIL::UkraineIAPMMLviv},\ref{AFFIL::ItalyOPalermo}}$
  G.~Piano$^{\ref{AFFIL::ItalyIAPS}}$
  E.~Pierre$^{\ref{AFFIL::FranceLPNHEUSorbonne}}$
  E.~Pietropaolo$^{\ref{AFFIL::ItalyUandINFNAquila}}$
  M.~Pihet$^{\ref{AFFIL::ItalyINFNPadova}}$
  F.~Pintore$^{\ref{AFFIL::ItalyIASFPalermo}}$
  G.~Pirola$^{\ref{AFFIL::GermanyMPP}}$
  C.~Pittori$^{\ref{AFFIL::ItalyORoma}}$
  C.~Plard$^{\ref{AFFIL::FranceLAPPUSavoieMontBlanc}}$
  F.~Podobnik$^{\ref{AFFIL::ItalyUSienaandINFN}}$
  M.~Pohl$^{\ref{AFFIL::GermanyUPotsdam},\ref{AFFIL::GermanyDESY}}$
  V.~Pollet$^{\ref{AFFIL::FranceLAPPUSavoieMontBlanc}}$
  G.~Ponti$^{\ref{AFFIL::ItalyOBrera}}$
  E.~Prandini$^{\ref{AFFIL::ItalyUPadovaandINFN}}$
  G.~Principe$^{\ref{AFFIL::ItalyUandINFNTrieste}}$
  C.~Priyadarshi$^{\ref{AFFIL::SpainIFAEBIST}}$
  N.~Produit$^{\ref{AFFIL::SwitzerlandUGenevaISDC}}$
  M.~Prouza$^{\ref{AFFIL::CzechRepublicFZU}}$
  G.~P\"uhlhofer$^{\ref{AFFIL::GermanyIAAT}}$
  M.~L.~Pumo$^{\ref{AFFIL::ItalyUCatania},\ref{AFFIL::ItalyINFNCatania}}$
  M.~Punch$^{\ref{AFFIL::FranceAPCUParisCite}}$
  A.~Quirrenbach$^{\ref{AFFIL::GermanyLSW}}$
  S.~Rain\`o$^{\ref{AFFIL::ItalyUandINFNBari}}$
  R.~Rando$^{\ref{AFFIL::ItalyUPadovaandINFN}}$
  S.~Razzaque$^{\ref{AFFIL::SouthAfricaUJohannesburg},\ref{AFFIL::USAGWUWashingtonDC}}$
  S.~Recchia$^{\ref{AFFIL::ItalyOBrera}}$
  M.~Regeard$^{\ref{AFFIL::FranceAPCUParisCite}}$
  P.~Reichherzer$^{\ref{AFFIL::UnitedKingdomUOxford},\ref{AFFIL::GermanyUBochum}}$
  A.~Reimer$^{\ref{AFFIL::AustriaUInnsbruck}}$
  O.~Reimer$^{\ref{AFFIL::AustriaUInnsbruck}}$
  I.~Reis$^{\ref{AFFIL::BrazilIFSCUSaoPaulo},\ref{AFFIL::FranceCEAIRFUDPhP}}$
  A.~Reisenegger$^{\ref{AFFIL::ChileUPontificiaCatolicadeChile},\ref{AFFIL::ChileUMCE}}$
  T.~Reposeur$^{\ref{AFFIL::FranceLP2IUBordeaux}}$
  W.~Rhode$^{\ref{AFFIL::GermanyUDortmundTU}}$
  D.~Ribeiro$^{\ref{AFFIL::USAUMinnesota}}$
  M.~Rib\'o$^{\ref{AFFIL::SpainICCUB}}$
  C.~Ricci$^{\ref{AFFIL::ChileUniversidadDiegoPortales}}$
  T.~Richtler$^{\ref{AFFIL::ChileUdeConcepcion}}$
  J.~Rico$^{\ref{AFFIL::SpainIFAEBIST}}$
  F.~Rieger$^{\ref{AFFIL::GermanyMPIK}}$
  M.~Rigoselli$^{\ref{AFFIL::ItalyIASFMilano}}$
  L.~Riitano$^{\ref{AFFIL::USAUWisconsin}}$
  V.~Rizi$^{\ref{AFFIL::ItalyUandINFNAquila}}$
  E.~Roache$^{\ref{AFFIL::USACfAHarvardSmithsonian}}$
  L.~S.~Rocha$^{\ref{AFFIL::BrazilIFUSaoPaulo}}$
  G.~Rodriguez~Fernandez$^{\ref{AFFIL::ItalyINFNRomaTorVergata}}$
  M.~D.~Rodr{\'\i}guez~Fr{\'\i}as$^{\ref{AFFIL::SpainUAlcala}}$
  J.~Rodriguez$^{\ref{AFFIL::FranceCEAIRFUDAp}}$
  J.~J.~Rodr{\'\i}guez-V\'azquez$^{\ref{AFFIL::SpainCIEMAT}}$
  P.~Romano$^{\ref{AFFIL::ItalyOBrera}}$
  G.~Romeo$^{\ref{AFFIL::ItalyOCatania}}$
  J.~Rosado$^{\ref{AFFIL::SpainUCMAltasEnergias}}$
  A.~Rosales~de~Leon$^{\ref{AFFIL::FranceLPNHEUSorbonne}}$
  G.~Rowell$^{\ref{AFFIL::AustraliaUAdelaide}}$
  B.~Rudak$^{\ref{AFFIL::PolandNicolausCopernicusAstronomicalCenter}}$
  A.~J.~Ruiter$^{\ref{AFFIL::AustraliaUNewSouthWalesCanberra}}$
  C.~B.~Rulten$^{\ref{AFFIL::UnitedKingdomUDurham}}$
  I.~Sadeh$^{\ref{AFFIL::GermanyDESY}}$
  L.~Saha$^{\ref{AFFIL::USACfAHarvardSmithsonian}}$
  T.~Saito$^{\ref{AFFIL::JapanUTokyoICRR}}$
  H.~Salzmann$^{\ref{AFFIL::GermanyIAAT}}$
  M.~S\'anchez-Conde$^{\ref{AFFIL::SpainIFTUAMCSIC}}$
  P.~Sangiorgi$^{\ref{AFFIL::ItalyIASFPalermo}}$
  H.~Sano$^{\ref{AFFIL::JapanUGifu},\ref{AFFIL::JapanUTokyoICRR}}$
  R.~Santos-Lima$^{\ref{AFFIL::BrazilIAGUSaoPaulo}}$
  V.~Sapienza$^{\ref{AFFIL::ItalyOPalermo},\ref{AFFIL::ItalyUPalermo}}$
  T.~\v{S}ari\'c$^{\ref{AFFIL::CroatiaFESB}}$
  A.~Sarkar$^{\ref{AFFIL::GermanyDESY}}$
  S.~Sarkar$^{\ref{AFFIL::UnitedKingdomUOxford}}$
  F.~G.~Saturni$^{\ref{AFFIL::ItalyORoma}}$
  S.~Savarese$^{\ref{AFFIL::ItalyINAF}}$
  A.~Scherer$^{\ref{AFFIL::ChileUniversidaddeSantiagodeChile}}$
  F.~Schiavone$^{\ref{AFFIL::ItalyUandINFNBari}}$
  P.~Schipani$^{\ref{AFFIL::ItalyOCapodimonte}}$
  B.~Schleicher$^{\ref{AFFIL::GermanyUWurzburg},\ref{AFFIL::SwitzerlandETHZurich}}$
  P.~Schovanek$^{\ref{AFFIL::CzechRepublicFZU}}$
  J.~L.~Schubert$^{\ref{AFFIL::GermanyUDortmundTU}}$
  F.~Schussler$^{\ref{AFFIL::FranceCEAIRFUDPhP}}$
  M.~Seglar~Arroyo$^{\ref{AFFIL::SpainIFAEBIST}}$
  I.~R.~Seitenzahl$^{\ref{AFFIL::AustraliaUNewSouthWalesCanberra}}$
  O.~Sergijenko$^{\ref{AFFIL::UkraineAstObsofUKyiv},\ref{AFFIL::UkraineObsNASUkraine},\ref{AFFIL::PolandAGHCracowSTC}}$
  M.~Servillat$^{\ref{AFFIL::FranceObservatoiredeParis}}$
  V.~Sguera$^{\ref{AFFIL::ItalyOASBologna}}$
  L.~Sidoli$^{\ref{AFFIL::ItalyIASFMilano}}$
  H.~Siejkowski$^{\ref{AFFIL::PolandCYFRONETAGH}}$
  V.~Sliusar$^{\ref{AFFIL::SwitzerlandUGenevaISDC}}$
  A.~Slowikowska$^{\ref{AFFIL::PolandTorunInstituteofAstronomy}}$
  H.~Sol$^{\ref{AFFIL::FranceObservatoiredeParis}}$
  S.~T.~Spencer$^{\ref{AFFIL::GermanyUErlangenECAP},\ref{AFFIL::UnitedKingdomUOxford}}$
  D.~Spiga$^{\ref{AFFIL::ItalyOBrera}}$
  A.~Spolon$^{\ref{AFFIL::ItalyOPadova}}$
  A.~Stamerra$^{\ref{AFFIL::ItalyORoma},\ref{AFFIL::ItalyCTAOBologna}}$
  S.~Stani\v{c}$^{\ref{AFFIL::SloveniaUNovaGoricaCAC}}$
  T.~Starecki$^{\ref{AFFIL::PolandWUTElectronics}}$
  R.~Starling$^{\ref{AFFIL::UnitedKingdomULeicester}}$
  {\L}.~Stawarz$^{\ref{AFFIL::PolandUJagiellonian}}$
  S.~Steinmassl$^{\ref{AFFIL::GermanyMPIK}}$
  C.~Steppa$^{\ref{AFFIL::GermanyUPotsdam}}$
  T.~Stolarczyk$^{\ref{AFFIL::FranceCEAIRFUDAp}}$
  J.~Stri\v{s}kovi\'c$^{\ref{AFFIL::CroatiaUOsijek}}$
  Y.~Suda$^{\ref{AFFIL::JapanUHiroshima}}$
  T.~Suomij\"arvi$^{\ref{AFFIL::FranceIJCLab}}$
  D.~Tak$^{\ref{AFFIL::GermanyDESY}}$
  M.~Takahashi$^{\ref{AFFIL::JapanUNagoyaISEE}}$
  R.~Takeishi$^{\ref{AFFIL::JapanUTokyoICRR}}$
  S.~J.~Tanaka$^{\ref{AFFIL::JapanUAoyamaGakuin}}$
  F.~Tavecchio$^{\ref{AFFIL::ItalyOBrera}}$
  T.~Tavernier$^{\ref{AFFIL::CzechRepublicFZU}}$
  A.~Taylor$^{\ref{AFFIL::GermanyDESY}}$
  L.~A.~Tejedor$^{\ref{AFFIL::SpainUCMAltasEnergias}}$
  K.~Terauchi$^{\ref{AFFIL::JapanUKyotoPhysicsandAstronomy}}$
  R.~Terrier$^{\ref{AFFIL::FranceAPCUParisCite}}$
  M.~Teshima$^{\ref{AFFIL::GermanyMPP}}$
  V.~Testa$^{\ref{AFFIL::ItalyORoma}}$
  W.~W.~Tian$^{\ref{AFFIL::JapanUTokyoICRR}}$
  L.~Tibaldo$^{\ref{AFFIL::FranceIRAPUToulouse}}$
  O.~Tibolla$^{\ref{AFFIL::UnitedKingdomUDurham}}$
  F.~Tombesi$^{\ref{AFFIL::ItalyINFNRomaTorVergata},\ref{AFFIL::ItalyORoma}}$
  D.~Tonev$^{\ref{AFFIL::BulgariaINRNEBAS}}$
  D.~F.~Torres$^{\ref{AFFIL::SpainICECSIC}}$
  G.~Tosti$^{\ref{AFFIL::ItalyOBrera},\ref{AFFIL::ItalyUPerugiaandINFN}}$
  N.~Tothill$^{\ref{AFFIL::AustraliaUWesternSydney}}$
  F.~Toussenel$^{\ref{AFFIL::FranceLPNHEUSorbonne}}$
  G.~Tovmassian$^{\ref{AFFIL::MexicoUNAMMexico},\ref{AFFIL::ItalyOBrera}}$
  A.~Tramacere$^{\ref{AFFIL::SwitzerlandUGenevaISDC}}$
  P.~Travnicek$^{\ref{AFFIL::CzechRepublicFZU}}$
  A.~Trois$^{\ref{AFFIL::ItalyINAFCagliari}}$
  S.~Truzzi$^{\ref{AFFIL::ItalyUSienaandINFN}}$
  A.~Tutone$^{\ref{AFFIL::ItalyIASFPalermo}}$
  L.~Vaclavek$^{\ref{AFFIL::CzechRepublicUOlomouc},\ref{AFFIL::CzechRepublicFZU}}$
  M.~Vacula$^{\ref{AFFIL::CzechRepublicUOlomouc},\ref{AFFIL::CzechRepublicFZU}}$
  P.~Vallania$^{\ref{AFFIL::ItalyINFNTorino},\ref{AFFIL::ItalyOTorino}}$
  C.~van~Eldik$^{\ref{AFFIL::GermanyUErlangenECAP}}$
  J.~van~Scherpenberg$^{\ref{AFFIL::GermanyMPP}}$
  J.~Vandenbroucke$^{\ref{AFFIL::USAUWisconsin}}$
  V.~Vassiliev$^{\ref{AFFIL::USAUCLA}}$
  M.~V\'azquez~Acosta$^{\ref{AFFIL::SpainIAC}}$
  M.~Vecchi$^{\ref{AFFIL::NetherlandsUGroningen}}$
  S.~Ventura$^{\ref{AFFIL::ItalyINFNPisa}}$
  S.~Vercellone$^{\ref{AFFIL::ItalyOBrera}}$
  G.~Verna$^{\ref{AFFIL::ItalyUSienaandINFN}}$
  I.~Viale$^{\ref{AFFIL::ItalyUPadovaandINFN}}$
  A.~Viana$^{\ref{AFFIL::BrazilIFSCUSaoPaulo}}$
  N.~Viaux$^{\ref{AFFIL::ChileDepFisUTecnicaFedericoSantaMaria}}$
  A.~Vigliano$^{\ref{AFFIL::ItalyUUdineandINFNTrieste}}$
  J.~Vignatti$^{\ref{AFFIL::ChileDepFisUTecnicaFedericoSantaMaria}}$
  C.~F.~Vigorito$^{\ref{AFFIL::ItalyINFNTorino},\ref{AFFIL::ItalyUTorino}}$
  J.~Villanueva$^{\ref{AFFIL::ChileUdeValparaiso}}$
  E.~Visentin$^{\ref{AFFIL::ItalyINFNTorino},\ref{AFFIL::ItalyUTorino}}$
  V.~Vitale$^{\ref{AFFIL::ItalyINFNRomaTorVergata}}$
  V.~Voisin$^{\ref{AFFIL::FranceLPNHEUSorbonne}}$
  V.~Voitsekhovskyi$^{\ref{AFFIL::SwitzerlandUGenevaDPNC}}$
  S.~Vorobiov$^{\ref{AFFIL::SloveniaUNovaGoricaCAC}}$
  G.~Voutsinas$^{\ref{AFFIL::SwitzerlandUGenevaDPNC}}$
  I.~Vovk$^{\ref{AFFIL::JapanUTokyoICRR}}$
  T.~Vuillaume$^{\ref{AFFIL::FranceLAPPUSavoieMontBlanc}}$
  S.~J.~Wagner$^{\ref{AFFIL::GermanyLSW}}$
  R.~Walter$^{\ref{AFFIL::SwitzerlandUGenevaISDC}}$
  M.~Wechakama$^{\ref{AFFIL::ThailandUKasetsart},\ref{AFFIL::ThailandNARIT}}$
  M.~White$^{\ref{AFFIL::AustraliaUAdelaide}}$
  A.~Wierzcholska$^{\ref{AFFIL::PolandIFJ}}$
  M.~Will$^{\ref{AFFIL::GermanyMPP}}$
  F.~Wohlleben$^{\ref{AFFIL::GermanyMPIK}}$
  T.~Yamamoto$^{\ref{AFFIL::JapanUKonan}}$
  R.~Yamazaki$^{\ref{AFFIL::JapanUAoyamaGakuin}}$
  L.~Yang$^{\ref{AFFIL::SouthAfricaUJohannesburg},\ref{AFFIL::ChinaUSunYatsen}}$
  T.~Yoshikoshi$^{\ref{AFFIL::JapanUTokyoICRR}}$
  M.~Zacharias$^{\ref{AFFIL::GermanyLSW},\ref{AFFIL::SouthAfricaNWU}}$
  G.~Zaharijas$^{\ref{AFFIL::SloveniaUNovaGoricaCAC}}$
  L.~Zampieri$^{\ref{AFFIL::ItalyOPadova}}$
  D.~Zavrtanik$^{\ref{AFFIL::SloveniaUNovaGoricaCAC}}$
  M.~Zavrtanik$^{\ref{AFFIL::SloveniaUNovaGoricaCAC}}$
  A.~A.~Zdziarski$^{\ref{AFFIL::PolandNicolausCopernicusAstronomicalCenter}}$
  W.~Zhang$^{\ref{AFFIL::SpainICECSIC}}$
  V.~I.~Zhdanov$^{\ref{AFFIL::UkraineAstObsofUKyiv}}$
  K.~Zi\k{e}tara$^{\ref{AFFIL::PolandUJagiellonian}}$
  M.~\v{Z}ivec$^{\ref{AFFIL::SloveniaUNovaGoricaCAC}}$
  J.~Zuriaga-Puig$^{\ref{AFFIL::SpainIFTUAMCSIC}}$
\newline\newline
\emph{Affiliations can be found at the end of the article}}}
\vspace{2.5cm}
\newpage

\date{Accepted XXX. Received YYY; in original form ZZZ}

\pubyear{2024}

\def \diff_flx {ph $\mathrm{cm}^{-2}$ $\mathrm{s}^{-1}$ $\mathrm{MeV}^{-1}$}

\begin{document}
\linenumbers %
\label{firstpage}
\pagerange{\pageref{firstpage}--\pageref{lastpage}}
\maketitle

\footnotetext[1]{\url{aloramas@iac.es} (A.~L\'opez-Oramas)\label{CONTACTAOUTHOR::1}}

\begin{abstract}
  A wide variety of Galactic sources show transient emission at soft and hard X-ray energies: low-mass and high-mass X-ray binaries containing compact objects
  , isolated neutron stars exhibiting extreme variability as magnetars as well as pulsar wind nebulae. Although most of them can show emission up to MeV and/or GeV energies, many have not yet been detected in the TeV domain by Imaging {Atmospheric} Cherenkov Telescopes. In this paper, we explore the feasibility of detecting new Galactic transients with the Cherenkov Telescope Array {Observatory} (CTAO) and the prospects for studying them with Target of Opportunity observations. We {show} that CTAO will likely detect new sources in the TeV regime, such as the massive microquasars in the Cygnus region, low-mass X-ray binaries with low-viewing angle, flaring emission from the Crab pulsar-wind nebula or {other} novae explosions, among others. {Since some of these  sources could also exhibit emission at larger timescales, we additionally test their detectability at longer exposures}. We {finally} discuss the multi-wavelength synergies with other instruments and large astronomical facilities.
\end{abstract}




\section{Introduction}


Timing astronomy and variability studies have proven to be a powerful tool to study extreme astrophysical processes at very high energies (VHE, E$>$ 100 GeV). The improvement of the Imaging Atmospheric Cherenkov Technique (IACT) over the past decade has revealed new transient phenomena with variability timescales from seconds to several weeks. The last generation of IACTs have discovered several classes of transient TeV sources such as gamma-ray bursts (GRBs) \citep{2019Natur.575..455M, 2019Natur.575..464A}, flaring blazars associated with high-energy neutrino sources \citep{2018Sci...361.1378I} or Galactic novae \citep{2022Sci...376...77H,2022NatAs...6..689A}, among others, unveiling new types of VHE emitters with highly variable fluxes \citep{2024Univ...10..163C}.

The Cherenkov Telescope Array Observatory (CTAO) will be the next generation ground-based observatory for VHE astronomy. It will allow the detection of gamma rays in the 20 GeV-300 TeV domain, with two observatory sites, one in the Northern hemisphere (CTAO-N; Observatorio Roque de los Muchachos, La Palma, Spain) and another in the Southern one (CTAO-S; Paranal, Chile). It will provide an improved sensitivity with respect to the current generation of IACTs of {of about} an order of magnitude \citep{2019sCTA.book.....C}. Of special importance will be the sensitivity of CTAO to short-timescale phenomena\footnote{For CTAO performance, see: \url{https://www.CTAO.org/for-scientists/performance/}}. 

CTAO will have 10$^4$-10$^5$ better sensitivity than the LAT instrument onboard the \textit{Fermi} satellite for the detection of short-duration transient events {\citep{2013APh....43..348F}}. 

The low energy threshold of $\sim$20 GeV of the largest telescopes of the array, the {Large-Sized} Telescopes (LSTs; \citealp{2023arXiv230612960P}) is key for the detection of new transient sources at the lower end of the VHE regime. This capability, together with the fast slewing response of the LSTs, which can be re-pointed in about 20 seconds, will allow a swift reaction to transient events. {The Medium and Small Sized telescopes (MSTs and SSTs) will also be key to understand the emission of this sources at higher energies. Finally, since the CTAO observatory will consist of two arrays located in two hemispheres, it will provide a better and more continuous coverage of many transient events accessible from both sites.}


The core program of CTAO will consist of 
different \textit{Key Science Projects} (KSPs) which {were considered to}
address the science questions of CTAO (see \citealp{2019sCTA.book.....C} for more details). The \textit{Transients} KSP {is proposed to encompass} the follow-up observations of several classes of targets such as GRBs, gravitational waves (GWs), high-energy neutrinos, core-collapse supernovae (CCSNe) and Galactic transients.  


In this paper, we focus on {Galactic} sources hosting compact objects whose emission is not periodic and{/or} that display unexpected flaring events, outflows or jets as described in {the Galactic transients section of the Transients KSP as defined in} \cite{2019sCTA.book.....C}. {Extragalactic transient events such as GRBs, core-collapse SNe or GWs will be addressed in separate publications.} We discuss the capabilities of CTAO to detect new transient phenomena {at VHE} from sources of Galactic origin, ranging from microquasars, to pulsar-wind nebulae (PWNe) flares, to novae, transitional millisecond pulsars or magnetars among others.  {Some of these sources could also exhibit persistent emission, hence we additionally test the detectability at longer exposures in some specific cases.}  {Since the nature of the source classes of study and hence the physical processes are different, the simulated timescales of the expected VHE emission also vary. We have used timescales ranging from as low as 10 min to few hours for transient detection and up to 50-200 h to test persistent emission in certain sources of interest.} For our simulations, we have used the software packages \verb+ctools+ 
\VerbatimFootnotes
\footnote{\verb+ctools+ is a software specifically developed for the scientific analysis of gamma-ray data, see \url{http://CTAO.irap.omp.eu/ctools/index.html}} \citep{2016A&A...593A...1K} and \verb+Gammapy+\VerbatimFootnotes
\footnote{ \verb|Gammapy| is an open-source Python package developed for gamma-ray astronomy, see \url{https://gammapy.org/} } \citep{gammapy:2023,gammapy:zenodo-1.1} with the official CTAO observatory instrument response functions (IRFs).\footnote{The IRF version \verb+prod3b-v2+ is the one used throughout the manuscript, unless otherwise specified. 
The newest \verb+prod5+ version corresponding to the \textit{Alpha Configuration}, {which corresponds to the  first stage of CTAO observatory construction} has also been tested in some science cases and are specified in the text.} For a full description of CTAO observatory IRFs and configurations see \cite{maier_g_2023_8050921}.

{The source {classes} of our interest are described in the following Subsections \ref{intro_mic}-\ref{intro_mag}. We present the sensitivity of CTAO to Galactic transient detection in Section \ref{sec:sensitivity} and population studies in Section \ref{sec:population}. The simulations, analysis results and discussion for each type of transient are collected in Section \ref{sec:simulations}. }Section \ref{sec:synergies} {describes} the synergies with multi-wavelength and multi-messenger astronomical facilities. The summary and final conclusions are listed in Section \ref{sec:summary}.

\subsection{Microquasars} \label{intro_mic} 
Microquasars are binary systems with a compact object (NS or a BH) orbiting around and  accreting material from a companion star. The matter lost from the star can lead to formation of an accretion disk around the compact object and a relativistic collimated jet {\citep{Mirabel1998Natur.392..673M}}. 

At the moment more than 20 microquasars are known in the Galaxy {(see i.e. \citealp{Corral2016A&A...587A..61C})}.  
Observations demonstrated correlations between the mass of the compact object,  radio (5 GHz) and X-ray (2–10 keV) luminosities \citep[e.g.][]{Falcke2004}, strengthening the link between active galactic nuclei (AGNs) and microquasars.  In AGNs, jets are known to be places of efficient particle acceleration and produce broad band non-thermal emission. The resulting radiation can extend from radio up to the VHE band. According to TeVCat \footnote{\href{http://tevcat2.uchicago.edu/}{http://tevcat2.uchicago.edu/}} more than 65 AGNs have been already detected by current IACTs. If similar jet production and particle acceleration mechanisms operate in microquasars and AGNs, this might imply that microquasars should be sources of  VHE $\gamma$-ray emission as well.

Up to now, only three microquasars have been detected in the high-energy (HE, E$>$100\,MeV) domain: Cygnus X-1 \citep{Bulgarelli2010, Sabatini2010, Sabatini2013, Malyshev2013, Zanin2016,Zdziarski2017}, Cygnus X-3 \citep{Tavani2009,Fermi-LAT2009,Zdziarski2018},  and SS 433 
\citep{SS433_bordas15,SS433_extFermi19,SS433_varFermi19,SS433_Fermi19,2020NatAs...4.1177L}, {all of them hosting a massive companion star.} {In the case of low-mass microquasars, the only one that has displayed a strong hint of gamma-ray emission (at high energies) was }the binary V404 Cyg during its 2015 outburst \citep{Loh2016,Piano2017}.
{Steady VHE emission was first detected from the interaction regions between the jet and the surrounding nebula for the time in SS 433 \citep{SS433_HAWC,SS433_2024}. Other microquasars have been recently reported to be sources of persistent of TeV and PeV emission \citep[for details see i.e.][]{2024Natur.634..557A,LHAASO2024arXiv241008988L}.
Regarding flaring emission, the strongest hint reported was the} 4.1$\sigma$ transient signal (post-trial) found at VHE in Cygnus X-1 \citep{Albert2007}.The expectations for the detection of both {massive microquasars and low-mass X-ray binaries (LMXBs) with CTAO are presented in Sections \ref{massive_micro}-\ref{CTAO_lmxbs}.}

{The relevance for studying binary systems in the VHE regime has already been addressed {by} \citet{2013APh....43..301P,2019A&A...631A.177C}.} In this paper, we do not focus on gamma-ray binaries displaying periodic orbital variability {and likely powered by non-accreting pulsars}, but only on {systems powered by accretion and displaying jets}, to better investigate the potential VHE {emission} of this {specific} class of  binaries.
{We discuss high-mass microquasars (Section \ref{massive_micro}) separately to low-mass X-ray binaries (Section \ref{CTAO_lmxbs}).}

\subsection{Transitional millisecond pulsars}\label{intro_tmsp} 
Transitional millisecond pulsars (tMSPs) are a class of neutron star binaries that has emerged in the last decade with the discoveries of three confirmed systems: PSR J1023+0038 
\citep{2009Sci...324.1411A,2014ApJ...781L...3P}, XSS J1227-4853 \citep{deMartino10_tMSP, bassa14_tMSP} and IGR J1824-2452 in the globular cluster M28 \citep{Papitto13_tMSP}. 
{Additionally, a handful of candidate tMSPs
have been recently discovered in the X-ray and GeV ranges (see review by \citealp{2022ASSL..465..157P}). tMSPs alternate between a radio-loud MSP state (RMSP, showing radio pulsations and no sign of an accretion disk) and a sub-luminous LMXB state (forming an accretion disk and showing X-ray pulsations).}
These sources are the direct link between the LMXB and the radio MSP phases {in} which neutron stars are spun up to ms periods during the LMXB-phase. 
Sudden transitions between the two states occur on a timescale of a few days to weeks, and are accompanied by drastic changes across the electromagnetic spectrum. The transition from the RMSP to LMXB state is accompanied by brightening of optical, UV \citep{Patruno14, Takata14}, X-ray and gamma-ray {\citep{Stappers14}} 
emission with the disappearance of radio pulsations. The origin of these transitions is still debated and, for this, intense multi-wavelength campaigns are on-going to understand the phenomenology in both the RMSP and LMXB states. {tMSPs were so far not detected in the VHE regime.} {The constrains to the VHE emission from tMSPs during the LMXB state} are discussed in Section \ref{CTAO_tmsps}.


\subsection{Pulsar wind nebulae}\label{intro_pwn} 
Pulsar wind nebulae (PWNe) are bubbles or diffuse structures of relativistic plasma powered by a central highly-magnetized rotating neutron star. {They represent one of the largest Galactic populations at VHE}. Recently, several PWNe have been suggested to be PeV particle (leptons) accelerators, with the detection of gamma rays at E$>$100 TeV \citep{2021Natur.594...33C,2023arXiv230517030C}. 
The Crab nebula is the standard candle at VHE and both the nebula and the pulsar have been {intensively} studied. Pulsations have been measured up to TeV energies \citep{2016A&A...585A.133A} and the nebula spectrum has been detected up to 100 TeV by IACTs \citep{2020A&A...635A.158M} and recently extended to PeV \citep{2021Sci...373..425L}. Unexpectedly, the Crab nebula displays rapid flaring emission over daily timescales at HE
as reported by AGILE and \textit{Fermi}-LAT \citep{2011Sci...331..736T,2011Sci...331..739A}. The enhanced fluxes measured over different flaring episodes were a factor three-to-six times larger than the standard flux. {These episodes of enhanced HE emission have been detected up to 10 GeV (as reported by \citealp{2011Sci...331..736T}) and can last up to few weeks}. {A detection of the synchrotron tail at higher energies or a additional inverse Compont component in the TeV domain could be expected.} So far, no signs of variability have been reported at VHE \citep{2010ATel.2967....1M,2010ATel.2968....1O,2014A&A...562L...4H,2019ICRC...36..812V}. The characterization of the expected VHE emission to be {putatively} detected by CTAO is shown in Section \ref{CTAO_pwne}.

\subsection{Novae}\label{intro_nov} 
Novae are thermonuclear runaway explosions on the surface of a white dwarf star in binary systems involving a white dwarf accreting matter, often through an accretion disc, usually from 
a late-type star \citep{gallagher1978a}. They are detected as transient events exhibiting huge and sudden increase of brightness. Though novae have been studied both observationally and theoretically for many decades, a comprehensive understanding of nova physics is still lacking {\citep{Iben1982a,Yaron2005a,2008clno.book.....B,2014ApJ...793..136K,chomiuk2021a}}. 
Particle acceleration in novae was predicted before the launch of the Fermi Gamma-ray space telescope \citep[see][]{2007ApJ...663L.101Tatischeff}. Shortly after, 
GeV emission 
from the outburst of the symbiotic binary system V407 Cygni, comprised of a white dwarf and an evolved red giant companion, was first detected.
Subsequently, classical novae with main sequence donor stars were also detected \citep{2010Sci...329..817Abdo,2014Sci...345..554Ackermann}.\footnote{\cite{gomez-gomar1998a} (and references therein) predicted gamma-ray emission from novae but of nuclear origin, in the keV-MeV domain. }
More recently, VHE emission in novae has been {predicted} and searched for in a handful of sources \citep[see e.g.][]{2012ApJ...754...77A,Ahnen2015a}, with the first detection at VHE gamma rays occurring in 2021
in the recurrent nova RS Ophiuchi \citep[RS Oph,][]{2022NatAs...6..689A,2022Sci...376...77H,2025A&A...695A.152A}.  

Since the first detection at 
HE gamma rays from nova Cygni 2010, 19 novae\footnote{According to \href{https://asd.gsfc.nasa.gov/Koji.Mukai/novae/latnovae.html}{https://asd.gsfc.nasa.gov/Koji.Mukai/novae/latnovae.html} (as of February 2024).} have been detected 
in this energy band (only RS Oph at VHE) with a rate of about 
one outburst detection per year. All novae so far detected at HE have been bright 
in the visible band ($\leq 10\,\rm{mag}$), and the vast majority are nearby sources with distances within $5\,\rm{kpc}$ \citep{2018A&A...609A.120F}. Non-thermal emission is expected to arise from leptonic and hadronic interactions by particles accelerated in radiative expanding shocks 
\citep{2010Sci...329..817Abdo,2012BaltA..21...62H}, which can originate from the interaction of the ejecta during the initial stage of the outburst and the circumbinary material, {or with} the fast wind produced by
nuclear burning in later stages of the outburst \citep{2010Sci...329..817Abdo,2014Sci...345..554Ackermann,2018A&A...612A..38Martin}. {The VHE signal reported by \citealp{2022NatAs...6..689A,2022Sci...376...77H,2025A&A...695A.152A} is suggested to be of hadronic origin, due to protons accelerated in the nova shock.}

Based on observations of novae in the nearby M31 Galaxy, as well as binary population synthesis models for the Milky Way, a rate of approximately 30 nova events per year is expected (see Section \ref{CTAO_novae}). However, a significant proportion of these will be obscured 
by intervening dust in the Galactic plane, {preventing multi-wavelength follow-up observations}. The number of nova events that will be detectable at HE and VHE gamma rays will be further constrained by properties of the system, such as 
the shock velocity and the target material density. This dependence on the parameter space and {prospects for detection} of novae at VHE will be 
characterised for CTAO in Section \ref{CTAO_novae}.



\subsection{Magnetars} \label{intro_mag} 


Magnetars are isolated neutron stars in which the main energy source is the magnetic field (e.g. \citealt{2015SSRv..191..315M,2017ARA&A..55..261K}, for reviews). They are observed as pulsed X-ray sources, with typical spin periods of a few seconds and strong spin-down rates {(typically 10$^{-12}$--10$^{-10}$~s~s$^{-1}$) \citep{1999ApJ...525L.125H}}, and/or through the detection of short bursts and flares in the hard X-ray/soft gamma-ray range. This led to their historical subdivision in the Anomalous X-ray Pulsars and Soft Gamma-ray Repeater classes \citep{2008A&ARv..15..225M}, but it is now clear that these are just two different manifestations of the same underlying object: a strongly magnetized neutron star powered by magnetic energy, as   proposed by \citet{1992AcA....42..145P} and  \citet{1992ApJ...392L...9D}. 

About 30 magnetars are known so far. With the exception of two sources in the Magellanic Clouds, all of them lie in the Galactic Plane. The majority of the  magnetars are transient X-ray sources that have been discovered when they became active, with an increase of their X-ray luminosity (from a quiescent level of $\sim10^{33}$ erg s$^{-1}$ up to $\sim10^{36-37}$ erg s$^{-1}$), accompanied by the emission of luminous and rapid bursts. This means that the total Galactic population of magnetars is larger than the currently observed sample, and more sources of this class will be known at the time of CTAO observations. Furthermore, magnetar-like behavior has recently been observed in some sources originally presumed to be of a different kind, such as rotation-powered (radio) pulsars \citep{2008Sci...319.1802G,2016ApJ...829L..25G}, and even in the gamma-ray binary LS I +61 303 \citep{2012ApJ...744..106T,2022NatAs...6..698W}.  

For what concerns the persistent emission, magnetars have not been detected above few hundred keV \citep{2010ApJ...725L..73A,2013A&A...549A..23A}. Their X-ray emission typically comprises a  soft thermal component that dominates in the 1-10 keV range and a hard power-law component that is believed to originate from multiple resonant scattering in the magnetosphere. The {upper limits (ULs) in the MeV range \citep{2017ApJ...835...30L} indicate a turn-off of this component implying that {their detectability is} below the CTAO capabilities, unless a different spectral component is present at higher energies. 
On the other hand, magnetar bursts and flares (in particular the so called Giant Flares) are potentially very interesting targets for CTAO, with the only disadvantage of their unprediCTAOble time of occurrence. Giant flares are extremely energetic and bright events, reaching isotropic peak luminosities as high as a few 10$^{47}$ erg s$^{-1}$ for a fraction of a second. However, they occur very rarely: only three have been seen from local magnetars in 40 years. The high luminosity of their short ($<$ 1 s) initial peaks implies that they can be detected, with properties resembling those of short GRBs,  up to distances of tens of Mpc by current hard X-ray instruments. Indeed, a few candidate extragalactic giant flares have been identified \citep{2008ApJ...680..545M, 2007AstL...33...19F,2021Natur.589..211S,2021Natur.589..207R}. {Of particular interest regarding CTAO's perspective to detect giant flares} is the case of the flare located in the Sculptor galaxy (NGC 253, at 3.5 Mpc) for which 
\textit{Fermi}-LAT  observation led to the detection of two high-energy photons with energies of 1.3 GeV and 1.7 GeV, likely produced via synchrotron mechanism \citep{2021NatAs.tmp...11F}. However, no emission from a magnetar has been yet detected at TeV energies \citep{2021ApJ...919..106A,Lopez-Oramas:2021zd}. For further discussion, see Section \ref{CTAO_mag}.

\section{Sensitivity of CTAO to transient detection in the Galactic plane} \label{sec:sensitivity}

%
\begin{figure*} 
    \begin{center}
        \includegraphics[trim=0 50mm 0 0,clip,width=\textwidth]{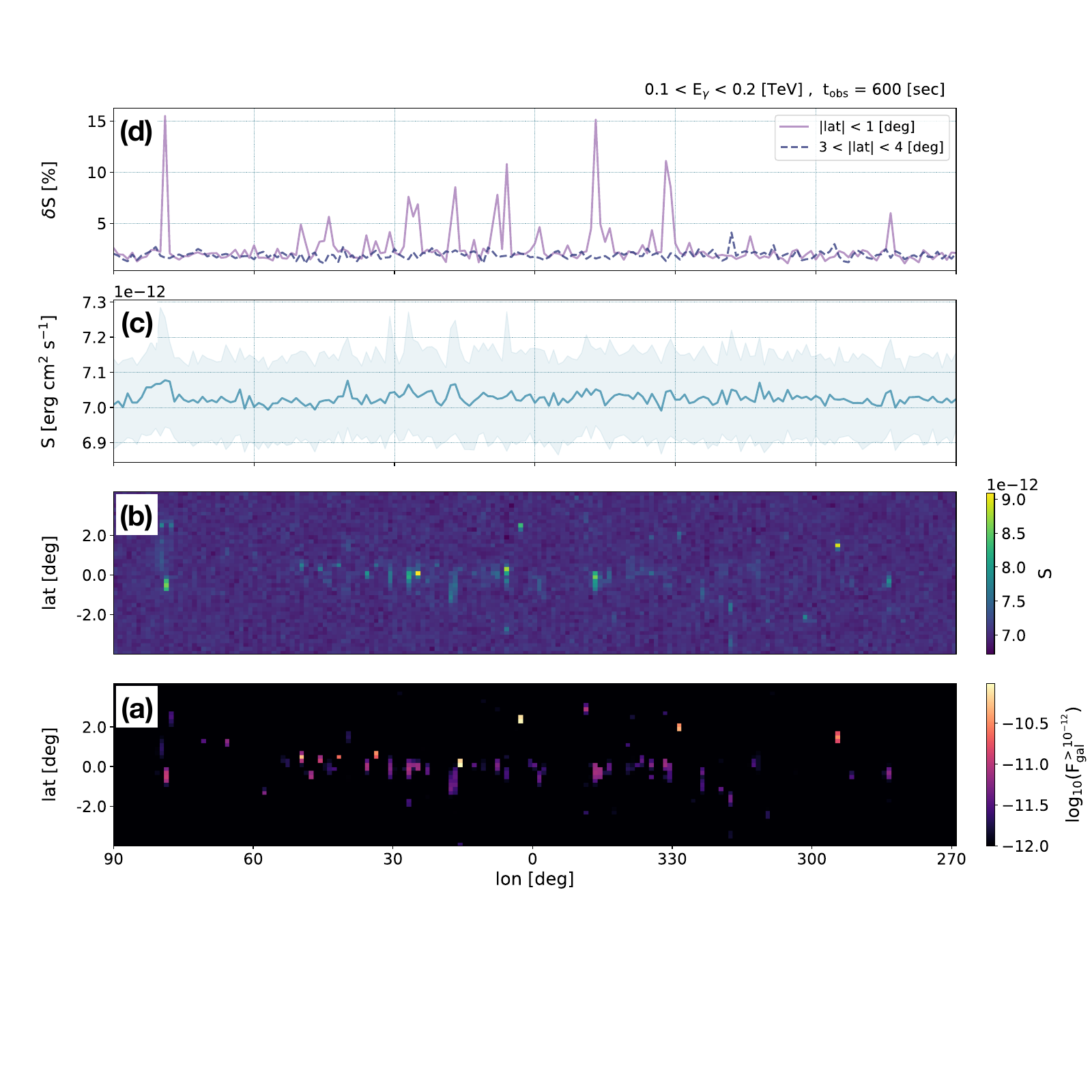}

        \caption{ Differential flux sensitivity, $S$, of the Southern CTAO array within 100--200~GeV for 10~min observation intervals, considering different putative source locations along the Galactic plane. 
        \textit{Panel (a)} shows a simulation of $F_{\text{gal}}^{>10^{-12}}$, 
        the Galactic emission {(Galactic diffuse emission and a simulated population of Galactic sources)} above a threshold of ${10^{-12}~\text{erg}\;\text{cm}^2\;\text{s}^{-1}}$, which is derived
        for different Galactic longitudes, \textit{lon}, and latitudes, \textit{lat}. 
        \textit{Panel (b)} shows the corresponding CTAO sensitivity. \textit{In panel (c)} we present the median of $S$ for different longitudes within the range, $-4 < \text{lat} < +4$~deg, where the shaded uncertainty region represents the $1\sigma$ variance of $S$. Finally, 
        \textit{panel (d)} shows the relative $1\sigma$ variance, $\delta{S}$,  (compared to the median) derived for two ranges in latitude, as indicated. 
        The variance away from the Galactic {plane} ($3 < |\text{lat}| < 4$~deg)
        represents the intrinsic statistical uncertainty of the sensitivity calculation.
        The variance in the inner Galactic region ($|\text{lat}| < 1$~deg) includes the intrinsic uncertainty, as well as the additional effect of the Galactic foregrounds, which are concentrated in this region. }
        \label{fig:CTAOsensitivityMap}
    \end{center}
\end{figure*}

CTAO will have unprecedented sensitivity over a broad energy range and will devote a large amount of time to sources in the Galactic plane, both with a dedicated Galactic Plane Survey (GPS) {(for details on the pointing strategy and expected results see \citealp{GPS2023arXiv231002828C})} and with pointed observations on specific targets. These are ideal capabilities for the discovery of new Galactic transients at TeV energies.

\begin{figure*} 
    \begin{minipage}[c]{1\textwidth}
        \begin{minipage}[c]{0.49\textwidth}
            \begin{center}
                \includegraphics[trim=6mm 30mm 18mm     0mm,clip,width=1\textwidth]{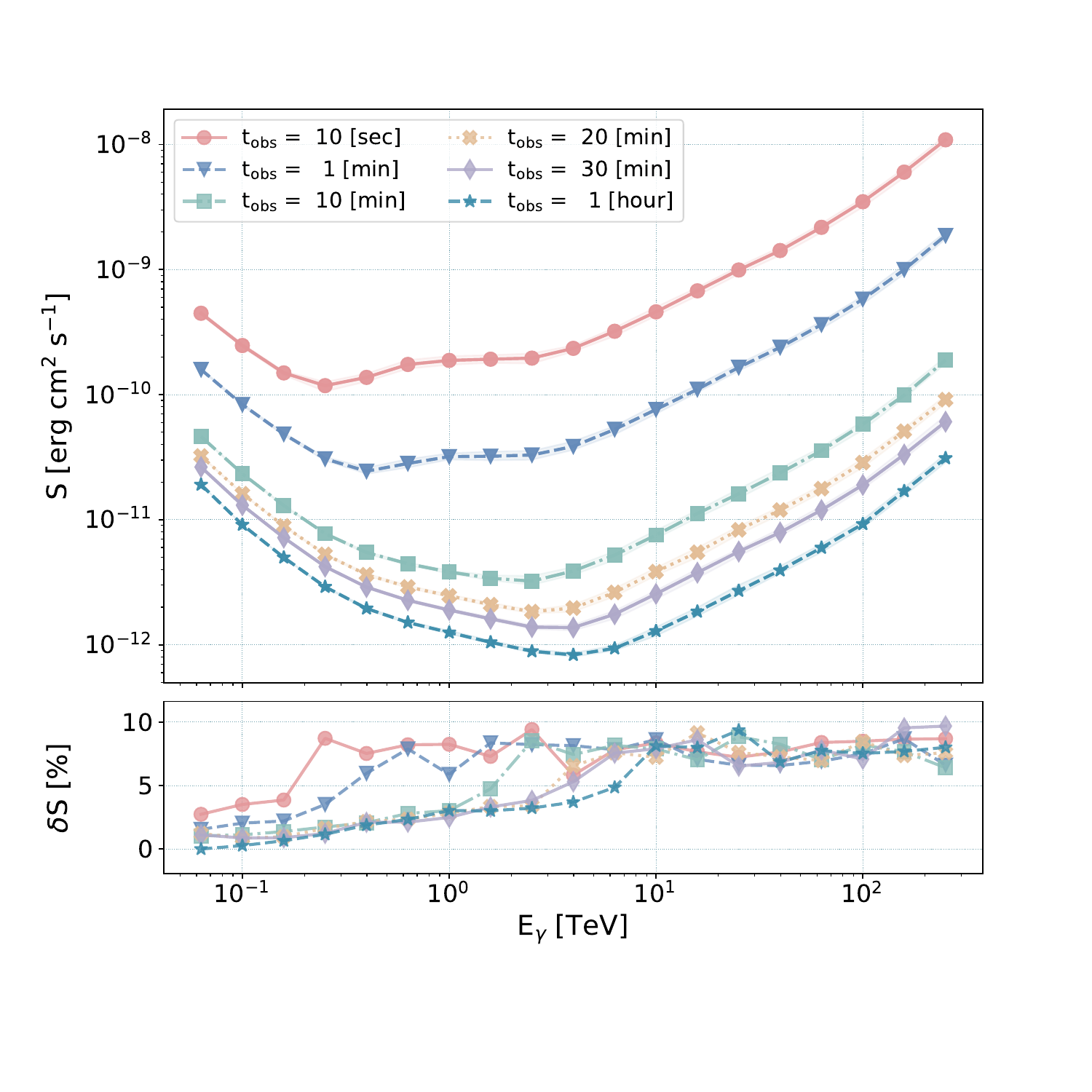}
            \end{center}
        \end{minipage}\hfill
        \begin{minipage}[c]{0.49\textwidth}
            \begin{center}
                \includegraphics[trim=6mm 30mm 18mm     0mm,clip,width=1\textwidth]{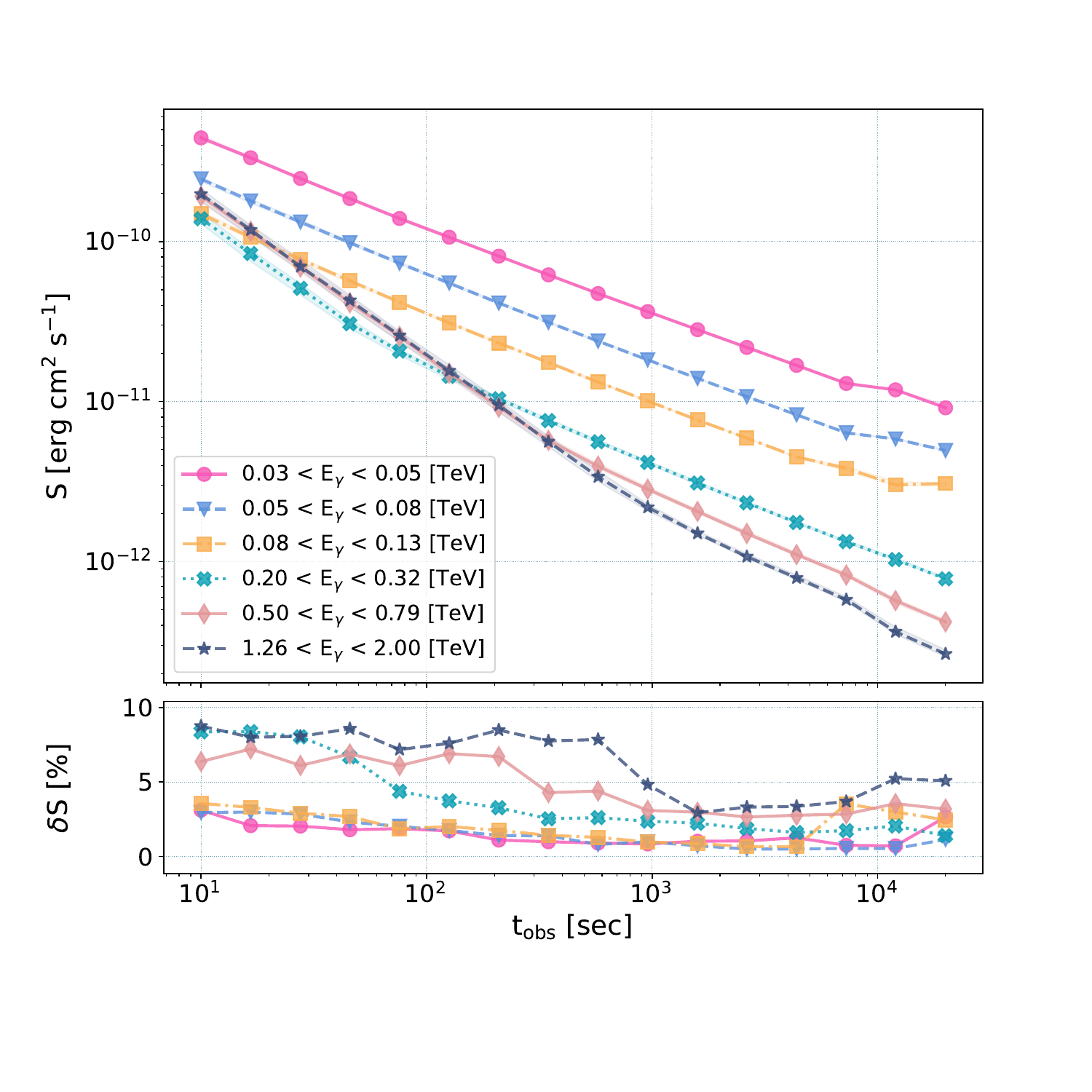}
            \end{center}
        \end{minipage}\hfill
        \begin{minipage}[c]{1\textwidth}
            \begin{center}
                \caption{\label{fig:sensitivityEnergiesTimes} The differential flux sensitivity, $S$, of the Southern CTAO array for different energy ranges, $0.03 < E_{\gamma} < 200~\text{TeV}$, and observation intervals, $10 < t_{\text{obs}} < 2\cdot10^{4}~\text{sec}$. The sensitivity is derived as the median value for various putative source positions, considering an area of ${4~\text{deg}^{2}}$ close to the Galactic centre. The bottom panels show the  relative $1\sigma$ variance of the sensitivity, $\delta{S}$, compared to the median. The variance accounts for both the intrinsic statistical uncertainty of the sensitivity calculation, and the degradation of performance due to the presence of steady sources. }
            \end{center}
        \end{minipage}\hfill

    \end{minipage}\hfill
\end{figure*} 

It is then important to characterize the sensitivity of CTAO in the Galactic plane. The differential sensitivity of CTAO for detecting a new source is defined as the minimal flux of a source, multiplied by the mean energy squared within the given energy interval, such that the source is detectable at the $5\sigma$ significance level. It is defined within a given energy range, and for a given observation interval or exposure. 
We assume a test point source power-law spectral model,
\begin{equation}
F(E)=P_{f}\left(\frac{E}{E_{0}}\right)^{-\gamma} \label{eq:1}
\end{equation}
%

%

where we set the pivot energy, ${E_{0}=1~\mathrm{TeV}}$, and the spectral index, ${\gamma=2.5}$, using {typical} values for {Galactic} VHE sources. 
{The prefactor, P$_{f}$} is varied as part of the sensitivity calculation, in order to find the minimal flux value for a $5\sigma$ detection.
In order to be compatible with previous analyses (e.g. {\citealp{2019ICRC...36..673F}}), we also require that the source emits at least 10~gamma-ray photons. 
In addition, we validate that this number of events is larger than~${5\%}$ of the {corresponding contribution from backgrounds (cosmic rays; electrons) and foregrounds (other coincident gamma-ray sources).} 

We explore the performance of CTAO in the Galactic plane region for various short observation intervals.
For illustration, the sensitivity of the Southern array is shown in Fig.~\ref{fig:CTAOsensitivityMap}, considering different putative source locations.
In this example, we estimated the performance for short observation intervals of 10~minutes within the energy range 100--200~GeV, exploring the detection potential of new sources in the low-energy range of CTAO.
We use a publicly-available Galactic sky model, based on observations of known gamma-ray sources and interstellar emission from cosmic-ray interactions in the Milky Way \citep{GPS2023arXiv231002828C}.\footnote{Galactic sky model available at \url{https://zenodo.org/records/10008527}} We simulate our putative transients on top of the emission derived from this sky model, such that the latter constitutes an additional background to the search.


As may be inferred from Fig.~\ref{fig:CTAOsensitivityMap}, upward fluctuations of the sensitivity (requiring brighter transient emission for detection), are correlated with the steady emission {from Galactic sources}.
In the selected energy range, the flux of the simulated Galactic foreground is mostly below the level of a few ${10^{-11}~\text{erg}\;\text{cm}^{-2}\;\text{s}^{-1}}$.
This is of the same order as the nominal sensitivity of the observatory in the absence of foregrounds. Correspondingly, the overall degradation in sensitivity to transients is not expected to be significant. 

In order to verify this, we calculated $\delta{S}$, the relative variation of the sensitivity (compared to the median value) for different Galactic longitudes.
The steady foreground sources are concentrated in the inner Galactic region. We therefore derived $\delta{S}$ for two regions in latitude, in order to enhance or suppress their effect.
Away from the Galactic {plane} we find $\delta{S}\sim2\text{--3}\%$ which amounts to the intrinsic statistical uncertainty of the sensitivity calculation. In the more crowded inner region, the variation is of the order of $5\text{--}15\%$.
This represents a mild increase in the flux threshold for a new transient source to be discovered, though only when coinciding with strong Galactic emitters.

We show the median sensitivity for various combinations of
energy ranges and observation intervals in Fig.~\ref{fig:sensitivityEnergiesTimes}. Here we consider an area of ${4~\text{deg}^{2}}$ next to the Galactic centre, where the steady emission is relatively strong. The observed variation in sensitivity is mild, of the order of $1\text{--}10\%$.

{As a test, the} presence of possible source variability is assessed in Section \ref{CTAO_micro_ss433}. {Other topics such as the study} of Galactic Centre sources and interstellar emission through observations of the Galactic Centre region and GPS and the prospects for the CTAO and its scientific results are covered in other KSPs (see \citealp{2019sCTA.book.....C,GPS2023arXiv231002828C})

We conclude that the performance of CTAO {in the Galactic plane} is consistent with the corresponding nominal extragalactic sensitivity. {(That is, the sensitivity in the absence of significant emission from other gamma-ray sources.)}


\section{Detectability of transients of unknown origin} \label{sec:population}

Apart from the transient sources of clearly identifiable type, others of unknown nature could also be serendipitously observed e.g. during a scan of the GPS. 
{The detailed study of serendipity and corresponding observational strategies for CTAO will be addressed in a dedicated separate publication. However, to assess capabilities of CTAO for the detection of Galactic transient sources of unknown origin, a population of generic sources can be used. A full study of such populations requires considering various models of sources, population sizes and observational setups, and therefore will be presented separately. Here, we illustrate the methodology with a specific, simplified example.} We simulate the populations of 100 generic transients. We consider the relatively short observation time of 1 hour {(compatible with the strategy defined in \citealp{GPS2023arXiv231002828C})} during which it would be possible to detect the source and make a decision about further observations. 


We simulated the variability of each source using the following lightcurve model:
\begin{equation}
F(t)=\frac{2F_0}{\exp(\frac{t_0-t}{T_r})+\exp(\frac{t-t_0}{T_d})}\nonumber
\end{equation}
where $T_r$ and $T_d$ stand for time rise and time decrease and $t_0$ is the time at which $F=F_0$; we normalize the lightcurve to $F=1$ at its maximum. Such a lightcurve describes the flux of a transient during its growth, at peak and when it falls, allowing the simulation of observations at each of these stages.
We assume $T_r$, $T_d$ and $t_0$ to be in ranges $1-86400$ s, $86400-604800$ s and $T_d - T_r$ respectively.

We used the model of \cite{2004A&A...422..545Y} for the radial distribution
of sources. 
For simplicity, we did not take into account the visibility constraints, assuming that all sources are visible to {either} array at the time of observation.



For each population, the parameters defining the spectrum and the lightcurve of each source are assigned randomly for each of them, 
assuming a log-uniform distribution for the prefactor and a uniform one for other parameters. The pivot energy for all sources is 1 TeV.

\begin{table}
\caption{Simulated populations. We consider sources with different spectral shapes and parameters.}
\label{sim_pops}
\begin{center}
\begin{small}
	\begin{tabular*}{\columnwidth}{cccc}
 \hline
    Population     & Spectrum   & Prefactor    & Spectral index    \\ \
  &     &     (ph cm$^{-2}$ s$^{-1}$ TeV$^{-1}$)    &    \\ 
  \hline
    1 & power-law  & $10^{-14}-10^{-09}$  & [-3.50, -1.50]  \\
    2 & power-law  & $10^{-18}-10^{-13}$  &  [-3.50, -1.50]    \\
    3 & log-parabola  & $10^{-14}-10^{-09}$  & [-3.50, -1.50]    \\
    4 & power-law (Alpha)  & $10^{-14}-10^{-09}$  & [-3.50, -1.50]     \\
    \hline
    \end{tabular*}
\end{small}
\end{center}
\end{table}


Four simulated populations are summarized in Table \ref{sim_pops}. They include different spectral shapes and parameters. {For the log-parabola model we assume the range of curvature [-0.25, 0.25]}. {Note that populations 1 and 4 are two different populations sharing only the spatial distribution of the sources.}  
For each population we employed the 0.5 h IRFs {for both CTAO-N and CTAO-S, and also tested the Alpha configuration in the case of population 4}. Both IRF sets contain three zenith angle observation options at 20$^{\circ}$, 40$^{\circ}$ and 60$^{\circ}$; and they also account for the azimuth dependence coming from the geomagnetic field pointing direction: North, South or an average over the azimuth direction.

For each source, we simulate the photon events list for 1 hour both for the CTAO-N and CTAO-S sites with a 5.0$^{\circ}$ ROI centered at a source, without any other sources within it, accounting only for the IRF background {as seen in Fig. \ref{fig:CTAOsensitivityMap} and Fig. \ref{fig:sensitivityEnergiesTimes}}. 
 The energy ranges used for both configurations at each site are collected in Table \ref{tab_energies}.
 The energy dispersion effect has been also taken into account (according to the IRFs). We then performed an unbinned maximum likelihood fitting. The test statistic (TS) equal or higher than 25 is used as criterion for a source detection. {The TS for different values of prefactor and spectral index are shown in Figs. \ref{fig:ats} and \ref{fig:gts} respectively.}

\begin{table}
\caption{Energies {(TeV)} assumed in the simulations depending on the array location, configuration and zenith angle. Different energy ranges were assumed depending on the geomagnetic field (average, North, South) for CTAO-N Alpha Configuration, as produced in the dedicated IRFs \citep{cherenkov_telescope_array_observatory_2021_5499840}.}
\label{tab_energies}
\begin{center}
\begin{small}
	\begin{tabular}{|c|c|c|c|}
 \hline
    Site     & 20$^{\circ}$    & 40$^{\circ}$    &60$^{\circ}$    \\ \hline
    CTAO-N        & 0.03-200       & 0.04-200     & 0.11-200   \\ 
    CTAO-S        & 0.03-200       & 0.04-200     & 0.11-200   \\ 
    CTAO-N (Alpha)       & 0.03-200       & 0.04-200     &  \begin{tabular}{@{}c@{}@{}}0.06-200 (A) \\ 0.12-200 (N) \\ 0.08-200 (S)\end{tabular}     \\ 

    CTAO-S (Alpha)       & 0.04-200       & 0.06-200     &  0.18-200  \\ \hline
    \end{tabular}
\end{small}
\end{center}
\end{table}




\begin{figure*}
    \centering
    \includegraphics[width=0.8\textwidth]{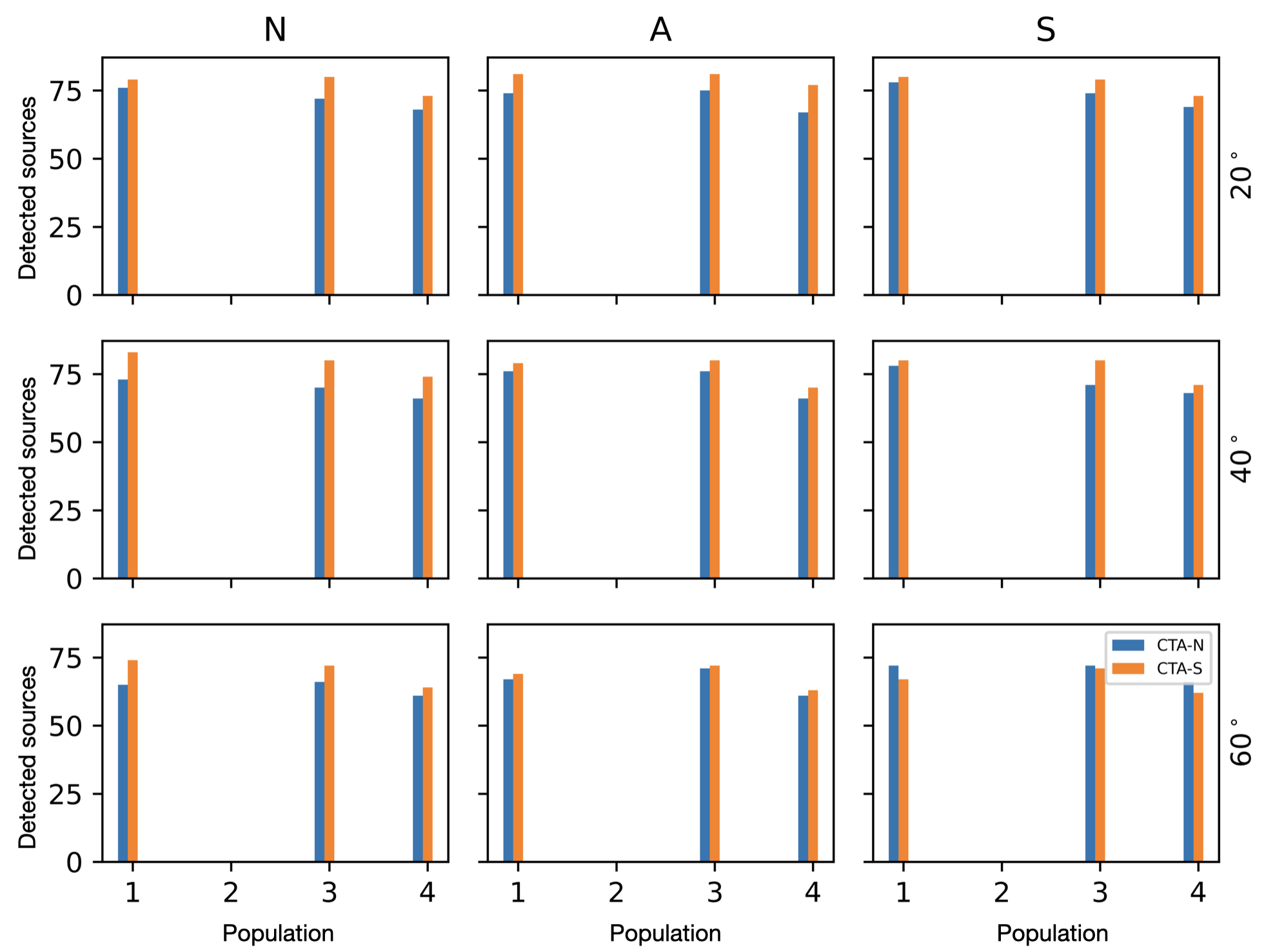}
    \caption{Number of detected sources in populations 1 to 4 {for CTAO-N (blue) and CTAO-S (orange), including the Alpha configuration in population 4}. From left to right: different configurations of the geomagnetic field (North, average and South). From top to bottom: different zenith angles (20$^{\circ}$, 40$^{\circ}$ and 60$^{\circ}$). }
    \label{fig:det}
\end{figure*}

\begin{figure*}
    \centerline{
    \includegraphics[width=0.3\textwidth]{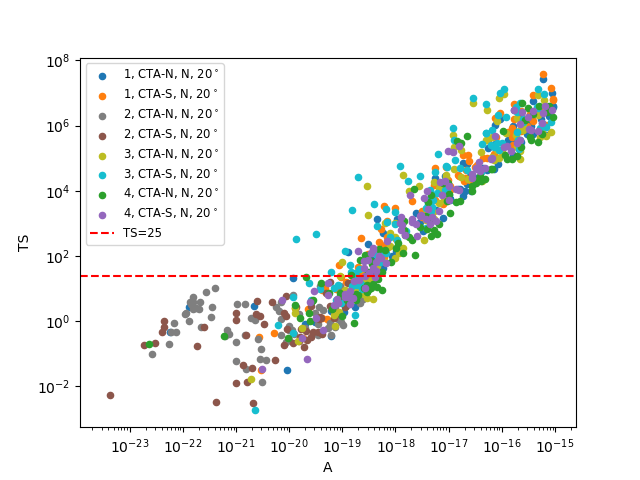} \includegraphics[width=0.3\textwidth]{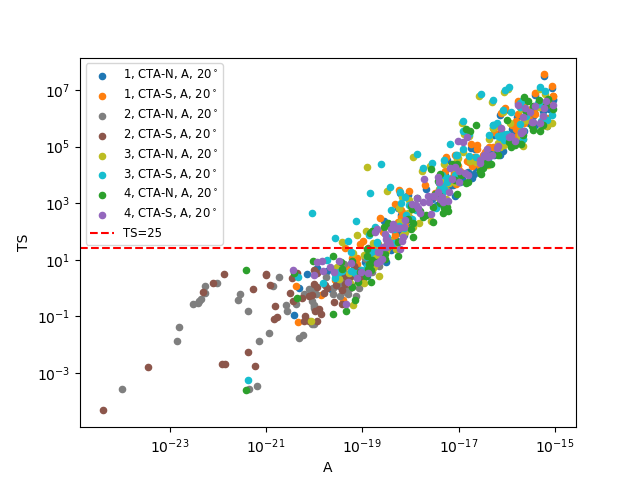} \includegraphics[width=0.3\textwidth]{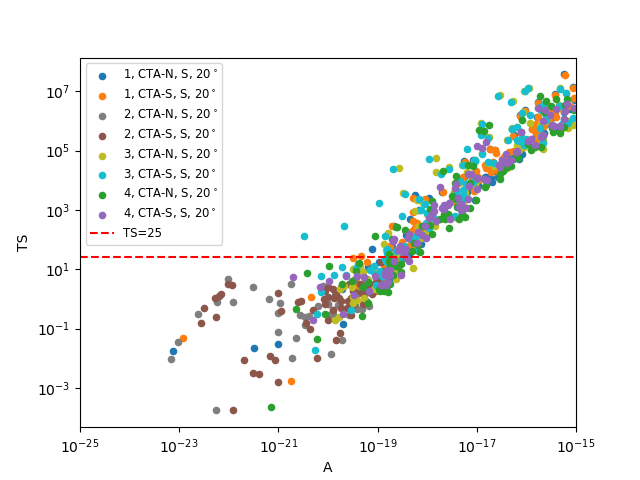}}
    \centerline{
    \includegraphics[width=0.3\textwidth]{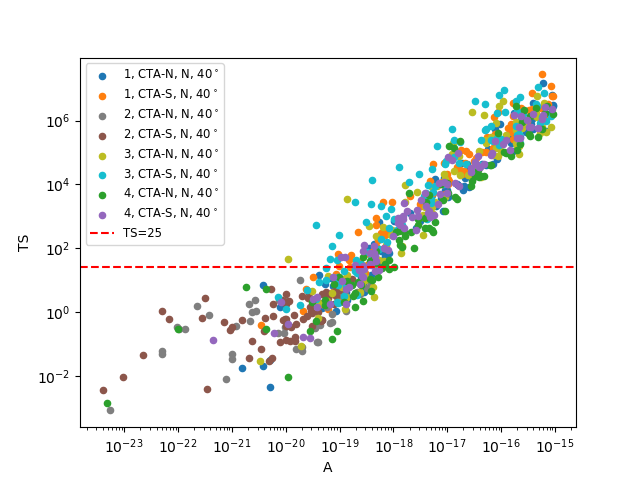} \includegraphics[width=0.3\textwidth]{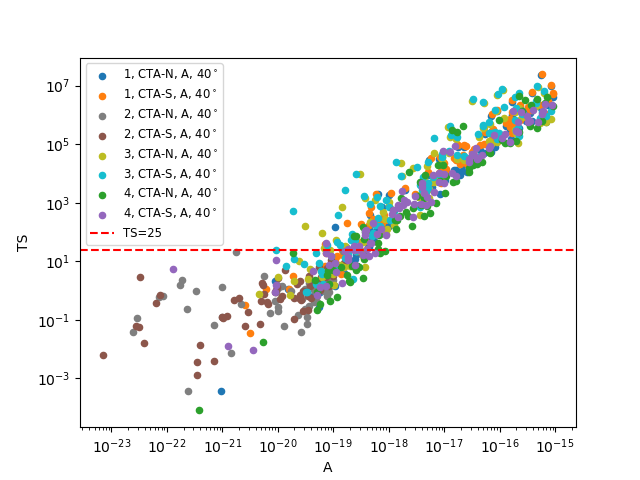} \includegraphics[width=0.3\textwidth]{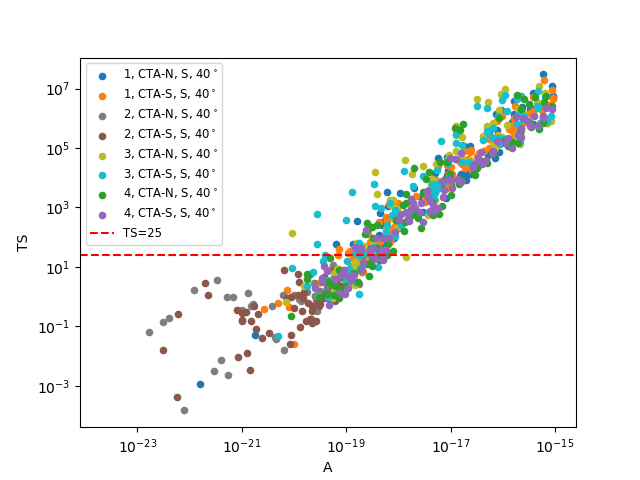}}
    \centerline{
    \includegraphics[width=0.3\textwidth]{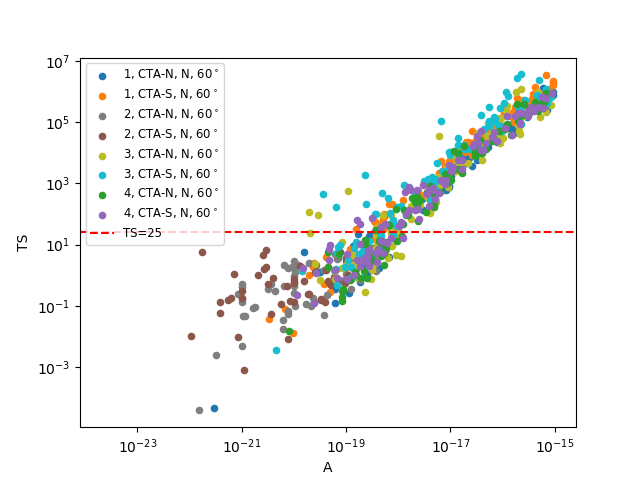} \includegraphics[width=0.3\textwidth]{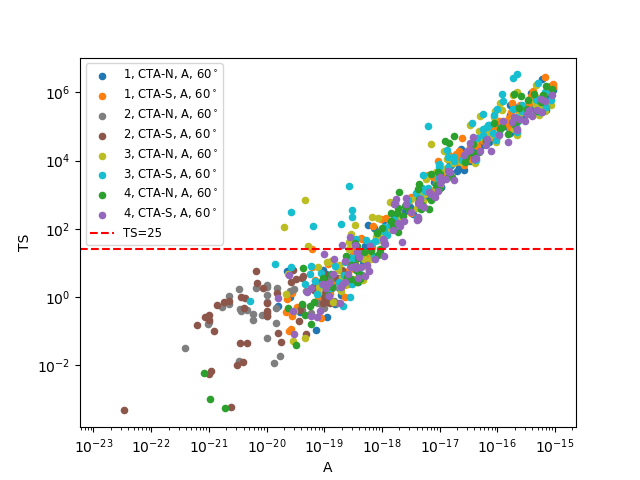} \includegraphics[width=0.3\textwidth]{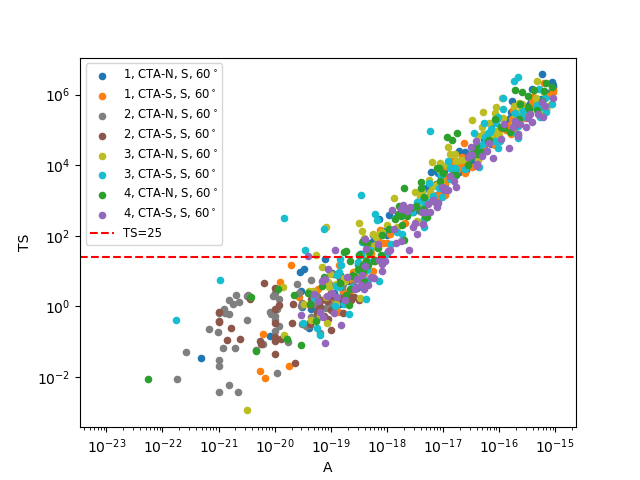}}
    \caption{{Test statistic for different values of prefactor in the simulated populations. From left to right: different configurations of the geomagnetic field (North, average and South). From top to bottom: different zenith angles (20$^{\circ}$, 40$^{\circ}$ and 60$^{\circ}$). The red dashed line marks TS = 25.} }
    \label{fig:ats}
\end{figure*}

\begin{figure*}
    \centerline{
    \includegraphics[width=0.3\textwidth]{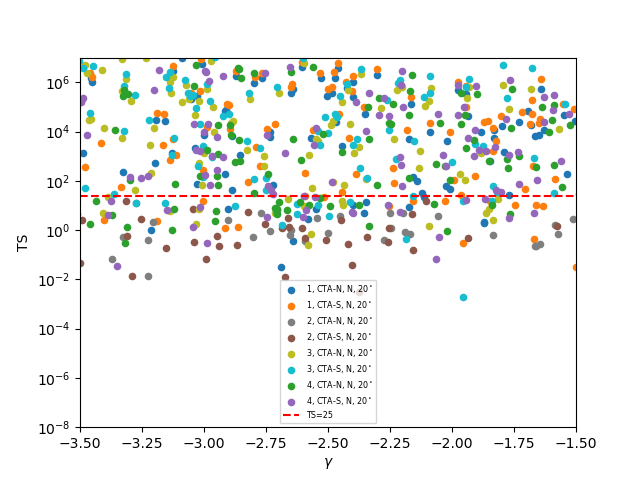} \includegraphics[width=0.3\textwidth]{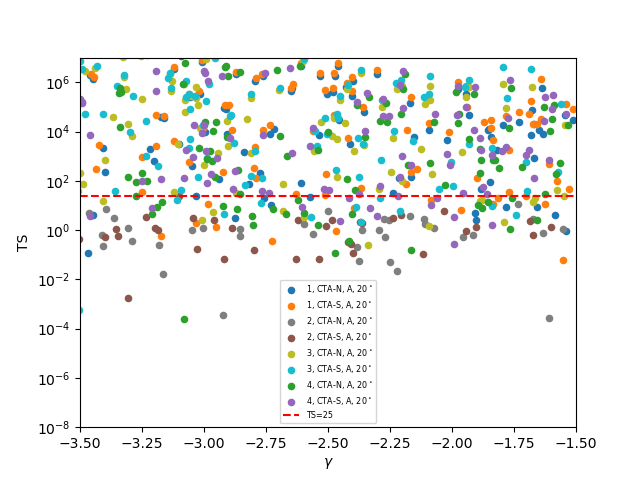} \includegraphics[width=0.3\textwidth]{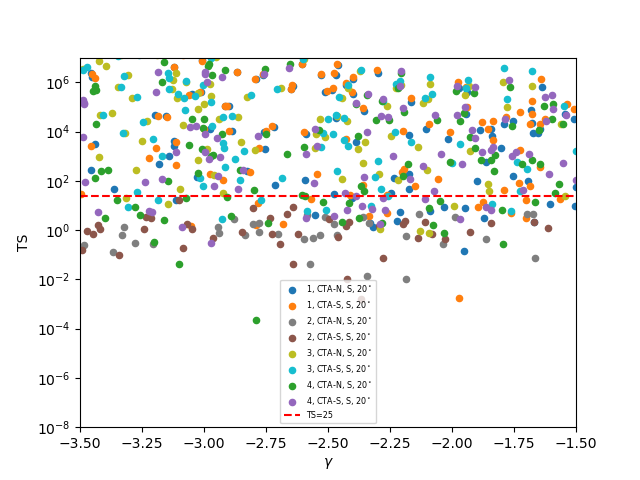}}
    \centerline{
    \includegraphics[width=0.3\textwidth]{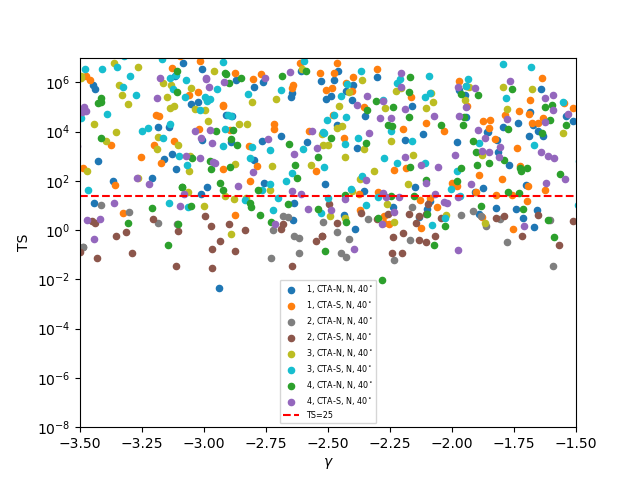} \includegraphics[width=0.3\textwidth]{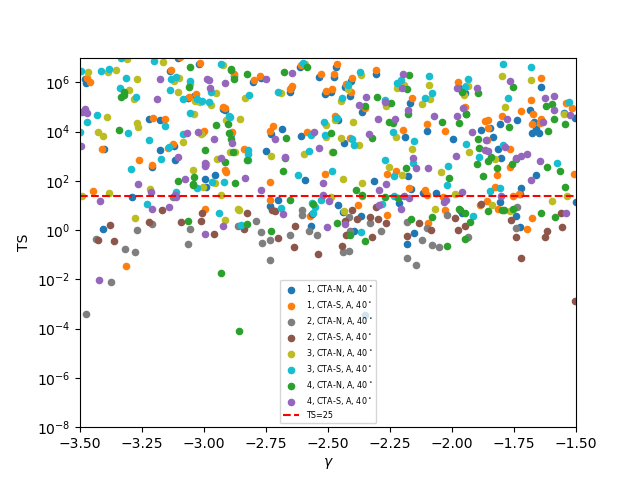} \includegraphics[width=0.3\textwidth]{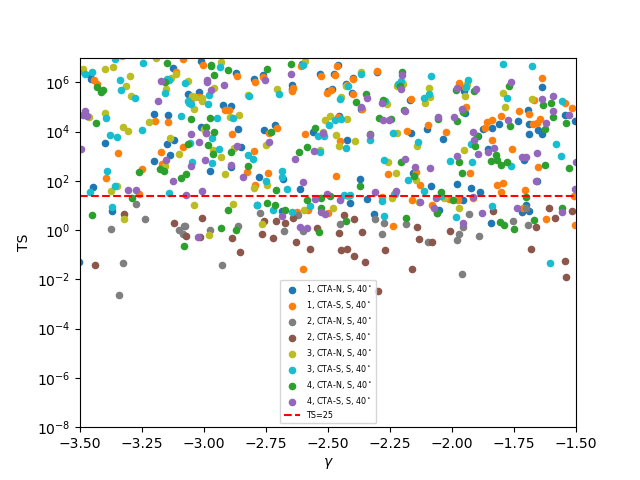}}
    \centerline{
    \includegraphics[width=0.3\textwidth]{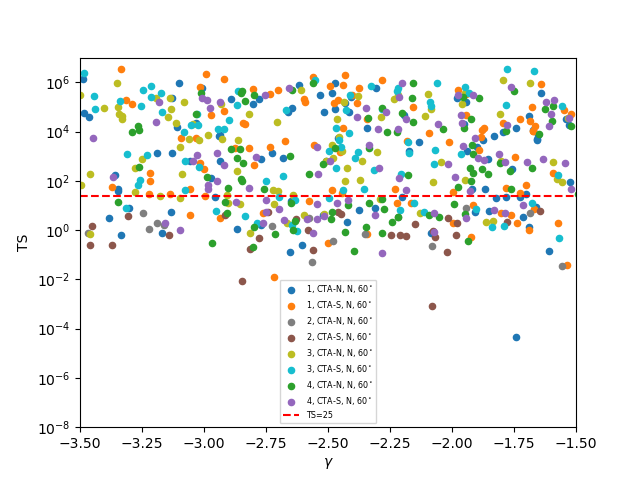} \includegraphics[width=0.3\textwidth]{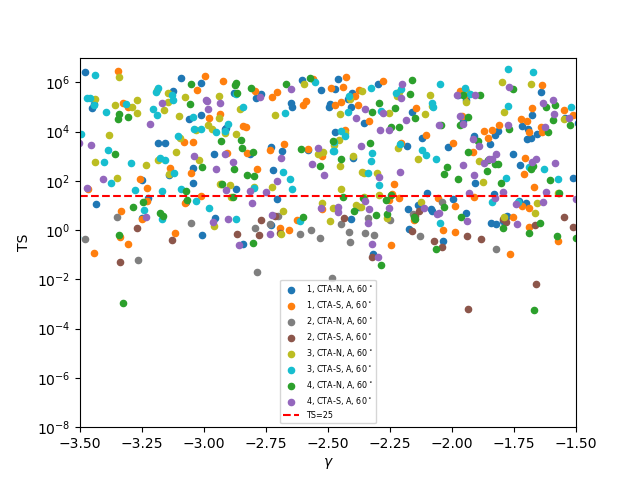} \includegraphics[width=0.3\textwidth]{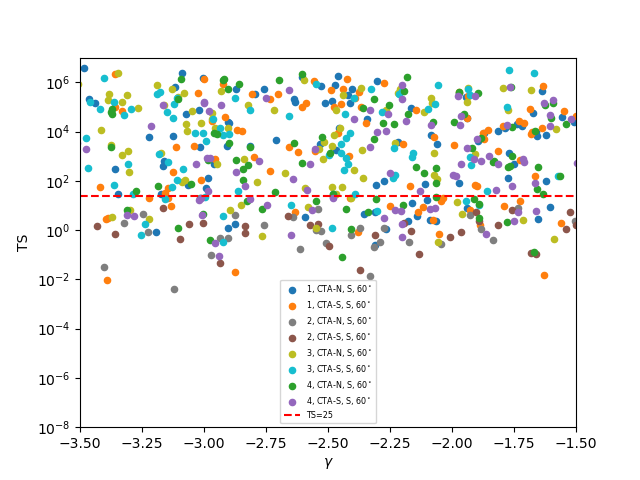}}
    \caption{{Test statistic for different values of spectral index in the simulated populations. From left to right: different configurations of the geomagnetic field (North, average and South). From top to bottom: different zenith angles (20$^{\circ}$, 40$^{\circ}$ and 60$^{\circ}$).} }
    \label{fig:gts}
\end{figure*}

\begin{figure*}
    \centerline{
    \includegraphics[width=0.3\textwidth]{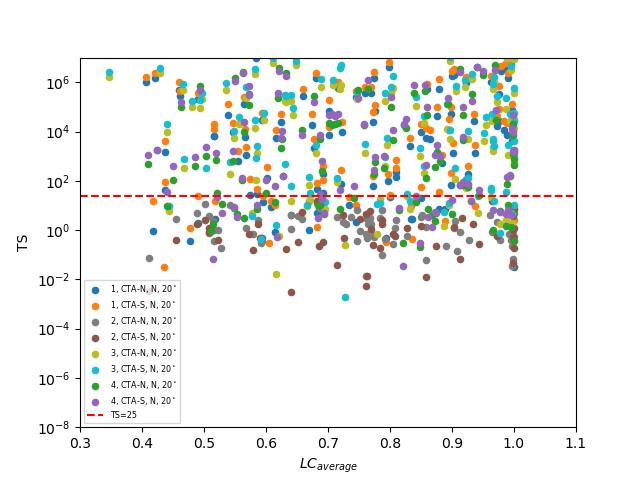} \includegraphics[width=0.3\textwidth]{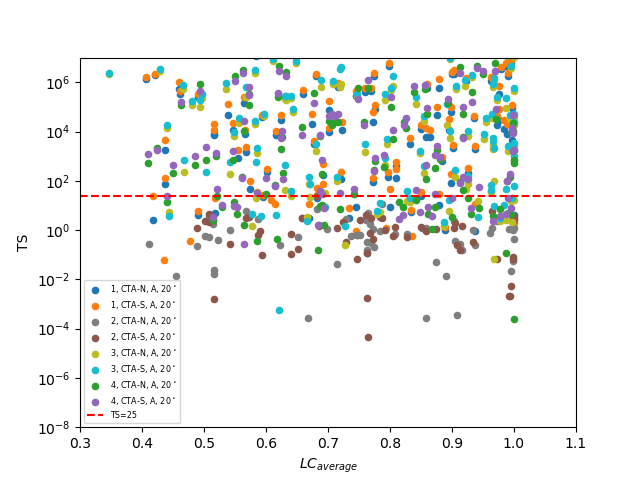} \includegraphics[width=0.3\textwidth]{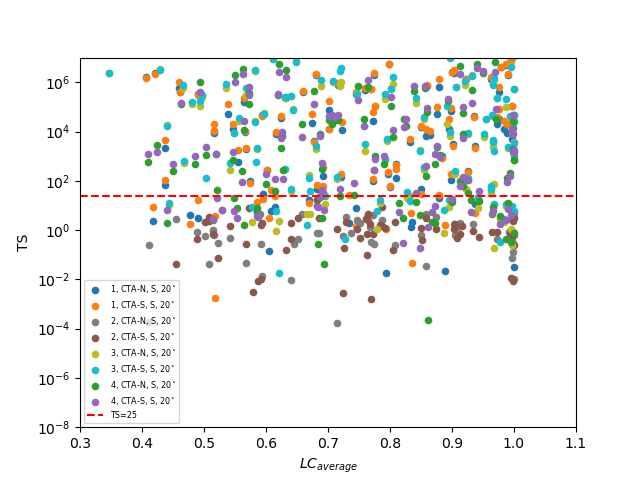}}
    \centerline{
    \includegraphics[width=0.3\textwidth]{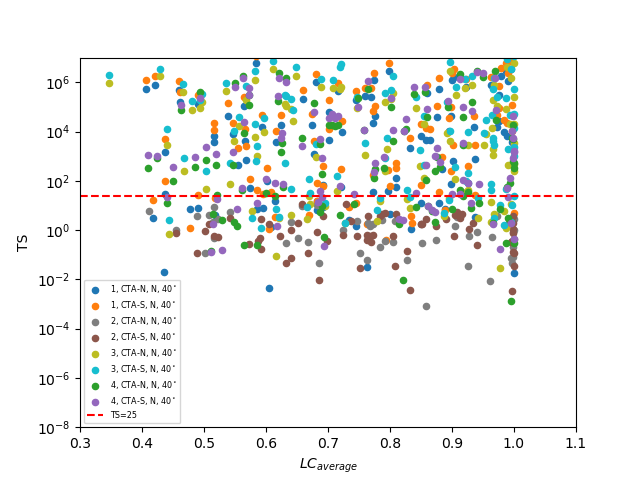} \includegraphics[width=0.3\textwidth]{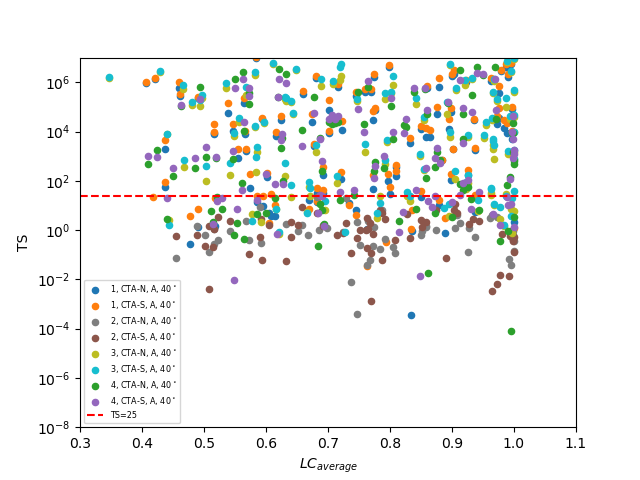} \includegraphics[width=0.3\textwidth]{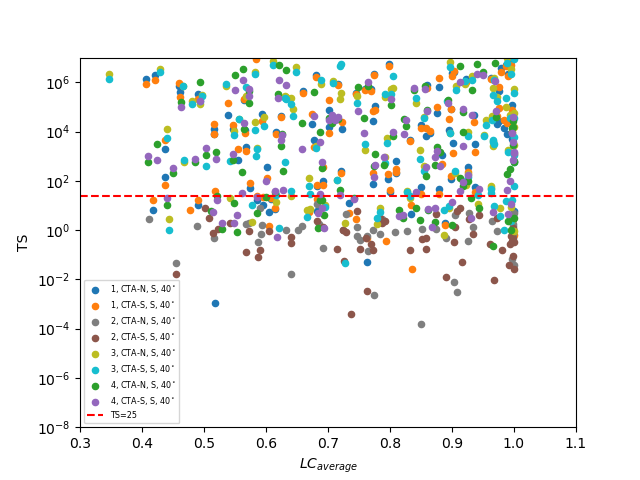}}
    \centerline{
    \includegraphics[width=0.3\textwidth]{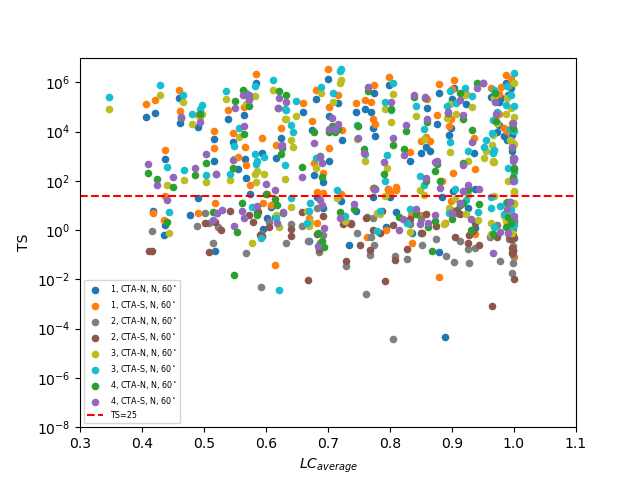} \includegraphics[width=0.3\textwidth]{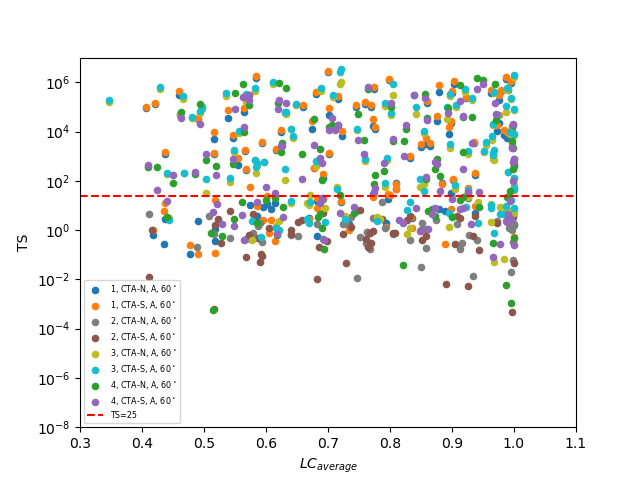} \includegraphics[width=0.3\textwidth]{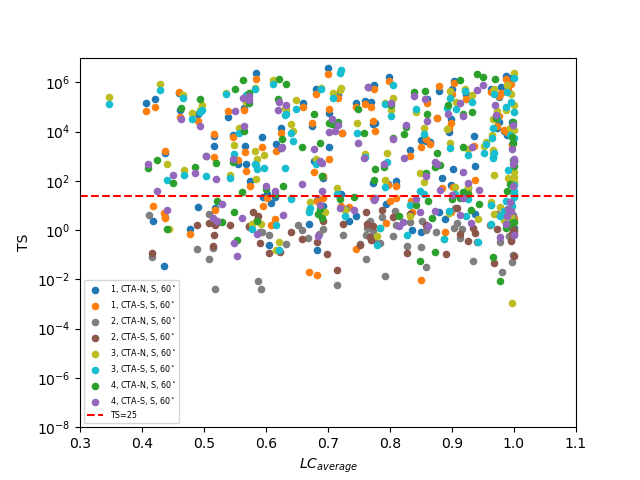}}
    \caption{{Test statistic for different values of the lightcurve averaged over the observation time in the simulated populations. From left to right: different configurations of the geomagnetic field (North, average and South). From top to bottom: different zenith angles (20$^{\circ}$, 40$^{\circ}$ and 60$^{\circ}$). The red dashed line marks TS = 25.} }
    \label{fig:lcts}
\end{figure*}

\begin{figure*}
    \centering
    \includegraphics[width=0.8\textwidth]{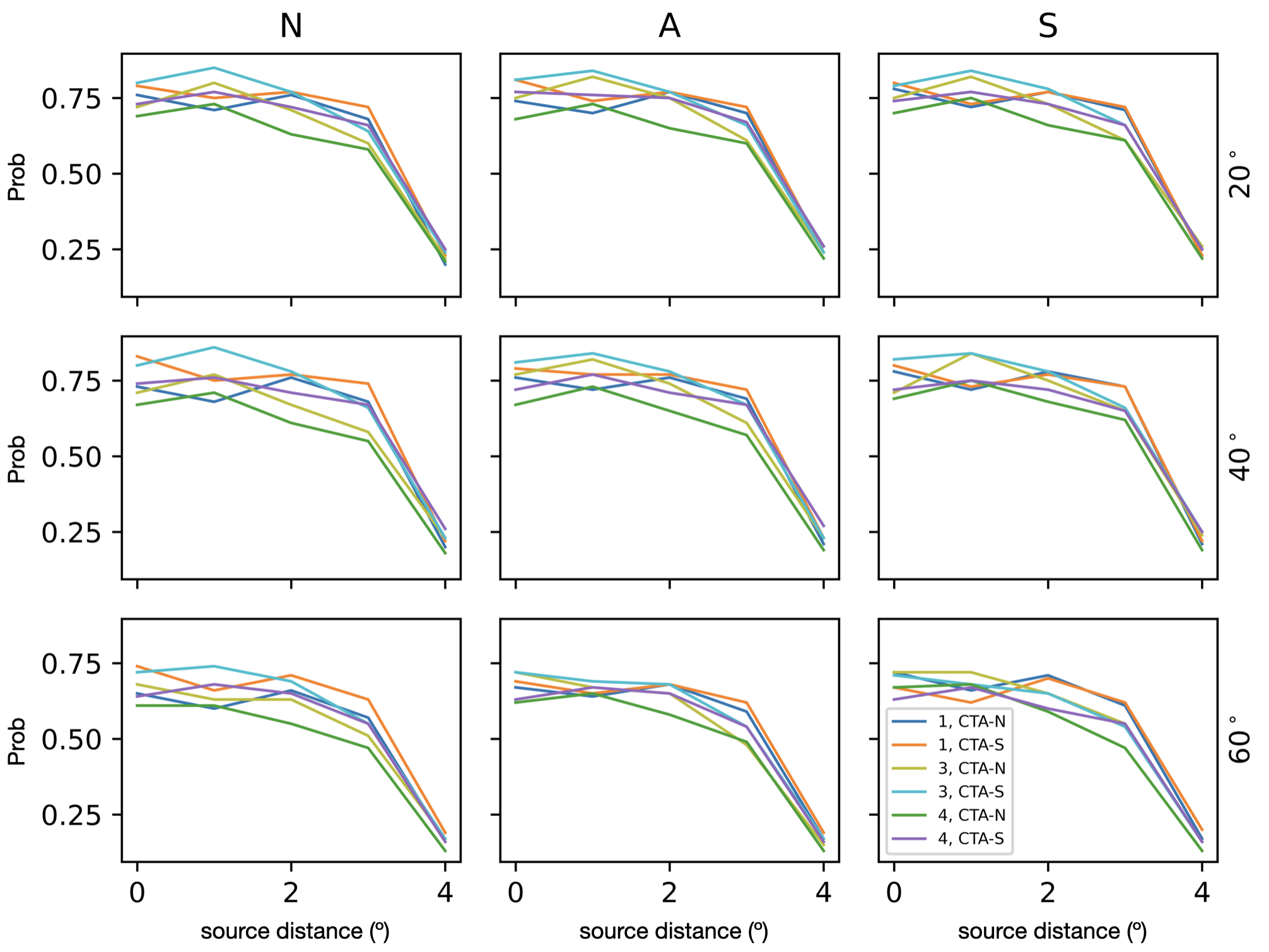}
    \caption{{Dependence of the CTAO detection probability {(vertical axis)} on the source distance {(horizontal, in degrees)} from the ROI center}. From left to right: different configurations of the geomagnetic field (North, average and South). From top to bottom: different zenith angles (20$^{\circ}$, 40$^{\circ}$ and 60$^{\circ}$). }
    \label{fig:det_off}
\end{figure*}

The results for all populations and observation configurations are presented in Fig. \ref{fig:det}-\ref{fig:lcts}. {Since the simulated lightcurves have different timescales and to have an idea at which lightcurve stage the observation takes place and how it affects the detectability of a source, we present in Fig. \ref{fig:lcts} the test statistic versus the value of lightcurve averaged over the observation time for each source in the simulated populations.} We found that in the case of having a power-law spectra and fluxes in the range of $10^{-14}-10^{-09}$ ph cm$^{-2}$ s$^{-1}$ TeV$^{-1}$, $73-83\%$ of sources for 20$^{\circ}$ and 40$^{\circ}$ zenith angles and $65-74\%$ of sources for 60$^{\circ}$ zenith angle will be detected, while for the population with fluxes $<10^{-13}$ ph cm$^{-2}$ s$^{-1}$ TeV$^{-1}$, CTAO will not detect any source during 1 hour of observation. In most cases CTAO-S performs marginally better than CTAO-N, with the larger difference for the North magnetic field configuration. 
Adding a curvature to the spectrum does not affect the detection rate in a statistically significant way. Using the Alpha Configuration (which corresponds to a first construction phase) slightly decreases the detectability, which is expected due to the reduced number of telescopes and the lack of LSTs in the CTAO-S Alpha configuration \citep{maier_g_2023_8050921}. 

In the case of ROIs not centered in a source, we present in Fig. \ref{fig:det_off} the dependence of the detection probability on the source distance (in degrees) from the ROI center, with a clear decrease in detectability for the offsets $>3^{\circ}$.

{Finally, to roughly estimate how  source visibility affects the presented observation probabilities, we compute, as an example, the number of detected sources for 1-h long observations starting at 2:00 UTC on three different nights taking into account the visibility constraints at each observatory site. For sources with zenith angle ranges of $[0^\circ,\;33^\circ]$, $[33^\circ,\;54^\circ]$ and $[54^\circ,\;66^\circ]$, we employed the IRFs corresponding to zenith angles of $20^\circ$, $40^\circ$ and $60^\circ$ respectively (observations were not considered for zenith angles exceeding than $66^\circ$). Each source was evaluated using the IRF appropriate for its azimuth. The results are presented in Table \ref{tab_pop_vis}. We see that while the detection probabilities without these constraints are high, there is a significant reduction in detection probabilities when visibility constraints are taken into account.}


\begin{table*}
\caption{{Number of detected sources in populations 1, 3 and 4 including the visibility constraints for observations taking place from 2:00 to 3:00 UTC.}}
\label{tab_pop_vis}
\begin{center}
\begin{small}
	\begin{tabular}{|c|c|c|c|c|c|c|c|c|c|c|c|c|c|c|c|c|c|c|c|}
 \hline
  Population&IRF&\multicolumn{6}{c}{2025-05-22}& \multicolumn{6}{c}{2025-08-22}&\multicolumn{6}{c}{2025-11-22} \\
 &zenith&\multicolumn{3}{c}{CTAO-N}&\multicolumn{3}{c}{CTAO-S}&\multicolumn{3}{c}{CTAO-N}&\multicolumn{3}{c}{CTAO-S}&\multicolumn{3}{c}{CTAO-N}&\multicolumn{3}{c}{CTAO-S}\\ 
&angle&N&A&S&N&A&S&N&A&S&N&A&S&N&A&S&N&A&S\\
 \hline\hline
1&20$^{\circ}$&0 &3 &1&0 &0 &0&0 &3 &0&11 &24 &2&0 &0 &0&0 &0 &0\\
&40$^{\circ}$&0 &17 &10&0 &14 &13&0 &13 &0&15 &4 &7&0 &1 &0&0 &0 &0\\
&60$^{\circ}$&0 &4 &11&0 &13 &1&0 &6 &0&3 &1 &2&0 &1 &0&0 &0 &0\\\hline
3&20$^{\circ}$&0 &4 &1&0 &0 &0&1 &0 &0&9 &25 &1&0 &0 &0&0 &0 &0\\
&40$^{\circ}$&0 &15 &8&0 &14 &16&0 &11 &0&11 &4 &11&1 &1 &0&0 &0 &0\\
&60$^{\circ}$&0 &2 &12&0 &14 &0&1 &7 &0&3 &1 &4&0 &1 &0&0 &1 &0\\\hline
4&20$^{\circ}$&0 &2 &1&0 &1 &0&1 &2 &0&13 &23 &2&0 &0 &0&0 &0 &0\\
&40$^{\circ}$&0 &17 &12&0 &12 &12&0 &9 &0&13 &2 &6&0 &0 &0&0 &0 &0\\
&60$^{\circ}$&0 &2 &12&0 &12 &0&0 &5 &0&1 &2 &2&0 &0 &1&0 &1 &0\\
\hline
    \end{tabular}
\end{small}
\end{center}
\end{table*}


 

\section{Source detection with CTAO} \label{sec:simulations}

Galactic transients that exhibit MeV-GeV emission are specially interesting to be studied with CTAO, since it is {already} known that non-thermal mechanisms leading to gamma-ray production are at work. We aim at understanding whether these sources of interest can also emit VHE radiation, which can be produced by the same HE mechanisms and be detected as a spectral extension, or be created by an additional component at TeV energies.

\subsection{High-mass microquasars} \label{massive_micro}

The microquasars of the Cygnus region, Cyg X-3, Cyg X-1 and the system SS 433 are
 the only microquasars that have been detected in the HE regime, hence they can be considered as potential targets for the CTAO observatory. After the discovery of {persistent} gamma-ray emission from SS 433 above 20 TeV by HAWC \citep{SS433_HAWC}, {{the latest LHAASO detections \citep{LHAASO2024arXiv241008988L} }} {and specially the first detection by an IACT \citep{SS433_2024}}, the CTAO observations of these microquasars will be crucial to shed light on the physical mechanism responsible for the VHE emission in this type of binary systems, {by investigating the limits of extreme particle acceleration in the jet.}


The importance of observing this subclass of binary systems with CTAO has  been {previously} discussed in \citet{2013APh....43..301P}. In particular, a detailed study on a possible detection of a TeV flare from Cyg X-1 was presented in that paper, showing conclusions similar 
to our findings (see Sect. \ref{Cyg X-1: transient}). {In this section, we show simulations on the {first} microquasar detected in the VHE regime, SS 433, and estimate the detectability {both} of transient and persistent emission from Cyg X-3 and Cyg X-1. {Even if the detection of persistent emission is not the scope of this paper, we perform this additional exercise to complement the expectations to detect microquasars with CTAO.} For the case of the microquasars in the Cygnus region, we} carried out several CTAO observation simulations by using the latest \verb+prod5-v0.1+ IRFs {to check if these two systems could already be detected in the first years of operation of CTAO}. Almost all the simulations have been carried out in the lowest range of energies for the CTAO observatory, where the bulk of the emission from these binary systems is expected. For each set of observations, besides the emission from the microquasars, we simulated the main field sources of the Cygnus region: 2HWC J2006+341 \citep{Araya2019}, VER J2016+371, VER J2019+368 \citep{Aliu2014}, Gamma Cygni SNR \citep{Ackermann2017, Abeysekara2018}, TeV J2032+4130 (emission model as detected by MAGIC before the periastron passage of November 2017, \citealp{Abeysekara2018c}). {This approach also applies to the case of the LMXB V404 Cyg located in the same region (see subsection \ref{V404 Cyg: CTAO simulations}).}

\subsubsection{SS 433} \label{CTAO_micro_ss433}

\begin{figure*}[ht!]
\centering
    \includegraphics[width=0.7\linewidth]{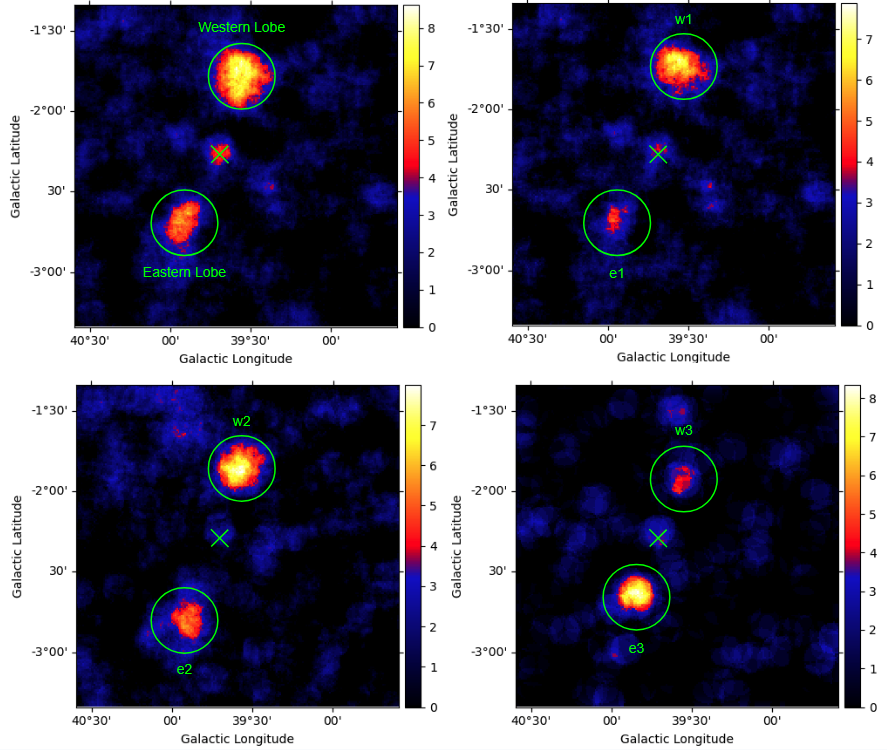}
    \caption{SS433 Simulations, taken with twenty total hours of exposure time spread across two precessional periods. Top right:  0.8-2.5 TeV. bottom left:  2.5-10 TeV. Bottom right: >10 TeV. Top left: total model from 0.8 TeV to 100 TeV. The position of the central source is marked with a cross.  }
    \label{fig:skymaps}
\end{figure*}

SS 433 is a binary system containing a supergiant star that is overflowing its Roche lobe with matter accreting onto a compact object, either a BH or a NS (see e.g. \citealt{1984ARA&A..22..507M,2004ASPRv..12....1F}). Two jets of ionised matter, with a bulk velocity of approximately one quarter of the speed of light in vacuum, extend from the binary, perpendicular to the line of sight, and terminate inside the supernova remnant W50 (e.g. \citealt{2004ASPRv..12....1F}). The lobes of W50 in which the jets
terminate about 40 parsecs from the central source, are  accelerating charged particles, as it follows from  radio and X-ray observations, consistent with electron synchrotron emission \citep{1980A&A....84..237G,2007A&A...463..611B}.

 \begin{table*}
    \centering
    \begin{tabular}{ccccccccc}
        \hline
        Source & $\phi_0$ [TeV$^{-1}$cm$^{-2}$s$^{-1}]$ & $\Gamma_1$ & $\Gamma_2$ & $\Gamma_3$ & E$_{B1}$ [TeV] & E$_{B2}$ [TeV] & $\beta_1$ & $\beta_2$\\
        \hline\hline
        Central Source &  $2.00\times10^{-14} $ & $ 2.00 $ & $ \ $ & $ \ $ & $ \ $ & $ \ $ & $ \ $ & $ \ $  \\
        \hline
        East 1 & $ 2.40\times10^{-13} $ & & $ 2.18 $ & $ 10.0 $  & & $ 2.00 $ & & $ 1.00 $  \\
        \hline
        East 2 & $ 1.00\times10^{-14} $ & $ -2.10 $ & $ 1.00 $ & $ 2.80 $ & $ 2.50 $ & $ 7.00 $ & $ 0.10 $ & $ 0.01 $ \\
        \hline
        East 3 & $ 1.00\times10^{-16} $ & $ -2.10 $ & $ 1.04 $ & $ 4.00 $ & $ 10.0 $ & $ 100 $ & $ 0.50 $ & $ 0.01 $  \\
        \hline
        West 1 &  $ 3.00\times10^{-13} $ & &$ 2.40 $ & $ 10.0 $ & & $ 2.00 $ &  $ \ $ & $ 1.00 $   \\
        \hline
        West 2 & $ 1.20\times10^{-14} $ & $ -2.12 $ & $ 1.15 $ & $ 2.80 $ & $ 2.50 $ & $ 7.00 $ & $ 0.10 $ & $ 0.10 $  \\
        \hline
        West 3 &  $ 1.00\times10^{-16} $ & $ -2.10 $ & $ 1.15 $ & $ 4.00 $ & $ 10.0 $ & $ 100 $ & $ 0.50 $ & $ 0.01 $  \\
        \hline

    \end{tabular}
    \caption{Spectral model parameters}
    \label{SS433_spec}
\end{table*}

At TeV energies SS 433 was detected by both the High Altitude Water Cherenkov (HAWC) Observatory  (1017 days of measurements, \cite{SS433_HAWC})  and H.E.S.S. (200 hours of observations, \cite{SS433_2024}).
 These observations demonstrate presence of two regions of gamma-ray emission {of leptonic nature} at the positions of the eastern and western jets. The reported H.E.S.S. fluxes at 1 TeV (($2.30\pm0.58)\times 10^{-13}$ TeV$^{-1}$ cm$^{-2}$ s$^{-1}$ and ($2.83\pm0.70)\times 10^{-13}$ TeV$^{-1}$ cm$^{-2}$ s$^{-1}$ for the eastern and western jets correspondingly) are inline with HAWC data. Quality of the H.E.S.S. data also allow to study the energy dependence of the source  morphology, demonstrating that while the gamma-ray emission above 10 TeV appears only at the base of the jets, the lower-energy gamma rays have their peak surface brightness at locations further along each jet, reflecting an energy-dependent particle energy loss timescale. 

 Analysis of  the \textit{Fermi}-LAT  data 
led to the discovery of the  significant HE gamma-ray emission from the region around SS 433 (see \citealt{SS433_bordas15,SS433_extFermi19,SS433_varFermi19,SS433_Fermi19,2020NatAs...4.1177L}).
However, the analysis is  model dependent and can lead to very different conclusions on the position and extension of the source. In \cite{SS433_varFermi19}, authors report evidence at 3 $\sigma$ level for the modulation of the $\gamma$-ray emission with the precession period of the jet of 162 days. This result suggests that at least some of the gamma-ray emission originates close to the base of the jet. \citet{2020NatAs...4.1177L} detected HE emission in the vicinity of SS 433 which shows periodic variation compatible with the processional period of the jets.

While we {do not} expect to detect variability in the VHE emission coming from the lobes, microquasars are known to have flaring emission on various timescales coming from its central source.  To test the possibility of CTAO to detect a central source and its putative variability we simulate the local region of SS 433 with the diffuse background and the nearby MGRO 1906+06 source, where SS 433 consists of both the aforementioned lobes and a central point source.

Following the H.E.S.S. observations, we have modelled the eastern and western lobes as a combination of three hotspots with Gaussian profiles (with the parameters summarised in Table S4 of \cite{SS433_2024}). Spectral models of the lobes were organised so that different hotspots appear in different energy bands. To represent the spectral model of each hotspot in agreement with \cite{SS433_2024}, we assumed that they follow a powerlaw distribution with a super-exponential cut-off at both high and low energies, to allow hotspots to arise at different energies: 

\begin{eqnarray*}
F(E) = & \phi_0  \left(\frac{E}{1 \textrm{TeV}}\right)^{-\Gamma_1}  \left( 1 + \left(\frac{E}{E_{B2}}\right)^{\left(\frac{\Gamma_3 - \Gamma_2}{\beta_2}\right)^{-\beta_2}} \right)
\\
&  \times \left(\frac{E}{1 \textrm{TeV}}\right)^{-\Gamma_2} 
\left(\frac{E}{E_{B1}}\right)^{\left(\frac{\Gamma_2 - \Gamma_1}{\beta_1}\right)^{-\beta_1}} 
\end{eqnarray*}

The parameters chosen to represent the H.E.S.S. spectrum are given in Table \ref{SS433_spec}. No low-energy cut-off was assumed for the e1 and w1 sources. The spectral model of the central source was obtained directly from the H.E.S.S. ULs, and was modelled using a simple power law model with a flux falling below the H.E.S.S. UL value. As it is seen in Figure \ref{fig:skymaps}, 20 hours of CTAO observations is enough to clearly measure the energy dependent source structure as well as to detect the central source at the assumed flux level. The dependence of the central source relative flux errors on the exposure time is shown in Figure \ref{SS433_exposure}. 

To study the CTAO possibility to detect possible variability {of about 15\%} with the precession and orbital periods at the level proposed by \cite{SS433_varFermi19} we have simulated a 500 hours observation of the source uniformly distributed along the precessional period, assuming   $F(\varphi) = \left( 0.99 + 0.14 \sin(2\pi(\varphi + 0.84))\right)\times F$ and 500 hours observation of the source uniformly distributed along the orbital period, assuming $F(\varphi) = \left(1.07 + 0.18 \sin(2\pi(\varphi + 0.81))\right)\times F$. The expected variability is shown in Figure \ref{SS433_precession}.

\begin{figure}
\includegraphics[width=\columnwidth]{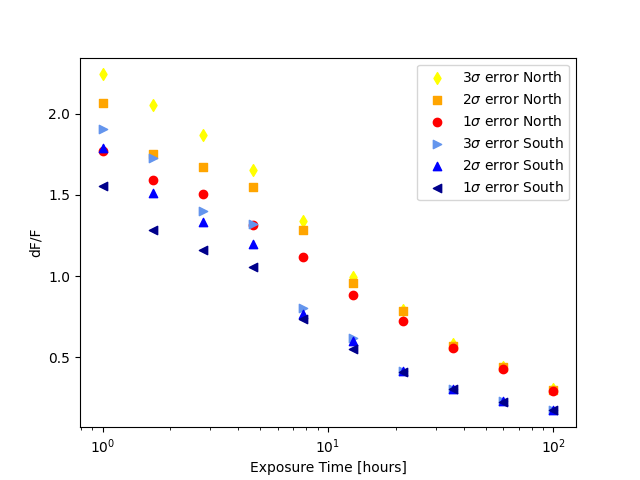}
\caption{Comparison of the relative flux errors for the SS 433 central source at different exposure times for the Northern and Southern arrays.}
\label{SS433_exposure}
\end{figure}

\begin{figure}
\includegraphics[width=\columnwidth]{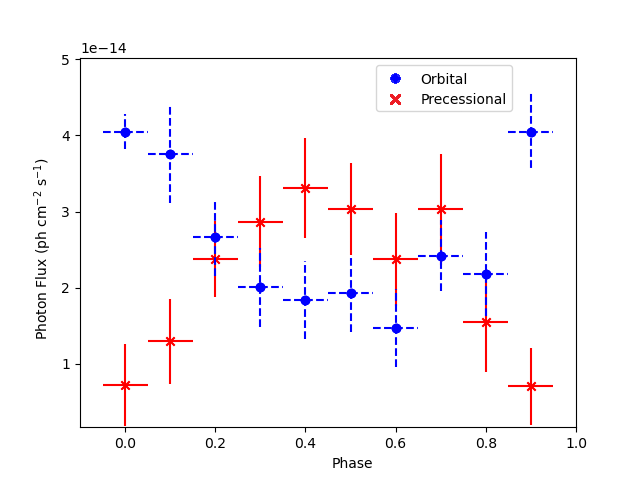}
\caption{500 hours observations of SS433 as observed with the Southern array  folded with the precessional and orbital periods.}
\label{SS433_precession}
\end{figure}

\begin{figure*}
\includegraphics[width=\columnwidth]{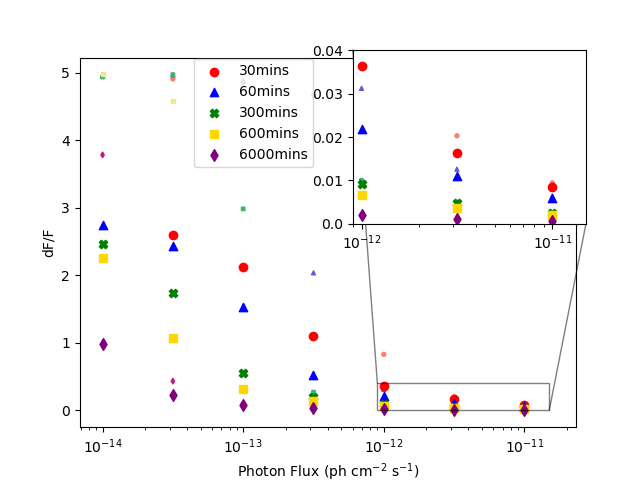}
\includegraphics[width=\columnwidth]{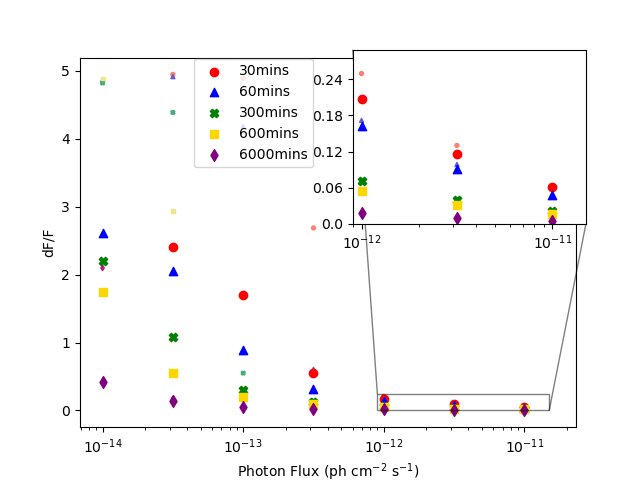}
\caption{{Dependence of the 1 and 3 $\sigma$ relative flux error ratio on the photon flux for  the Northern (left) and Southern (right) arrays. Different exposure times are given by different colours as indicated in the legend. 3 $\sigma$ relative flux errors are shown with smaller and fainter symbols.}}
\label{fig:SS433_flux_sigma}
\end{figure*}

 To systematically study the CTAO sensitivity to detect different level of variability for various levels of the source flux we run batches of {5000 simulations for low exposure times of 30 mins and 60 mins, and batches of 1000 simulations for higher exposure times of 300, 600 and 6000 mins using the North and South site IRFs }
 The flux was calculated after every simulation and the total data was compiled into histograms for each exposure time to determine the error range of the detections. As SS 433 could be viewed from both the north and south hemispheres the results of error measurements from both sites were compared to determine which can produce more sensitive detections. Figure  \ref{fig:SS433_flux_sigma} shows the comparison of this flux error ratio depending on the integrated source flux above 1~TeV and exposure for both Northern (left) and Southern arrays.  In this figure one can see also simulations for the values exceeding the H.E.S.S. upper limit on the central source ($\sim 10^{-13}$ ph cm$^{-2}$ s$^{-1} $). This was done  to test the effectiveness of CTAO on dim transient sources to determine what level of variability can be observed with short exposures.

Based on the simulations run at different fluxes, sources with a photon flux $<1\times10^{-13}$ ph cm$^{-2}$ s$^{-1}$ will require exposure times of more than 
10 hours in order to detect flux variability at about 50\% level. However, for sources with a flux $\geq 1\times10^{-13}$ ph cm$^{-2}$ s$^{-1}$, CTAO may be able to detect variability as low as $\sim 10\%$ observing from 5 to 10 hours. At a photon flux $\geq3\times10^{-12}$ ph cm$^{-2}$ s$^{-1}$ the 1 sigma ratio gets as low as $\sim 5\%$ with exposure times of an hour long, meaning that even low variability may be detectable from relatively bright sources with short observations.

To be sure that our results are applicable to the sources located in crowded regions we compared our results assuming that the source with the flux of  $\sim 10^{-13}$ ph cm$^{-2}$ s$^{-1} $ was located either in the uncrowded region, like SS 433, or in the region with multiple nearby TeV sources, like LS 5039.  It was found that the observed flux of the central source and its error agreed in these two cases within few percents and thus our results are valid for sources located in both crowded and uncrowded regions.

\subsubsection{Cyg X-3}\label{Cyg_x3:simulations}
Cyg X-3 is a HMXB located at a distance of {$\sim$9 kpc \citep{2023ApJ...959...85R}.} 
The companion star is a Wolf-Rayet (WR) with a strong wind {mainly composed of helium}. The nature of the compact object is still unknown, although a black hole scenario is favored {\citep{Zdziarski2013,2022ApJ...926..123A}}. The orbital period is very short, $\sim4.8$ hours, indicating that the compact {object} is very close to the WR {star}, totally enshrouded in its stellar wind (orbital distance $\sim 3 \times 10^{11}$ cm). {Recent observations with the Imaging X-ray Polarimetry Explorer (IXPE) show a high X-ray polarization degree from the system, during different spectral states, indicating the presence of collimated optically-thick outflows, which hide the central engine \citep{Valedina2024a,Valedina2024b}.} The binary system is known to produce giant radio flares (flux > 10 Jy), 
produced by synchrotron processes from a relativistic jet oriented very close to the line of sight. Transient gamma-ray activity above 100 MeV was reported for the first time in 2009 by AGILE \citep{Tavani2009} and \textit{Fermi}-LAT \citep{Fermi-LAT2009}, {and reported in several studies over the years, since its discovery (see \citealp{Prokhorov2023} for a recent study on the transient activities observed by Fermi-LAT).} The flaring activity {(typical duration: 1-2 days)} was observed in coincidence with a repetitive pattern of multi-frequency emission \citep{Piano2012}: the gamma-ray flares have been detected (i) during soft X-ray spectral states (around minima of the hard X-ray light curve), (ii) in the proximity of spectral transitions, and (iii) a few days before giant radio flares. In particular, transient gamma-ray emission was found when the system is moving into or out of the quenched state, a spectral state -- characterized by a very low (or undetectable) flux at radio and hard X-ray frequencies -- that is known to occur a few days before major radio flares. 

The quenching activity of Cyg X-3 {turns out} to be a key condition for the observed activity above 100 MeV. According to theoretical models, a simple leptonic scenario -- based on inverse Compton (IC) scattering between electrons/positrons accelerated in the jet and seed photons from the WR companion -- can account for the flaring gamma-ray fluxes and the 4.8 h modulation detected by \textit{Fermi}-LAT during the transient activity \citep{Dubus2010,Prokhorov2023}. A simple phenomenological picture, based on dominant leptonic processes in the jet, can account for the non-thermal emission pattern: around the quenching, the jet would consist of plasmoids, ejected with high Lorentz factor. This transient jet would be responsible for the HE flare (for IC processes), produced in the proximity of the binary system ($10^{10} - 10^{12}$ cm), and it would subsequently produce the major radio flares (synchrotron processes), by moving out from the central engine (distances $> 10^{14}$ cm). MAGIC repeatedly observed Cyg X-3, both during hard and soft spectral states, but never detected any significant VHE activity from the microquasar \citep{Aleksic2010}. 

\paragraph{Cyg X-3: transient emission}

We carried out simulations 
by assuming two different theoretical models based on IC processes in the jet \citep{Piano2012,Zdziarski2018}, in order to test the possibility of a CTAO detection of transient VHE gamma rays from Cyg X-3.

We performed a binned analysis in the energy range 100 GeV -- 1 TeV with \verb+ctools+, by simulating observations with the Alpha Configuration of the Northern array of the CTAO observatory (IRF: \verb+North_z20+), taking into account the energy dispersion. A multi-source input model with the main background TeV sources {(see Section \ref{massive_micro})} and the CTAO instrumental background (\verb+CTAOIrfBackground+) has been considered.


In the first case, we adopted a simple power-law spectrum {(see Eq. \ref{eq:1})} inferred from the leptonic \textit{Model A} from \citet{Piano2012}, 
where prefactor $P_f = 1.34 \times 10^{-21}$  \diff_flx , index $\gamma = 4.5$ and pivot energy $E_0 = 1$ TeV.
The leptonic model is based on IC scatterings between accelerated electrons in the jet and soft seed photons from the accretion disk (X-rays) and from the companion star (UV). We simulated 5 hour and 50 hour observations, and we investigated the resulting simulated data, by performing a binned analysis. The resulting spectra are shown in Fig.~\ref{fig:cyg_x-3_piano}, together with the X-ray ``hypersoft'' spectrum \citep{Koljonen2010}, the AGILE flaring spectrum \citep{Piano2012} and the MAGIC 
flux ULs observed during the soft states \citep{Aleksic2010}. All the spectra (not simultaneous observations) are referred to the same spectral state of Cyg X-3 when the transient gamma-ray activity is detected at MeV-GeV energies (quenched state). We show the reference theoretical model and the input simulated power-law, together with the CTAO simulated spectra. By assuming this input spectrum we found no detection with CTAO-N with 5-h observation and a weak hint of signal ($\sim$3$\sigma$) for 50 hours of observation time.

\begin{figure} 
\centering
\includegraphics[width=\columnwidth]{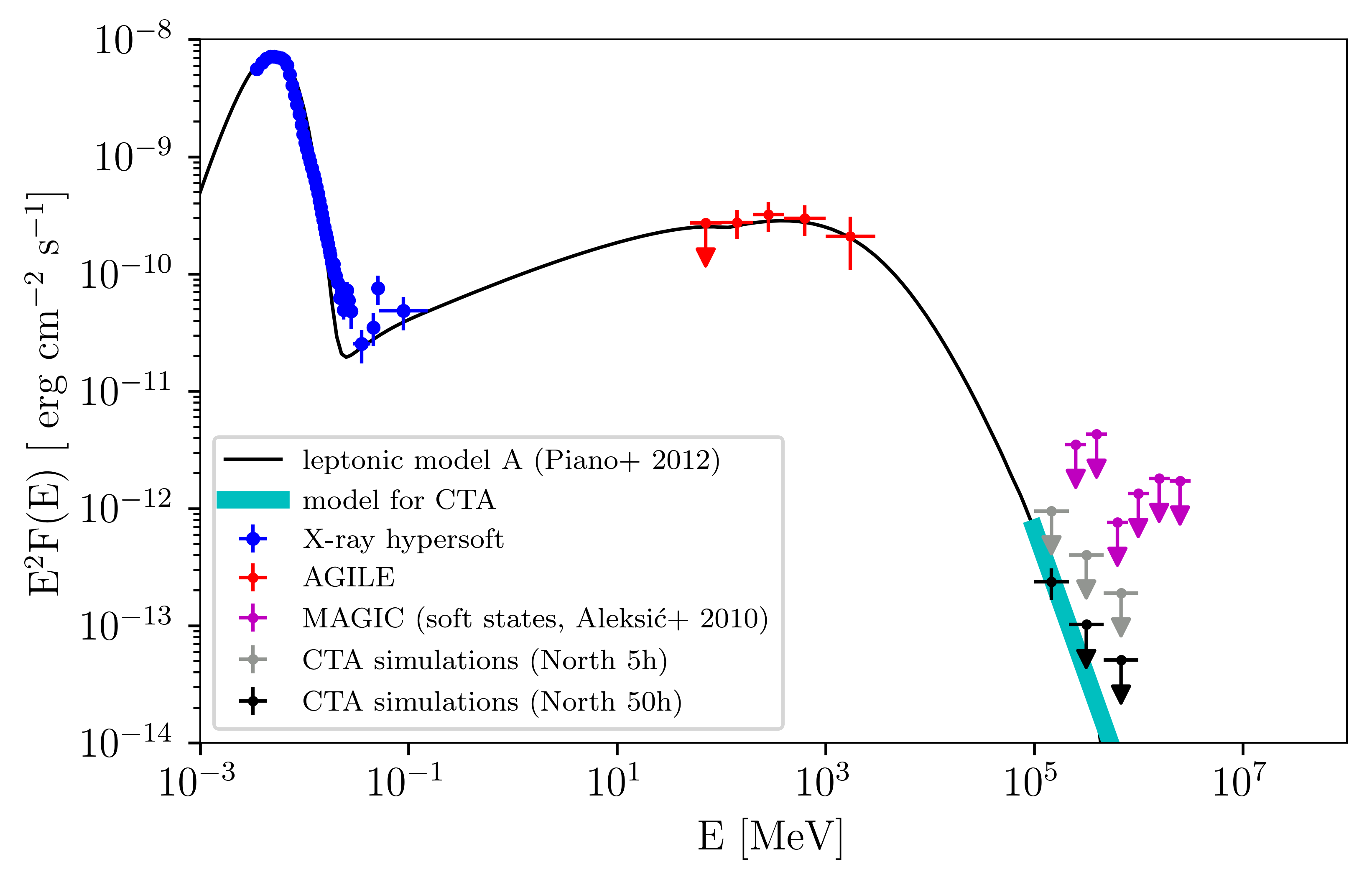} 
    \caption{Multi-frequency spectral energy distribution of Cyg X-3. Black solid curve: leptonic model A \citep{Piano2012}. Cyan solid curve: CTAO input model for the simulation. Blue points: ``hypersoft'' X-ray spectrum \citep{Koljonen2010}, RXTE-PCA and RXTE-HEXTE data ($\sim$3 to $\sim$150 keV). Red points: HE gamma-ray cumulative flaring spectrum \citep{Piano2012}, AGILE (50 MeV -- 3 GeV). Magenta points: VHE gamma-ray flux ULs (95\% C.L.) from \citet{Aleksic2010}, MAGIC (199 GeV -- 3.16 TeV). Gray points: CTAO {flux ULs} for a simulated observation of 5 ho. Black points: CTAO spectrum for a simulated observation of 50 h.}
    \label{fig:cyg_x-3_piano}
\end{figure}


{In the second case}, we assumed a different theoretical model, developed by \citet{Zdziarski2018} in order to fit the flaring spectrum from Cyg X-3 as detected by \textit{Fermi}-LAT (cumulative spectrum of 49 1-day flares detected between August 2008 and August 2017). The theoretical model presented in their paper is similar to the one presented in \citet{Piano2012}, but in \citet{Zdziarski2018},  the electrons in the jet scatter blackbody soft photons from the companion star only. The orbital and geometrical parameters are similar.
Also in this case, the model is focused on the HE emission from the microquasar (E $\leq$ 100 GeV). Thus, we assumed a simple power-law extension of the model up to TeV energies (assuming: prefactor $P_f = 2.15 \times 10^{-19}$  \diff_flx , index $\gamma = 2.85$ and pivot energy $E_0 = 1$ TeV). Similarly, we simulated 5 h and 50 h observations. The results of these simulations are shown in Fig.~\ref{fig:cyg_x-3_zdziarski}. In this case, by assuming a harder and brighter input spectrum, we found clear detections with CTAO-N: $\sim$10$\sigma$ with 5-h observation,  $\sim$30$\sigma$ with-50 h observation.

Thus, by assuming two simple power-law input spectra adapted from theoretical leptonic models -- both created ad hoc in order to account for the flaring activity observed by AGILE and \textit{Fermi}-LAT --  a possible detection with CTAO North is plausible {even with a few hours observations}. {It is important to note that these extrapolation do not take into account $\gamma \gamma$ absorption for pair production in the companion star's photon field, which could be not negligible between 100 GeV and 1 TeV \citep{Zdziarski2012}.} {Nevertheless, we cannot rule out to detect the 4.8 h orbital modulation, in the case of a prolonged TeV flare.} A CTAO detection of transient VHE gamma-ray activity would represent an unprecedented {result for this elusive system, never observed at TeV energies}. Nevertheless, a CTAO non-detection would give new strong constraints on theoretical models about microquasars. The lack of a transient VHE signal from Cyg X-3, correlated with non-thermal flaring activity, could indicate that: (i) the TeV signal, eventually produced in the jet, is absorbed for pair production by the companion star's UV photons; (ii) the acceleration efficiency in the jet is intrinsically low, the maximum energies of the jet particles are not sufficient to generate TeV photons.

\begin{figure} 
\centering
\includegraphics[width=\columnwidth]{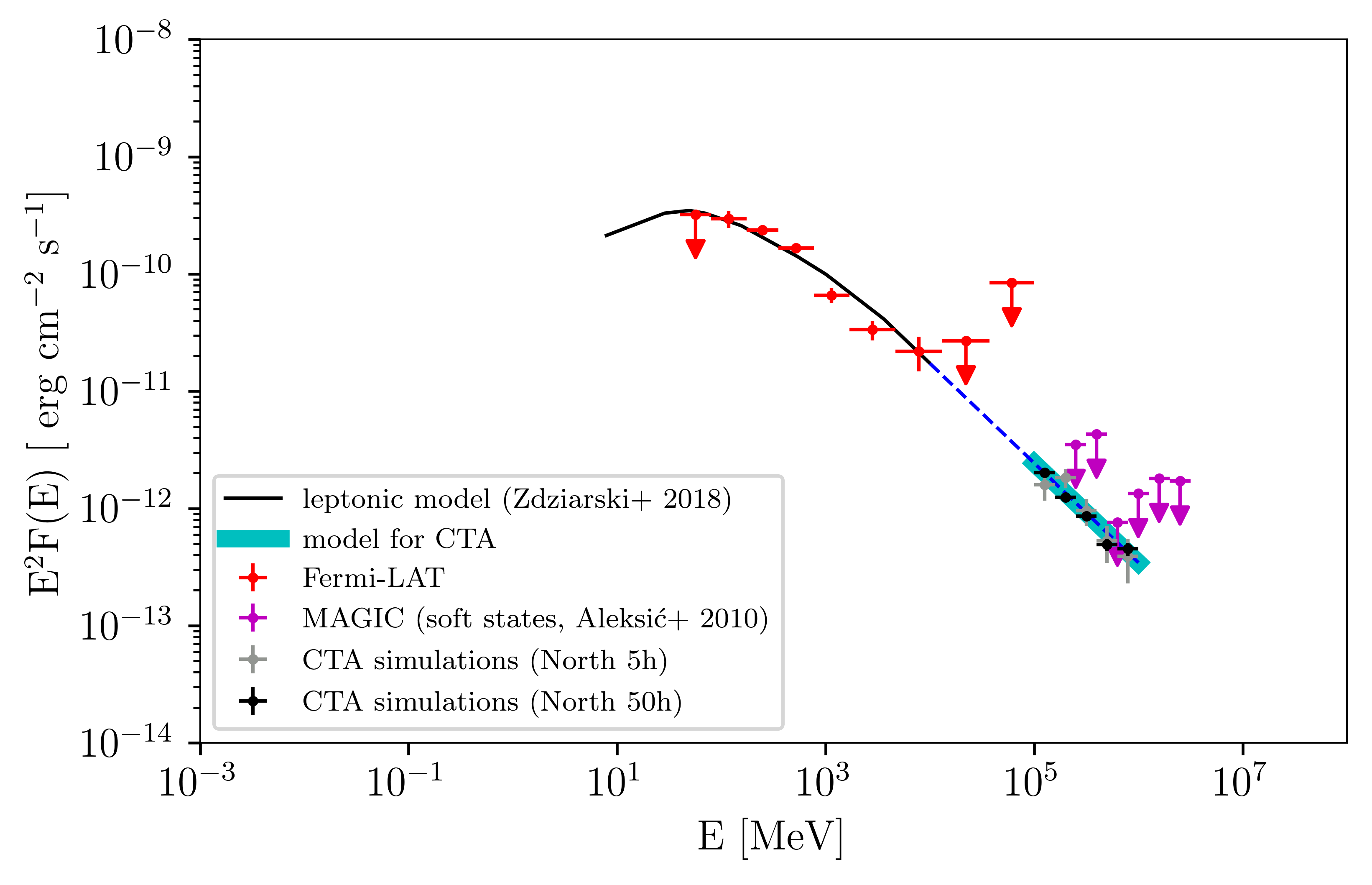} 
    \caption{Gamma-ray spectral energy distribution of Cyg X-3. Black solid curve: leptonic model \citep{Zdziarski2018}. Power-law extension of the model up to TeV energies. Cyan solid curve: CTAO input model for the simulation. Red points: HE gamma-ray cumulative flaring spectrum \citep{Zdziarski2018}, Fermi-LAT (50 MeV -- 100 GeV). Magenta points: VHE gamma-ray flux ULs (95\% C.L.) from \citet{Aleksic2010}, MAGIC (199 GeV -- 3.16 TeV). Gray points: CTAO spectrum for a simulated observation of 5 h. Black points: CTAO spectrum for a simulated observation of 50 h.}
    \label{fig:cyg_x-3_zdziarski}
\end{figure}

\subsubsection{Cyg X-1}\label{Cyg X-1: CTAO simulations}
Cyg X-1 is a HMXB, composed of a black hole ($M_{\rm X} = 21.2 \pm 2.2$ $M_{\odot}$) and a O9.7Iab supergiant companion star ($M_{\rm opt} = 40.6 ^{+7.7}_{-7.1}$ $M_{\odot}$, \citealp{Miller-Jones2021}). The system is located at a distance of $2.22^{+0.82}_{-0.17}$ kpc \citep{Miller-Jones2021}, and the orbital period is 5.6 days. The X-ray spectra can be accurately modeled by hybrid Componization models \citep{Coppi1999}. The soft state of Cyg X-1 is characterized by a strong disk blackbody component peaking at $kT\sim1$ keV and a power-law tail extending up to $\sim$10 MeV, related to Componization processes in the corona. In the hard state, the accretion disk is truncated and the emission from the corona is dominant. In this state, the coronal plasma is composed by a hot quasi-thermal population of electrons ($kT \sim 100$ keV) with a sharp cutoff at $\sim$200 keV. At sub-MeV energies, the microquasar exhibits a non-thermal power-law tail with a strong linear polarization \citep{Laurent2011, Jourdain2012}. This emission could be ascribed {either to synchrotron processes in the jet, by assuming a very efficient particle acceleration and strong jet magnetic fields \citep{Zdziarski2014}, or to the corona itself \citep{Romero2014}}. Recent studies investigate the physical origin of this power-law tail at sub-MeV energies, detected during both soft and hard spectral states \citep{Cangemi2021}.
Above 100 MeV, deep observations with \textit{Fermi}-LAT found evidences of persistent emission from Cyg X-1 only during hard X-ray spectral states \citep{Zanin2016,Zdziarski2017}. Transient HE emission was observed by AGILE \citep{Bulgarelli2010, Sabatini2010, Sabatini2013} 
on 1-2 day timescales, in coincidence with both hard and soft X-ray spectral states. At TeV energies, a hint of detection ($\sim$4$\sigma$) was {observed} by MAGIC on September 24, 2006 \citep{Albert2007}, during a hard X-ray flare of Cyg X-1. {A $\sim$4$\sigma$ persistent TeV signal was recently reported by LHAASO  \citep{LHAASO2024arXiv241008988L}}.

For Cyg X-1, we investigated the possibility that CTAO will detect both transient and persistent emission from the microquasar.

\paragraph{Cyg X-1: transient emission.} \label{Cyg X-1: transient}
In this case, we carried out a simulated short-term observation of Cyg X-1, during a possible VHE gamma-ray flare. We simulated a 30-minute observation with the same setup reported in Section~\ref{Cyg_x3:simulations}: a multi-source simulation with photon energies between 100 GeV and 1 TeV. We assumed, as input spectrum for the simulation, the same power-law observed by MAGIC in September 24, 2006 (\citet{Albert2007}; prefactor $P_f = 2.3 \times 10^{-18}$  \diff_flx , index $\gamma = 3.2$, pivot energy $E_0 = 1$ TeV). We obtained an overall detection of {the source at a significance level of} $\sim$38$\sigma$. The resulting spectrum is shown in Fig.~\ref{fig:cyg_x1_flaring_spectrum}, together with the observed flaring spectrum observed by MAGIC. Our results confirm that CTAO will be able to detect a flare similar to the one reported by MAGIC in 2006 in a few minute observation, with unprecedented spectral accuracy. {A fainter TeV flare - weaker than the one reported by MAGIC - would require a longer CTAO observation (a few hours) to be significantly detected. This possibility will be properly assessed in a potential ToO observation, on the basis of the triggering flare flux in other wavelengths.}

\begin{figure} 
\centering
\includegraphics[width=\columnwidth]{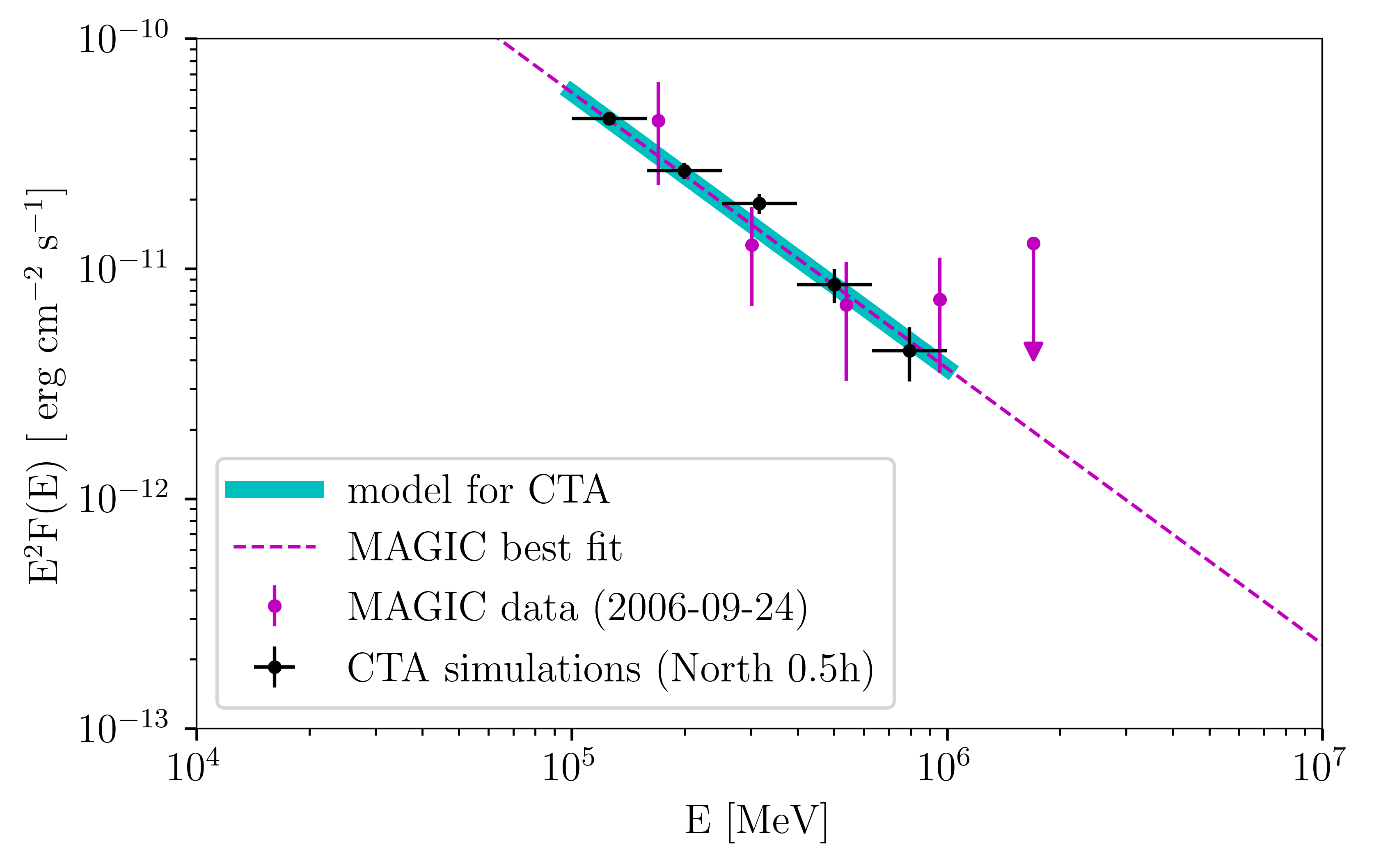} 
    \caption{VHE gamma-ray spectral energy distribution of Cyg X-1, related to the 2006-September flaring activity. Magenta points: VHE gamma-ray spectrum from \citet{Albert2007}, MAGIC (150 GeV -- 1.9 TeV), {accounting for} 78.9 minutes of observation. Magenta dashed line: MAGIC best fit. Cyan solid curve: CTAO input model for the simulation. Black points: CTAO spectrum for a simulated observation of 30 minutes.}
    \label{fig:cyg_x1_flaring_spectrum}
\end{figure}

\paragraph{CygX-1: persistent emission.} Cyg X-1 exhibits persistent HE emission during the hard state, as observed by \textit{Fermi}-LAT \citep{Zanin2016,Zdziarski2017}. Thus, we investigated the possibility of a CTAO detection of VHE persistent emission above 100 GeV. Again, we assumed the same setup as reported in  Section~\ref{Cyg_x3:simulations}. We analyzed three different scenarios. In the first one, we assumed as input spectral model for CTAO, a simple extension of the power-law spectral shape reported in the \textit{Fermi}-LAT 4FGL Catalog, without any cut off around 100 GeV \citep{Abdollahi2020}. In the second scenario, we assumed a spectral shape based on a purely leptonic theoretical model, {in which gamma-ray emission is produced due to IC scatterings} in the persistent jet during the hard state \citep{Zdziarski2017}. According to this model, a sharp cut off -- due to the Klein-Nishina effects -- is 
predicted at $\sim$100 GeV. In the third scenario, we assumed a spectral shape based on the lepto-hadronic theoretical model presented in \citet{Kantzas2021}. In that paper, the authors modeled the GeV persistent spectrum as detected by \textit{Fermi}-LAT during the hard state, by assuming that both electrons and protons are accelerated in the jet. A comprehensive model, based on a superposition of leptonic (IC scatterings) and hadronic processes (gamma rays from the decay of neutral mesons, produced in $p$$\gamma$ interactions) can properly fit the multi-wavelength spectrum up to the high-energy emission from Cyg X-1.

For the first hypothesis (4FGL-like spectrum), we assumed a simple power-law (assuming: prefactor $P_f = 3.2 \times 10^{-14}$  \diff_flx , index $\gamma = 2.15$ and pivot energy $E_0 = 4.15$ GeV). A multi-source simulation with photon energies between 100 GeV and 1 TeV has been carried out. With this spectrum, we obtained a detection with a significance of $\sim$17$\sigma$ for a 50h simulated observation. The resulting simulated spectrum is shown in Fig.~\ref{fig:cyg_x1_fsteady_spectrum_4fgl}, together with the \textit{Fermi}-LAT 4FGL spectrum \citep{Abdollahi2020}, and the MAGIC flux ULs during the hard state \citep{Ahnen2017a}.

For the second hypothesis (spectrum inferred from \citealp{Zdziarski2017}), we assumed a simple power-law (with prefactor $P_f = 9.5 \times 10^{-21}$  \diff_flx , index $\gamma = 3.2$ and pivot energy $E_0 = 1$ TeV). A multi-source simulation with photon energies between 100 GeV and 1 TeV has been carried out. In this case, we did not detect any significant emission with CTAO with a simulated observation of 50 h (significance $\sim$2$\sigma$). The resulting differential spectral ULs are shown in Fig.~\ref{fig:cyg_x1_fsteady_spectrum_zdz}, together with the theoretical model from \citet{Zdziarski2017}, the HE gamma-ray spectra as detected by \textit{Fermi}-LAT \citep{Zanin2016,Zdziarski2017}, and the MAGIC ULs related to the hard state \citep{Ahnen2017a}.

For the third hypothesis (spectrum inferred from \citealp{Kantzas2021}), we used as input the theoretical model itself, by simulating photon energies between 100 GeV and 100 TeV. We carried out the usual multi-source binned analysis, and we found a clear detection with a significance of $\sim$36$\sigma$ for a 50h simulated observation. The resulting simulated spectrum is shown in Fig.~\ref{fig:cyg_x1_fsteady_spectrum_lh}, together with the theoretical model from \citet{Kantzas2021}, the HE gamma-ray spectra as detected by \textit{Fermi}-LAT \citep{Zanin2016,Zdziarski2017}, the MAGIC flux ULs during the hard state \citep{Ahnen2017a} {and the HAWC flux ULs between 0.1 and 100.0 TeV \citep{Albert2021}. In particular, we note that above 10 TeV the HAWC ULs for a prolonged stacked observation (1523 days) are below our theoretical model. This spectral behavior could weaken the chance of a CTAO sharp detection at the highest energies.}

\begin{figure} 
\centering
\includegraphics[width=\columnwidth]{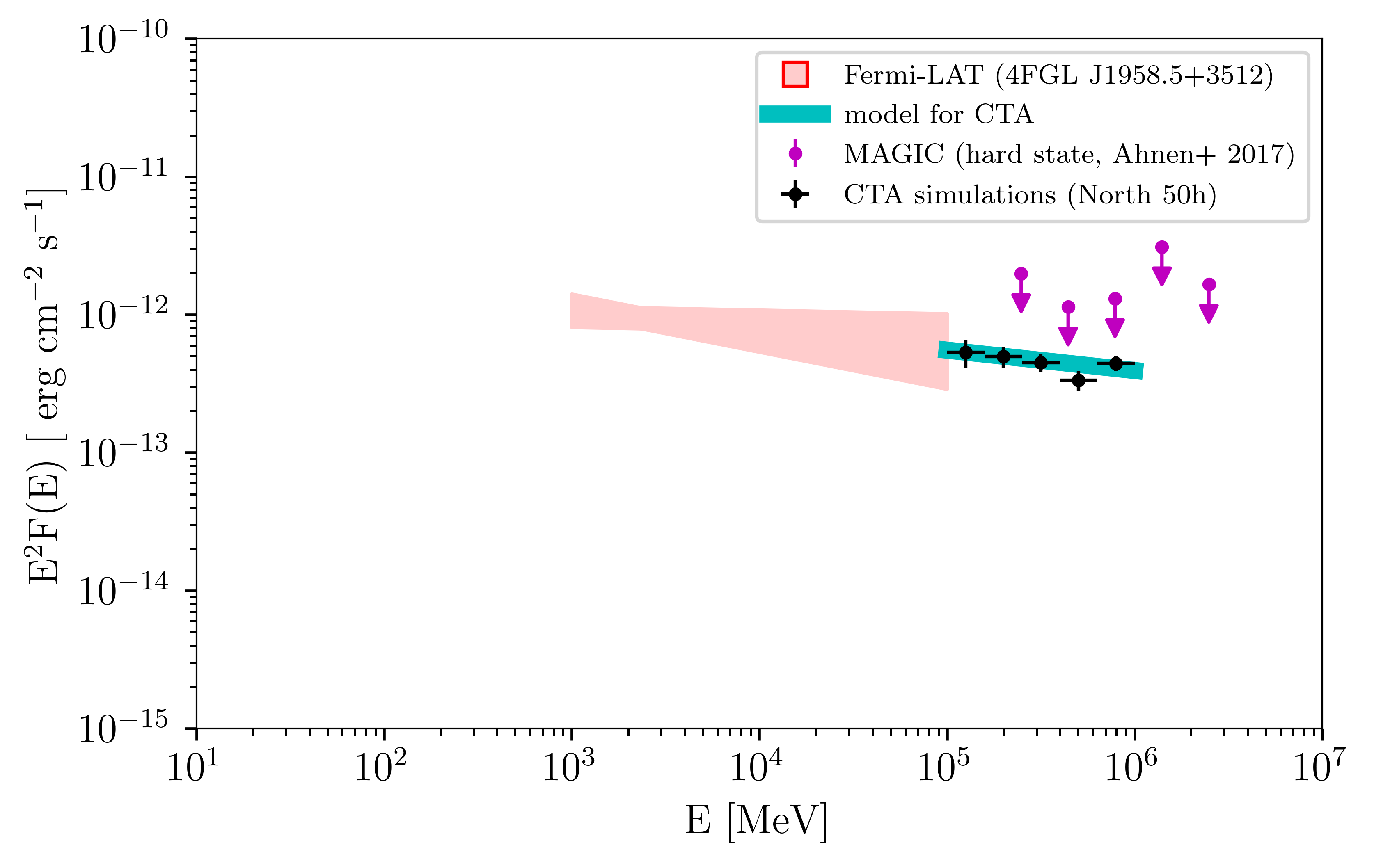} 
    \caption{Gamma-ray SED of Cyg X-1, for a possible steady emission up to VHE. Red shaded region: The \textit{Fermi}-LAT 4FGL Catalog HE steady spectrum,  \citep{Abdollahi2020}, \textit{Fermi}-LAT (1-100 GeV).  Cyan solid curve: CTAO input model for the simulation. Magenta points: MAGIC  (160 GeV -- 3.5 TeV) VHE ULs (95\% C.L.) from \citet{Ahnen2017a}. Black points: CTAO spectrum for a simulated observation of 50 h.}
    \label{fig:cyg_x1_fsteady_spectrum_4fgl}
\end{figure}

\begin{figure} 
\centering
\includegraphics[width=\columnwidth]{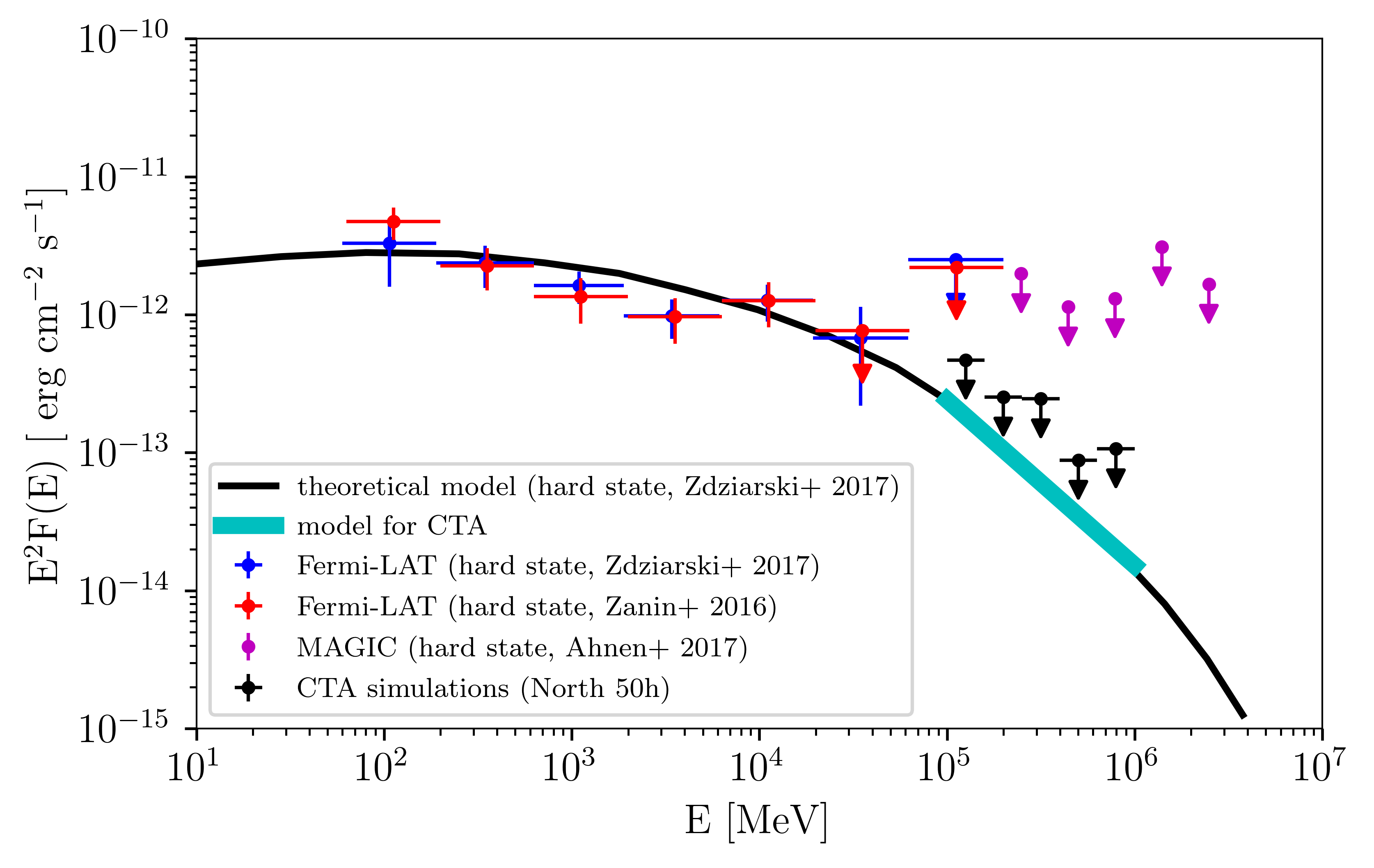} 
    \caption{Gamma-ray SED of Cyg X-1, for a possible steady emission up to VHE. Black solid curve: theoretical model from \citet{Zdziarski2017}, based on IC processes in the jet during the hard state. Blue points: HE steady spectrum during the hard state \citep{Zdziarski2017}, \textit{Fermi}-LAT (60 MeV -- 200 GeV). Red points: HE steady spectrum during the hard state \citep{Zanin2016}, \textit{Fermi}-LAT (60 MeV -- 200 GeV).  Magenta points: MAGIC (160 GeV -- 3.5 TeV) VHE flux ULs (95\% C.L.) from \citet{Ahnen2017a}. Cyan solid curve: input model used for the simulation. Black points: simulated spectrum for a CTAO observation of 50 h. }
    \label{fig:cyg_x1_fsteady_spectrum_zdz}
\end{figure}

\begin{figure} 
\centering
\includegraphics[width=\columnwidth]{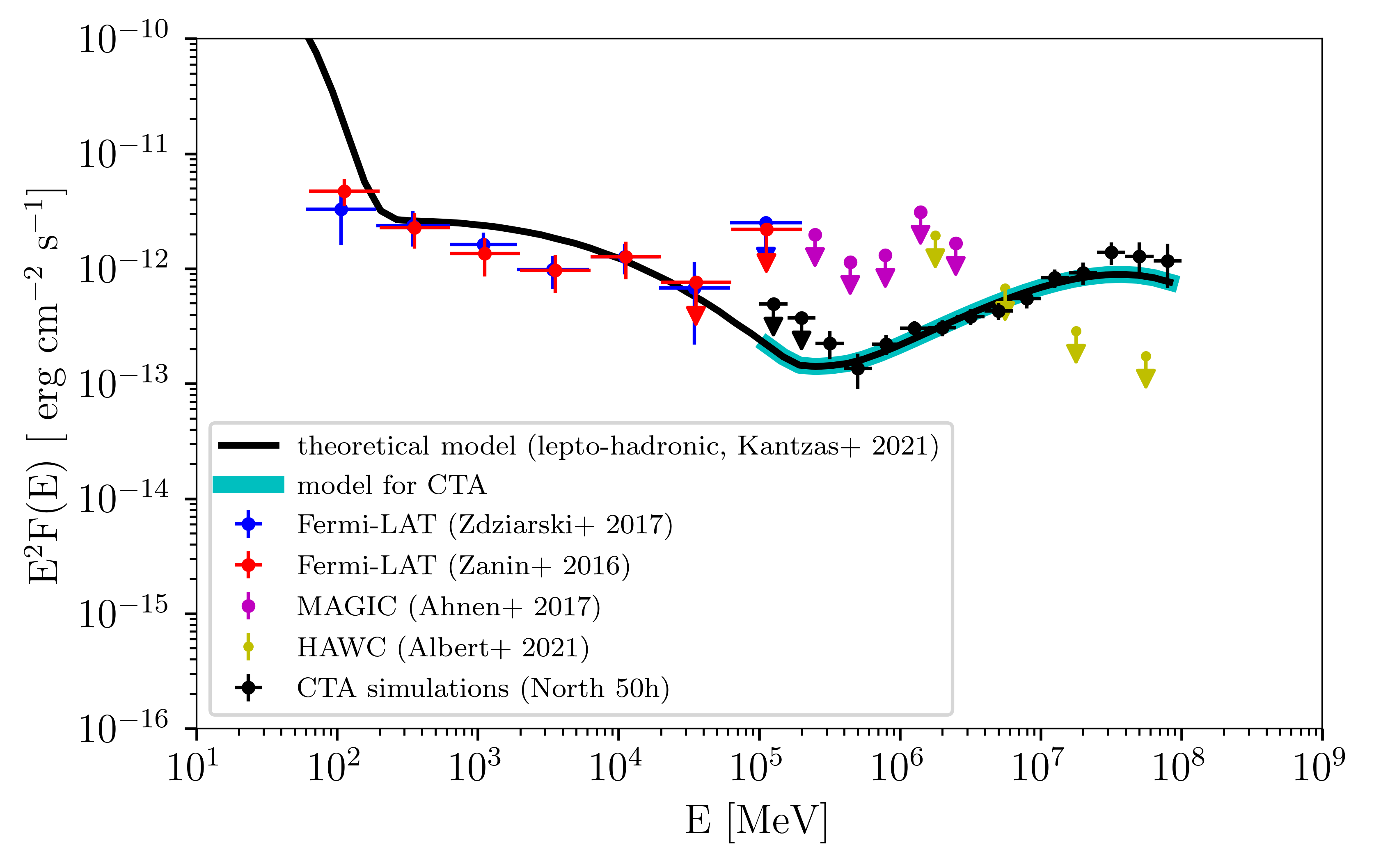} 
    \caption{Gamma-ray SED of Cyg X-1, for a possible steady emission up to VHE. {Blue, red and magenta points as reported in Fig.~\ref{fig:cyg_x1_fsteady_spectrum_zdz}}. {Yellow ULs: HAWC observations \citep{Albert2021}.} Black solid curve: theoretical model from \citet{Kantzas2021}, based on lepto-hadronic processes in the jet during the hard state. Cyan solid curve: input model used for the simulation. Black points: simulated spectrum for a CTAO observation of 50 h.}
    \label{fig:cyg_x1_fsteady_spectrum_lh}
\end{figure}

Thus, according to our simulations, CTAO will be able to detect a possible persistent VHE gamma-ray emission from the jet of Cyg X-1, if the spectrum is not characterized by a sharp cut off around 100 GeV. According to purely leptonic models, a sharp cut off is expected below 1 TeV. {On the contrary}, hadronic processes could be responsible for a bright emission above 1 TeV, which could be detected by the CTAO Observatory.

\subsection{Low-mass X-ray binaries}\label{CTAO_lmxbs}

LMXBs harbor a low-mass companion star and a black hole (or an accreting neutron star), object tightly connected to jet launching that are responsible for the non-thermal multi-wavelength emission (see review in \citealp{2022abn..book.....C}). {Up to now, no LMXB has been detected at HE (apart from tMSPs) and only strong hints of emission at HE have been reported in V404 Cyg. No LMXB has neither been detected at VHE by any IACT.} (see e.g. \citealt{2011ApJ...735L...5A,Ahnen2017b,2018A&A...612A..10H}). The most recent X-ray outburst of a black hole LMXB (BH-LMXB) which was followed up by the IACTs MAGIC, H.E.S.S. and VERITAS  was that of MAXI~J1820$+$070, without detecting any VHE emission \citep{2022MNRAS.517.4736A}. We examine here if CTAO will be able to detect 
such a similar exceptionally bright outburst but for a hypothetical source located within a maximal distance of 4\,kpc from Earth. Based on the theoretical lepto-hadronic model of \cite{kantzas2022gx}, used since the modeled LMXB can be considered a canonical source, we perform a number of simulations where we rescale the predicted VHE emission for a number of different {jet} inclination angles between 5 and $65^{\circ}$. We perform each simulation for a number of different hypothetical sources at different distances within 2 and 4\,kpc.  

In Fig.~\ref{fig:lmxbs 30 deg}, we show the predicted VHE flux for a BH-LMXB with inclination angle of $30^{\circ}$ assuming that the emission persists in three different energy bins between 0.1 and 10\,TeV, for at least 2~weeks, and compare it to the CTAO sensitivity curves (see Fig.~\ref{fig:CTAOsensitivityMap}). 

\begin{figure}
\centering
\includegraphics[width=\columnwidth]{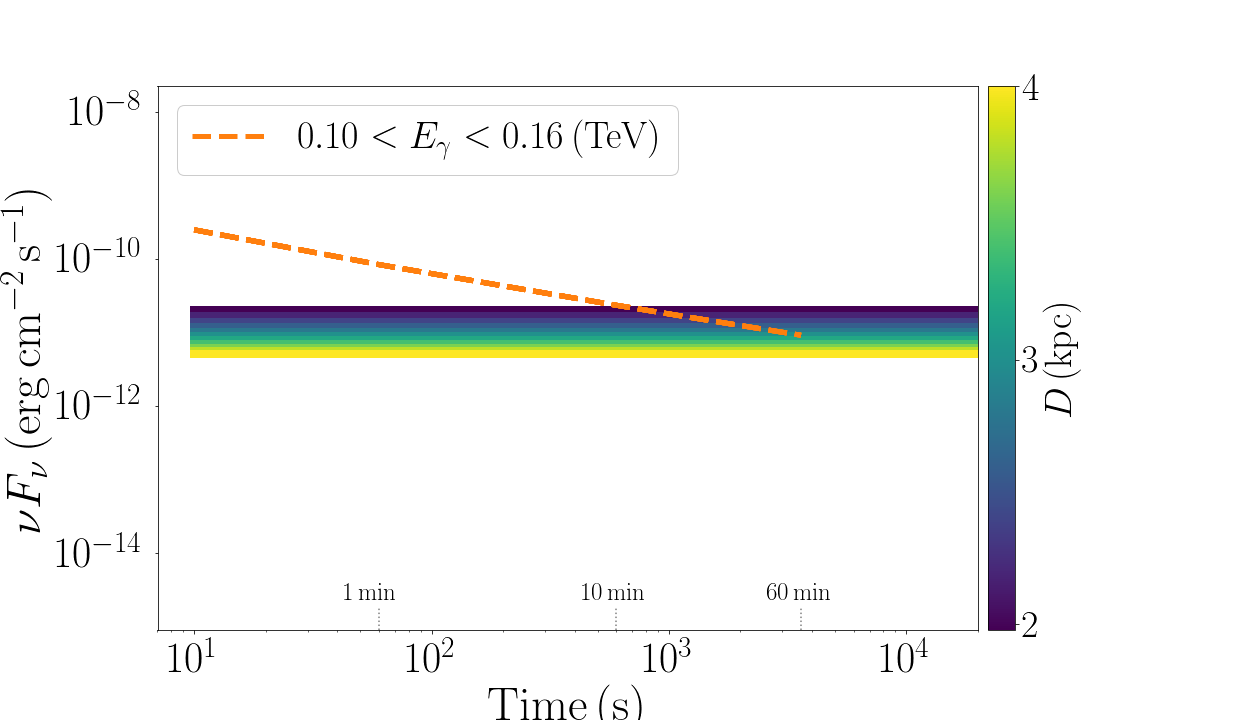} \\ \vspace{-10pt}
\includegraphics[width=\columnwidth]{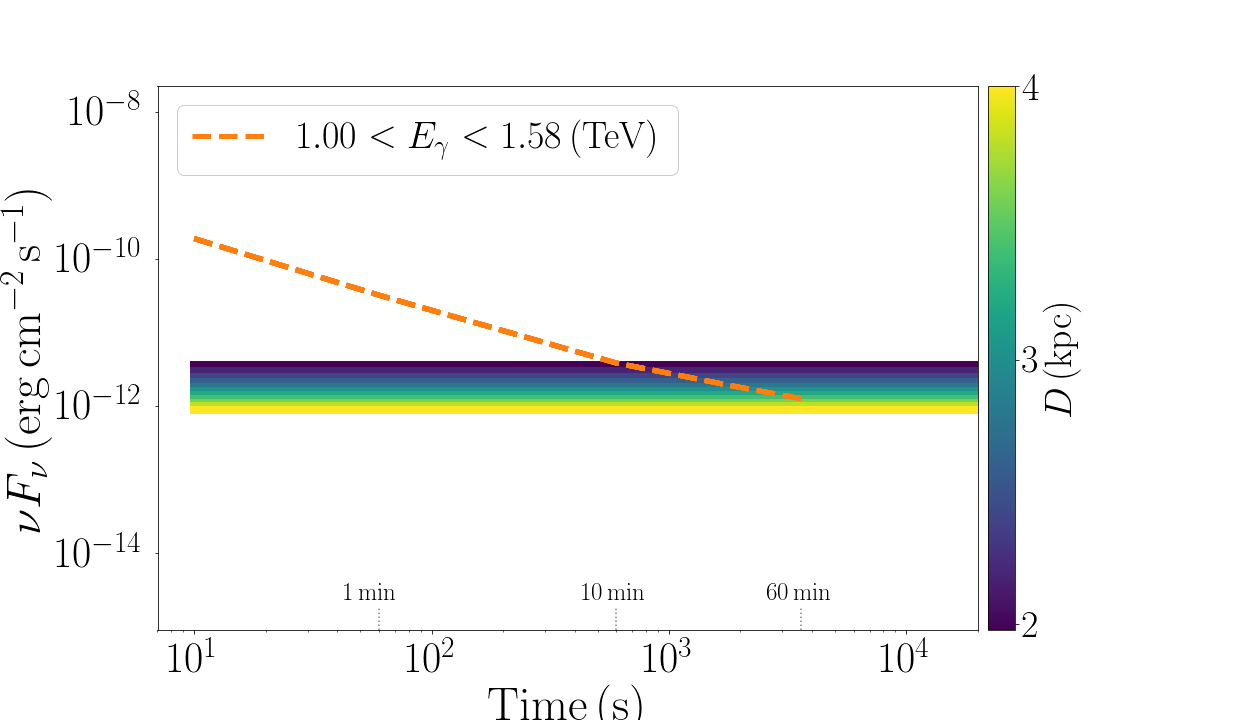} \\ \vspace{-10pt}
\includegraphics[width=\columnwidth]{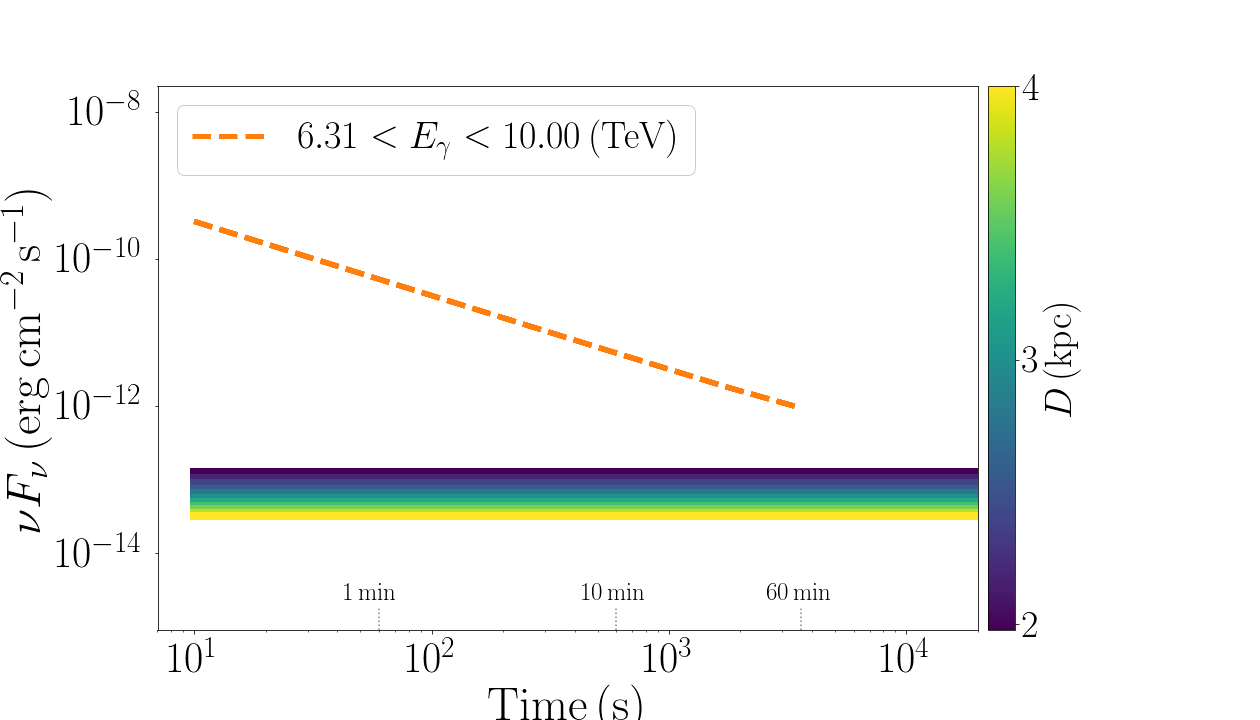} 
    \caption{Predicted VHE emission of a hypothetical BH-LMXB for three different energy bins, as shown in the legends. The BH-LMXB follows the recent outburst of MAXI~J1820$+$070, but with an inclination angle of $30^{\circ}$ instead, and its distance given by the colormap (lighter colors correspond to {more distant} sources). We assume an emission lasting at least two weeks. {The CTAO sensitivity for each energy bin is represented as a dashed orange line.}}
    \label{fig:lmxbs 30 deg}
\end{figure}

\begin{figure}
\centering
\includegraphics[width=\columnwidth]{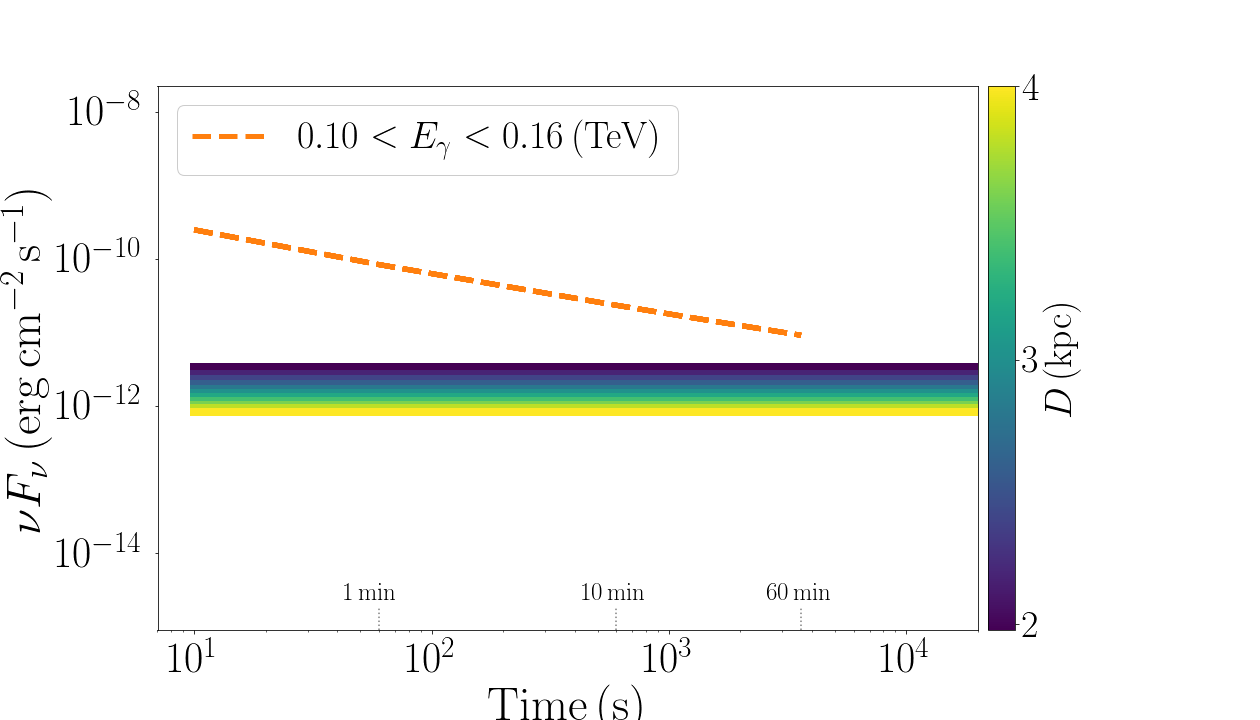} \\ \vspace{-10pt}
\includegraphics[width=\columnwidth]{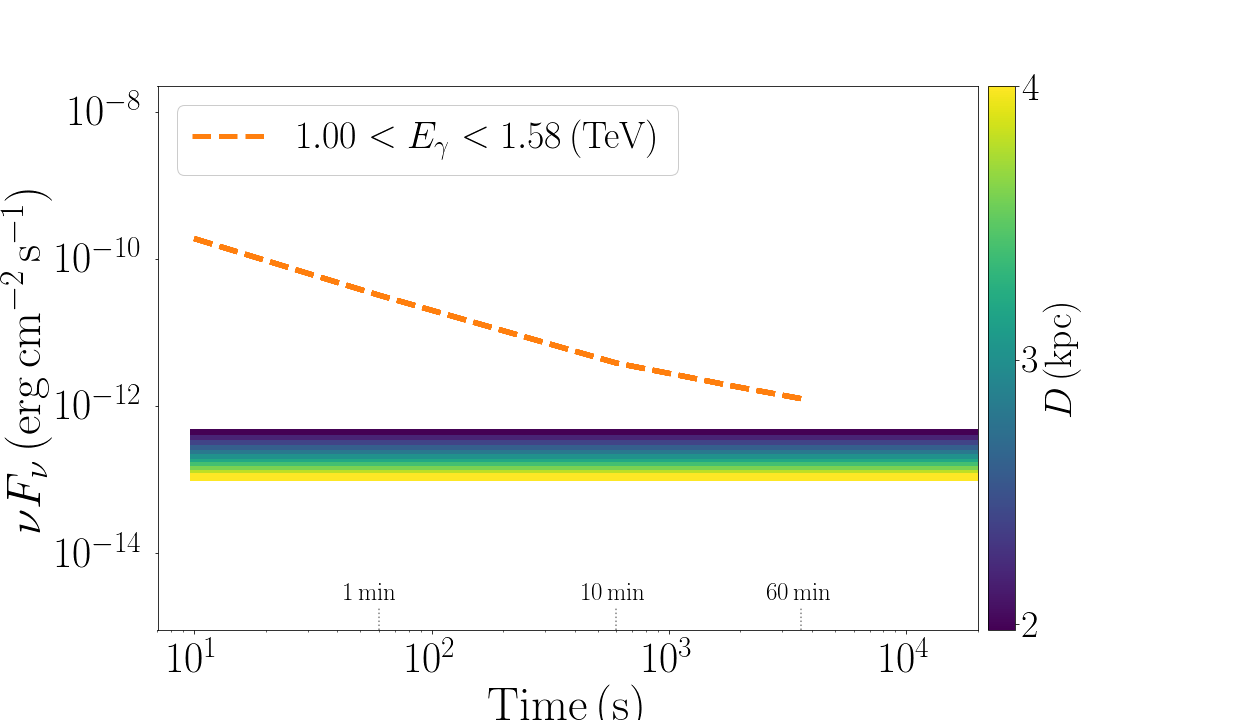} \\ \vspace{-10pt}
\includegraphics[width=\columnwidth]{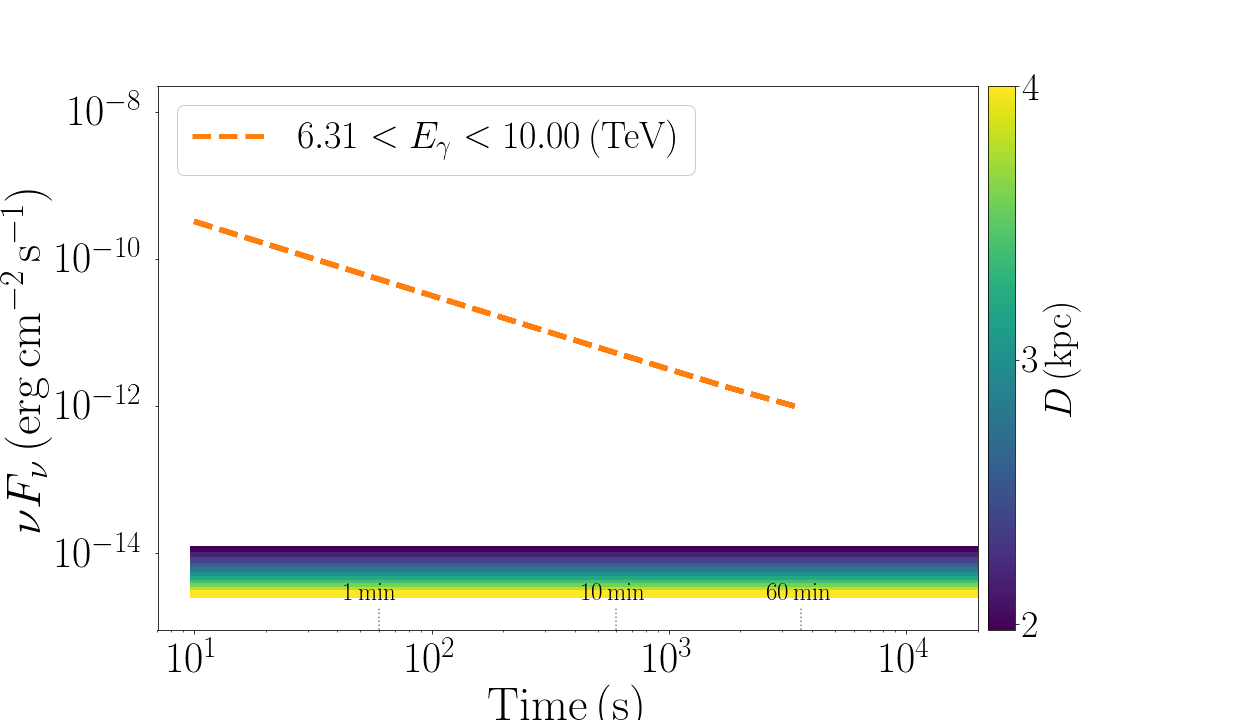} 
    \caption{Same as Fig.~\ref{fig:lmxbs 30 deg} but for an inclination angle of 40$^{\circ}$. }
    \label{fig:lmxbs 40 deg}
\end{figure}

We overall see that CTAO will be able to detect an outburst similar to MAXI~J1820$+$070 in the sub-TeV regime within a few tens of minutes if the LMXB is located closer than $\sim4\,$kpc at energies $<1.6$\,TeV. The inclination angle of the LMXB we assume here is relatively small compared to the average value between $40$ and $70^{\circ}$ \citep[see, e.g.,][]{Tetarenko2016}, but LMXBs with an inclination angle greater than $\sim30^{\circ}$ fail to be detected within the first hour of observations (Fig.~\ref{fig:lmxbs 40 deg}). Sources with an inclination angle less than $\sim20^{\circ}$ could be observed within a few minutes, 
such as the case of MAXI~J1836$-$194 or V4641 Sgr, both 
microblazar candidates \citep{Russell2014J1836,2014MNRAS.438L..41G}. 

\subsubsection{The case of V404 Cyg.}\label{V404 Cyg: CTAO simulations}
The system V404 Cyg is a LMXB located at a distance of $2.39 \pm 0.14$ kpc, inferred {through parallax measurements} \citep{Miller-Jones2009}. The system is composed of a $9 ^{+0.2}_{-0.6}$ $M_\odot$ black hole and a $0.7^{+0.3}_{-0.2}$ $M_\odot$ K3 III companion star with an orbital period of $6.4714 \pm 0.0001$ days \citep{Casares1992}. LMXBs are known to undergo long periods of quiescence (years) and rapid outburst states (weeks). After a $\sim$26 year-long quiescent phase, V404 Cyg entered in a bright active phase in the second half of June 2015. The outburst, lasting $\sim$2 weeks, was observed in all the bands of the electromagnetic spectrum, from radio to {GeV} energies. AGILE-GRID and \textit{Fermi}-LAT observed a strong hint of emission in gamma rays ($\approx$4$\sigma$), coincident with the brightest peak of luminosity observed in radio, hard X-ray and soft gamma-ray bands \citep{Loh2016,Piano2017}. The gamma-ray event was observed between June 24 and 26 and it is simultaneous with rapid transitions between the optically thin and the optically thick phases of the radio emission in the jet, and coincident with a bright peak of the 511 keV emission line detected by INTEGRAL \citep{Siegert2016}. As for other microquasars, the HE emission could be related to either leptonic (IC scattering on soft photons) or hadronic processes (decay of $\pi^0$ mesons produced in proton-proton collisions) in the jet. Nevertheless, in this case the companion is an old spectral type, cold and small star, and it does not provide a sufficiently high density of seed photons and hadronic material in the stellar wind. Thus, the HE emission is possibly related to interactions between the particles accelerated in the jet and the radiation (and the magnetic field) of the jet itself. MAGIC repetitively pointed at V404 Cyg between June 18 and 27, for more than 10 hours, but the observations did not show any significant emission at TeV energies \citep{Ahnen2017b}. 

\subparagraph{V404 Cyg: transient emission}
We carried out a 50-h CTAO simulated observation for V404 Cyg with the same setup described in Section~\ref{Cyg_x3:simulations}: 100 GeV -- 1 TeV simulated photons with a multi-source approach. The CTAO input spectral model for V404 Cyg is a simple extension of the power-law spectrum observed by \textit{Fermi}-LAT during the 2015-June flaring activity and reported by \citet{Loh2016} (assuming: prefactor $ = 8 \times 10^{-22}$  \diff_flx , index {$\gamma$} = 3.5 and pivot energy $E_0$ = 1 TeV). The resulting spectrum from our simulation is shown in Fig.~\ref{fig:v404_cyg_spectrum}, together with the non-thermal HE spectra observed during the 2015-June flare. Thus, {even} if we assume the same spectral trend as observed by the HE gamma-ray detectors, we expect no detection with CTAO in a 50-h observation. This is in agreement with the simulations on LMXBs {described} in Subsection \ref{CTAO_lmxbs}.

\begin{figure} 
\centering
\includegraphics[width=\columnwidth]{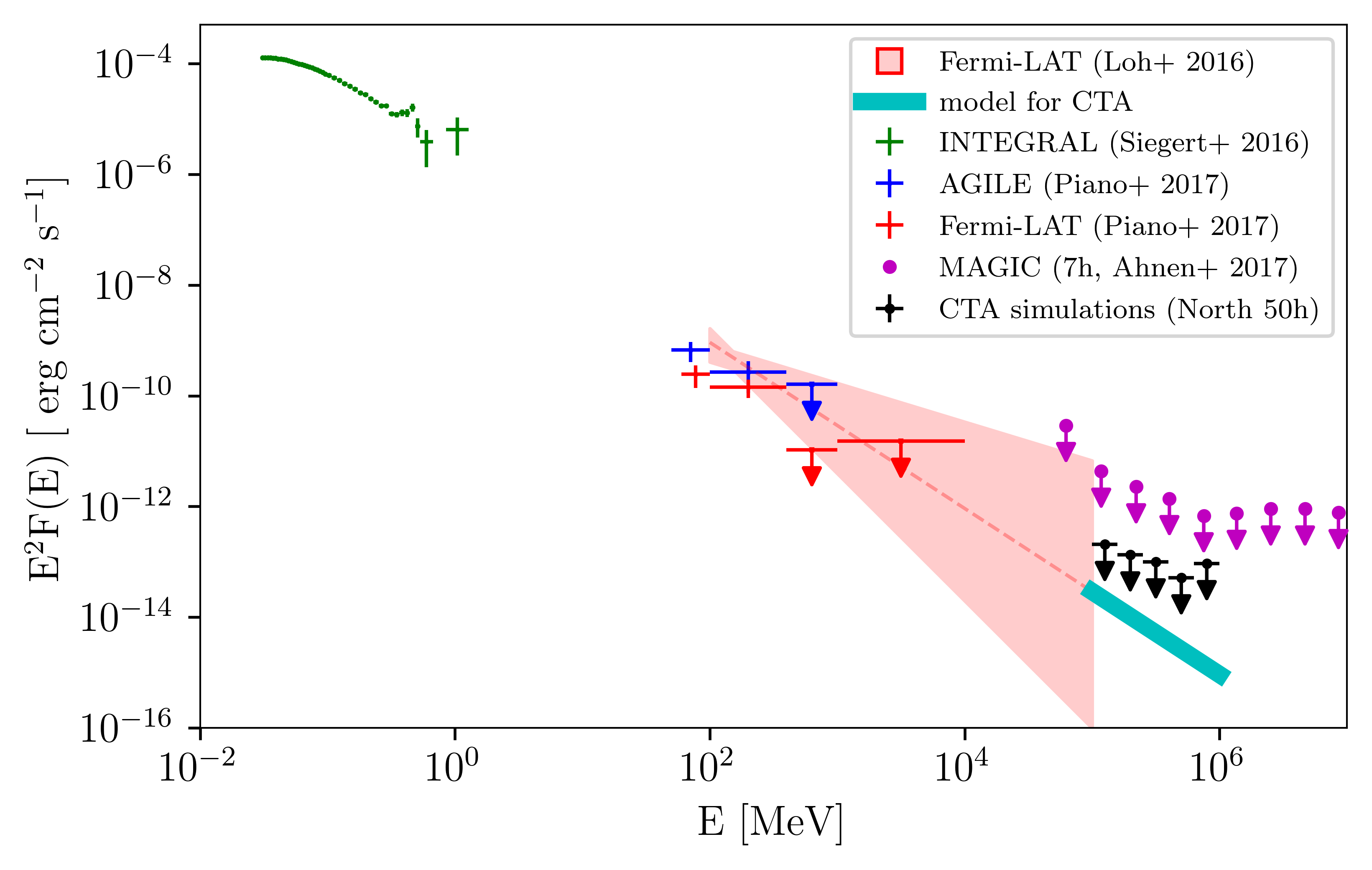} 
    \caption{Multi-frequency SED of V404 Cyg, related to the 2015-June flaring activity. Green points: hard X-ray spectrum \citep{Siegert2016}, INTEGRAL data (30 keV -- 2 MeV). Blue points: HE  flaring spectrum \citep{Piano2017}, AGILE (50 MeV -- 1 GeV). Red points: HE flaring spectrum \citep{Piano2017}, \textit{Fermi}-LAT (60 MeV -- 10 GeV). Red shaded region: HE flaring spectrum, \citep{Loh2016}, \textit{Fermi}-LAT (100 MeV -- 100 GeV).  Magenta points: MAGIC (50 GeV -- 10 TeV) VHE flux ULs (95\% C.L.) from \citet{Ahnen2017b}. Cyan solid curve: input model used for the simulation. Black points: simulated spectrum for a CTAO observation of 50 h.}
    \label{fig:v404_cyg_spectrum}
\end{figure}

\subsection{Transitional millisecond pulsars}\label{CTAO_tmsps}

Out of the three confirmed tMSPs, only PSR J1023+0038 is currently in the LMXB state, whereas XSS J1227$-$4853 and IGR J18245$-$2452 are currently in the RMSP state. As previously mentioned, other candidates were found through X-ray peculiar variability and association with \textit{Fermi}-LAT sources (see review by \citealp{2022ASSL..465..157P}).
Particularly interesting are the two confirmed tMSPs PSR J1023+0038 and XSS J1227-485 that when in LMXB state they were found by \textit{Fermi}-LAT with a {luminosity} of about 10$^{34}$ erg s$^{-1}$ in the energy range 0.1--10 GeV, which is up to ten times brighter than the levels observed during the RMSP state \citep{Papitto15_J1023,2017ApJ...836...68T}. 
This fact makes them particularly interesting for a possible detection with CTAO. In this Section, we estimate the chances of detecting these two tMSPs with CTAO given also their relatively close distance of about 1.5 kpc.


\subsubsection{PSR J1023+0038}
This tMSP was initially detected as a variable source in the radio band \citep{bond02_j1023} and
showing clear characteristics of an accretion disc around the compact object in the optical band. 
Later, \cite{Thorstensen05_j1023} suggested PSR J1023+0038 as an NS-LMXB. The observations did not reveal an accretion disc but the existence of a strong irradiation on the optical star from an unseen companion. 
The compact object was identified as a 1.69 millisecond radio pulsar in a 4.75 hr orbit around a 0.2 M$_{\odot}$ companion star \citep{Archibald09_j1023}. 
In June 2013 the source came back to a LMXB state, where it has remained until now, and the radio pulsar signal switched off. 
During the LMXB state, PSR J1023+0038 shows a peculiar behaviour in  X-rays: it exhibits frequent modes switching between three different X-ray 
levels, dubbed high, low and flaring \citep{Bogdanov15_j1023}. 
The HE gamma-ray emission detected by \textit{Fermi}-LAT has been reported to brighten by a factor of 5 after the transition. 
The average \textit{Fermi}-LAT spectrum is described by a power-law with index 1.8 and a cutoff at an energy of 2.3 GeV according to \cite{Takata14} and by a power-law with index 2 and an energy cutoff at 3.7 GeV, the significance of the cutoff is 4.3$\sigma$ level according to \cite{2017ApJ...836...68T}. Neither pulsations nor steady emission were found in the VHE regime \citep{Aliu16}. 
To test the capability of CTAO to detect emission from this source we first studied the HE gamma-ray emission from \textit{Fermi}-LAT during the LMXB state (2013-2021), in order to obtain the spectral parameters of the source. 
{For the \textit{Fermi}-LAT analysis, all photons in the energy range 0.1-300 GeV included in a circular region of 10 degrees centered on the source were considered}. The binned likelihood analysis was performed using 20 energy bins. 
The two spectral models that have been considered for the CTAO simulations are a logparabola and a broken power-law. 
{We considered only these two models, because the simulation with the power law with an exponential cut-off model did not return any detectable spectral bin in the VHE range (the cut-off is at very low energies, a few GeV). }
{On the other hand} a simple power-law extending in the energy range from GeV up to 1 TeV appears physically difficult to achieve. 
This is {compatible} by the low significance of the results by \cite{Takata14} and \cite{2017ApJ...836...68T}.

\noindent {The spectral input parameters inferred from the analysis of the \textit{Fermi}-LAT  data in the accretion phase are} for the broken power-law: {prefactor P$_{f}$ = (0.06$\pm$0.01) $\times$ 10$^{-10}$ photons cm$^{-2}$ s$^{-1}$ MeV$^{-1}$,  $\alpha_1$ = -2.12$\pm$0.03,  $\alpha_2$ = -2.91$\pm$0.06, Energy break = (1.15$\pm$0.09) GeV}; 
{while for} the logparabola: prefactor P$_{f}$ = (0.34$\pm$0.007) $\times$ 10$^{-10}$ photons cm$^{-2}$ s$^{-1}$ MeV$^{-1}$, $\alpha$ = 2.23$\pm$0.02, $\beta$ = 0.16$\pm$0.02, Energy break = 0.524 GeV. \\ Batches of 100 simulations were run using both CTAO-N and CTAO-S IRFs 
for 50, 100 and 200 hours of observations; we binned the simulated data into initial 20 logarithmic energy bins considering an energy range of 0.03-100 TeV. 
The resulting spectrum of PSR J1023+0038 is shown in Fig.\  \ref{fig:tMSPs} and, for simplicity, only the results with the highest statistic are reported (200 h observations). The full analysis results for the broken power-law model are reported in Table \ref{tab:j1023_simul}. {The parameters obtained from the unbinned analysis are:} 
{P$_{f}$ = 7.31 $\times$ 10$^{-12}$ photons cm$^{-2}$ s$^{-1}$ MeV$^{-1}$, $\alpha_1$ = -2.12, $\alpha_2$ = -2.91, Energy break = 1.15 GeV.} We find that long integration times are needed to detect this tMSP, with at least 100 h for CTAO-N and 50 h for CTAO-S. 
{We tried to also fit the simulation obtained with a logparabola model but the fit did not converge, despite several attempts changing various parameters (prefactor, ROI, etc.). This is likely caused by the very low statistics of the simulated spectrum.}

\begin{figure*} [htp!]
    \begin{minipage}[c]{1\textwidth}
        \begin{minipage}[c]{0.49\textwidth}
            \begin{center}
                \includegraphics[trim=0mm 0mm 0mm     0mm,clip,width=1\textwidth]{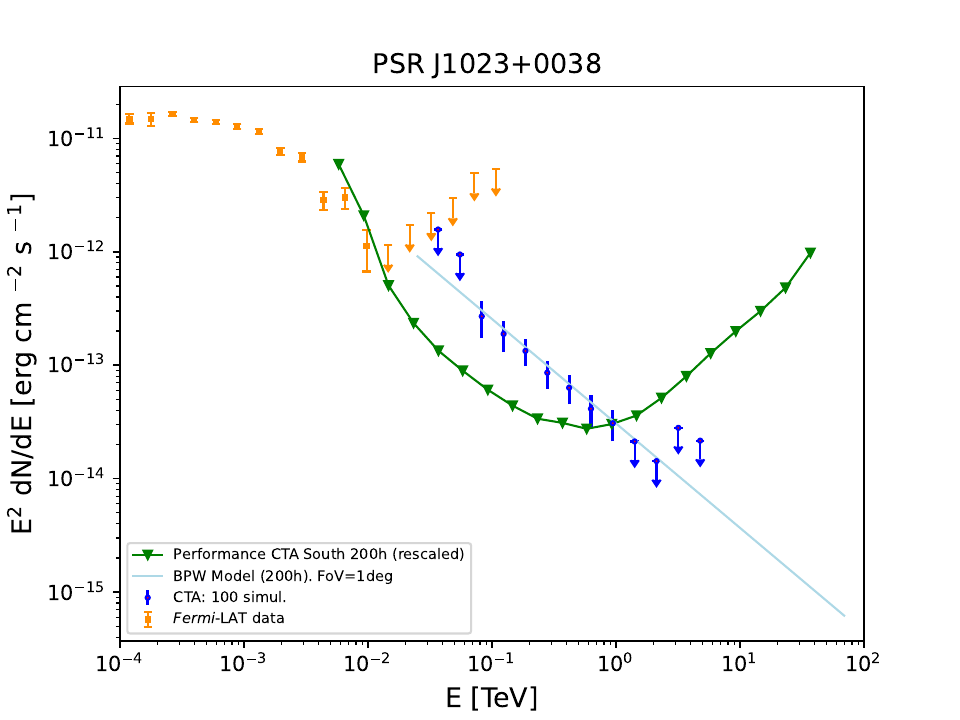}
            \end{center}
        \end{minipage}\hfill
        \begin{minipage}[c]{0.49\textwidth}
            \begin{center}
                \includegraphics[trim=0mm 0mm 0mm     0mm,clip,width=1\textwidth]{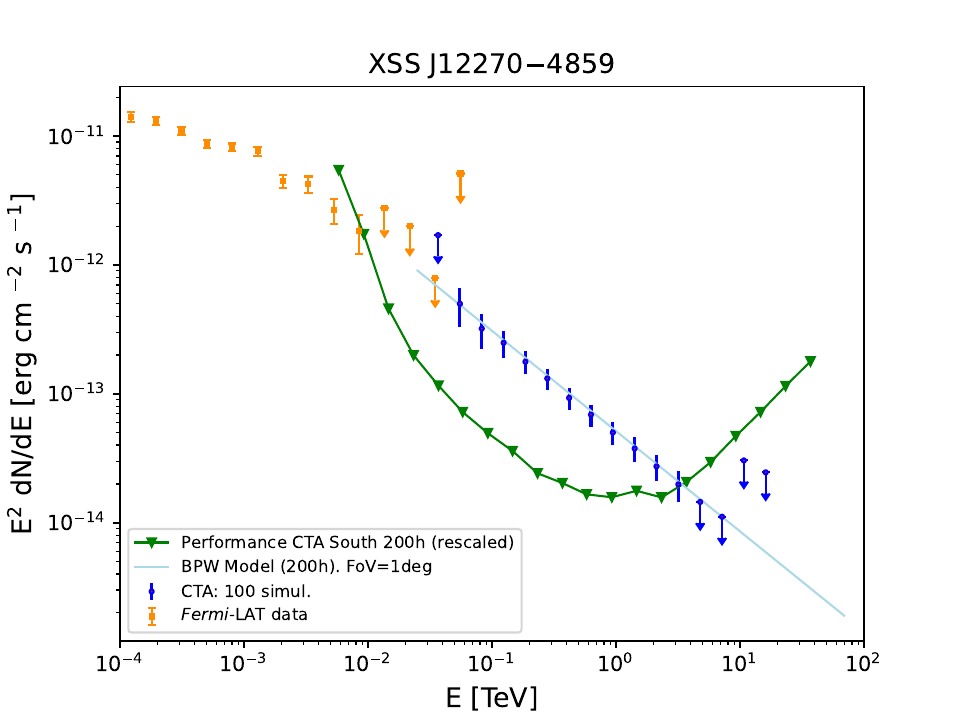}
            \end{center}
        \end{minipage}\hfill
        \begin{minipage}[c]{1\textwidth}
            \begin{center}
                \caption{\label{fig:tMSPs} CTAO-S simulations for the tMSP PSR J1023+0038 (on the left) and XSS J1227-48538 (on the right) considering the broken power-law  model. We consider 200 h of observations. The CTAO performance curve (green) is rescaled for 200 h. The \textit{Fermi}-LAT spectrum during the accretion phase is reported in orange. }
            \end{center}
        \end{minipage}\hfill

    \end{minipage}\hfill
\end{figure*} 
%


\begin{table}
\caption{Broken power-law model significance's results for North and South IRFs, considering 50, {100} and 200 h of observations of the tMSP PSR J1023+0038.  }
\centering
\footnotesize
\label{tab:j1023_simul}
\begin{tabular}{c|c|c|c}
\hline
    & Hours & TS & $\sigma$ \\
    \hline
    {North }& 50 & 18.50 & 4.30 \\
    & 100 & 39.86 & 6.31\\
    & 200 & 77.15 & 8.78\\
    \hline
    {South }& 50 & 29.36 & 5.42\\
    & 100 & 55.47 & 7.45\\
    & 200 & 114.77 & 10.71\\
\hline
\end{tabular}
\end{table}

\subsubsection{XSS J1227-48538}
XSS J1227-48538 was initially detected as a hard X-ray source and was tentatively identified as a cataclysmic variable, similarly to PSR J1023+0038, based on the double-peaked emission lines (typical of an accretion disc) in the optical spectrum \citep{Masetti06_j1227}. While the low X-ray luminosity was not in contrast with a cataclysmic variable interpretation, the peculiar X-ray variability with mode switching and the unexpected association with a \textit{Fermi}-LAT source proned to identify XSS J1227-48538 as an unusual LMXB \citep{deMartino10_tMSP}.
The system transitioned to a radio pulsar state between November 14 and December 21, 2012, characterized by the disappearance of the emission lines in the optical spectrum and the softening observed in the radio, optical, X-ray and gamma-ray bands \citep{bassa14_tMSP, 2017ApJ...836...68T}. 
Just after the transition, observations with the Giant Metrewave Radio Telescope allowed to detect a radio pulsar with a 1.69 ms spin period in a binary system with an orbital period of 6.9 hr \citep{roy15_j1227}. 
Before the transition to the radio state, the gamma-ray emission was a factor of 2 larger \citep{2017ApJ...836...68T}. The \textit{Fermi}-LAT analysis performed for the period in which the source was in the sub-luminous disk state (2008-2012) provides results consistent with those reported by \cite{Xing_2015j1227} and  \cite{2017ApJ...836...68T}: XSS J1227-48538 is best described by a power-law with a cutoff at E$_{cut}$= 5.3 GeV (at 3.4$\sigma$) and a spectral index of $\gamma$= 2. {
The \textit{Fermi}-LAT analysis procedure performed on this source is similar to that described before for the other tMSP.}
\\ 
Similarly to PSR J1023+0038, the two spectral models considered for the CTAO simulations are the logparabola and the broken power-law. As input models for the CTAO simulations we considered the output from the \textit{Fermi}-LAT study. For the broken power-law: P$_{f}$ = (2.71$\pm$0.71) $\times$ 10$^{-12}$ photons cm$^{-2}$ s$^{-1}$ MeV$^{-1}$, Index1 = -2.23$\pm$0.04, Index2 = -2.77$\pm$0.10, Energy break = (1.32$\pm$1.44) GeV. 
For the logparabola: P$_{f}$ = (3.26$\pm$0.91) $\times$ 10$^{-11}$ photons cm$^{-2}$ s$^{-1}$ MeV$^{-1}$, $\alpha$ = 2.28$\pm$0.05, $\beta$ = 0.09$\pm$0.02, Energy break = (4.45$\pm$5.42) GeV. We performed batches of 100 simulations using only the CTAO-S site IRFs and 50, 100 and 200 h of observations. The simulated data were binned into 20 logarithmic energy bins in an energy range of 0.03-100 TeV.\\
\noindent The resulting spectrum of XSS J1227-48538 is shown in Figure \ref{fig:tMSPs} and, for simplicity, we reported only the results with the highest statistics. The {complete results of} the analysis for the broken power law are reported in Table \ref{tab:j1227_simul} and {the parameters of the unbinned analysis are: 
P$_{f}$ = 3.15 $\times$ 10$^{-12}$ photons cm$^{-2}$ s$^{-1}$ MeV$^{-1}$, $\alpha_1$ = -2.23, $\alpha_2$
 = -2.78, Energy break = 1.33 GeV. }The source could be detected with CTAO-S at 8.14$\sigma$ with 50 h of observation. \\
\noindent 
{As for PSR J1023+0038, we interpret the non convergence of the logparabola model as caused by the fact that the flux inferred from the fit  falls marginally below the CTAO sensitivity curve for 200 hours.}


\noindent Our simulations prove that the detection {of the spectral component seen by \textit{Fermi}-LAT from of close-by tMSPs when in a disc state could be possible with long exposures, {provided that the emission has no cutoff at few GeVs}. While this will likely not allow to catch fully a transition if lasting less than several days, here we demonstrate that once a transition has occurred CTAO could be able to detect such type of sources, identifying tMSPs as VHE emitters for the first time. 
{ While during the rotation-powered state MSP binaries will not be at the reach of CTAO, the known tMSPs have demonstrated an increase of their HE flux by a factor of 3-10 when transitioning from the radio to the disc states (see \cite{2017ApJ...836...68T}). The possibility to look-back in the CTAO Galactic Plane and other higher latitude pointings, where many MSP binaries are located, will be crucial for assessing any possibly related VHE flux increase or for complementing the X-rays and lower-energy coverage in case only upper limits are obtained.} Also, if there are additional components (not considered here, such as magnetic reconnection of pulsar wind) it could be possible to detect changes in the VHE flux along transitions.


\begin{table}
\caption{Broken power-law model significance's results for South IRFs, considering 50, {100} and 200 h of observations of the tMSP XSS J1227-48538.  }
\centering
\footnotesize
\label{tab:j1227_simul}
\begin{tabular}{c|c|c|c}
\hline
    & Hours & TS & $\sigma$ \\
    \hline
    {South}& 50 & 66.26 & 8.14\\
    & 100 & 128.04 & 11.31\\
    & 200 & 253.99 & 15.94\\
\hline
\end{tabular}
\end{table}

\subsection{Flares in PWNe: the Crab Nebula}\label{CTAO_pwne}
The Crab Nebula is the best-studied PWN in the VHE regime. It is located at a distance of $\approx 2.2$ kpc with $\approx 3.8$ pc of size \citep{1973PASP...85..579T, 1985ARA&A..23..119D}. Since 2009, several rapid and bright flares have been detected from the nebula at HE with space-borne gamma-ray instruments \citep{2011Sci...331..736T,2011Sci...331..739A,2012ApJ...749...26B,2013ApJ...775L..37M,2013ApJ...765...52S,2020ApJ...897...33A}.
The observed flares presented variability timescales of hours. During these flaring periods, the nebula showed rapid variations of flux and large releases of energy \citep{2011Sci...331..736T,2011Sci...331..739A}. Several multi-wavelength campaigns involving \textit{Chandra} X-Ray Observatory, Keck Observatory and Very Large Array (VLA) \citep{2013ApJ...765...56W} and TeV searches by IACTs  \citep{2010ATel.2967....1M,2010ATel.2968....1O,2014A&A...562L...4H,2014ApJ...781L..11A,2019ICRC...36..812V} were carried out to follow these flares. 
However, none of them reported a correlation of the flares with morphological and/or spectral variations in the nebula.

CTAO will cover a fundamental energy range to understand the origin of these flares: on the one hand, the low energy threshold will allow the sampling of the \textit{Fermi}-LAT spectral shape at few tens of GeV of synchrotron nature,
providing important clues on the acceleration and emission processes; on the other hand, the excellent sensitivity in the TeV regime will serve to explore the IC component that might arise at a detectable level from the electron population behind the MeV flares, off-scattering soft photon fields.

To evaluate the capability of CTAO to detect {Crab flares}, we performed simulations of the SEDs both in flaring and steady (non-flaring) states of the nebula. We simulated flares of different spectral characteristics {starting} from a parent particle population, varying the physical properties of the environment. In particular, we simulated three types of flares: a very bright flare with a similar flux (at hundreds of MeV) to the one observed by \emph{Fermi}-LAT in April 2011 \citep{2012ApJ...749...26B,2013ApJ...765...52S}, 
which is the flare with the largest flux to date, and two dimmer flares corresponding to the first one re-scaled by a factor of 0.5 and 0.1. {Thus, the corresponding flux enhancement at the simulated flares' peaks (above 100 MeV of energy) ranges from 3 to 30 times the average flux of the nebula, as observed in tens of flares detected since 2007} \citep{2021ApJ...908...65H}. Since no spectral variability has been reported at HE, we assume the same spectral model for these dimmer flares.

The simulations of the nebula in flaring and steady states are performed for the CTAO-N with the methods and tools presented in previous works \citep{2020Mestre, 2021Mestre}. {\cite{2021Mestre} pointed to the crucial role of the LSTs, best sensitive at the sub-TeV energy regime, in the prospects for the Crab flares' detection (even in the early stages of CTAO-N operations). The latter, however, cast doubts on CTAO-S prospects due to the lack of LSTs in the Alpha configuration.
 }. The electron population was simulated with a fixed index ($\Gamma_{\rm e}$) of 2.5, to guarantee the detection in the TeV regime (e.g., from 1.25\,TeV to 50\,TeV) of the brightest model of flare in less than 10\,h {at 95\% confidence level (see Fig.\ 6 of \citealt{2021Mestre})}. The different flare models are computed for a magnetic field (B) in the acceleration region ranging from 100\,$\mu$G to 1\,mG and compared (see section 2 of \citealt{2021Mestre}) to the steady nebula SED in both tens of GeV (e.g., from 20\,GeV to 200\,GeV) and TeV regimes. We obtained the simulations of the Crab nebula SED in steady state from \citealt{2020Mestre}. To compare the flaring and steady nebula simulations, we computed the mean total expected excess (e.g., counts from the source after background subtraction), both in flaring and steady state, in 21 bins of energy from 12\,GeV to 200\,TeV with observation times ranging from a few minutes to 500\,h. We compared the excess distributions using the Pearson's chi-squared test, corresponding the null hypothesis ($\rm{H}_{0}$) to the steady state. Then, we consider the flare implies a detectable flux level if $\rm{H}_{0}$ is rejected at {$3\sigma$ significance (see Table \ref{tab:Flaredetectionhi.}; values in parentheses correspond to $5\sigma$ confidence)}.


The simulations performed show that the different models of flare are best detected in the GeV regime and in particular 
in less than an hour at energies below 200\,GeV, see Fig. \ref{FigCrabFlares}. In the TeV regime, flares dimmer than April 2011 flare by a factor 0.5 (at hundreds of MeV) would be detected in less than 10\,h for B $< 500$\,$\mu$G (see Table \ref{tab:Flaredetectionhi.}). {As a reference, the gamma-ray flares from the Crab detected with Fermi-LAT showed varied duration scales from a few days to weeks and complex substructures, with sub-flares of a few hours or day-long duration \citep{2021ApJ...908...65H}}. Some of the models considered imply an energy in TeV electrons larger than $\tau_{\rm syn} \times L_{\gamma} \approx 5\times 10^{43}$ erg, being $\tau_{\rm syn}$ the synchrotron cooling time, and $L_{\gamma} \sim 2\times 10^{35}$ erg s$^{-1}$, the luminosity of the nebula in gamma rays \citep{1998MNRAS.295..337R}. However, note that the energy in electrons available in the nebula is not limited to the one computed above if particle re-acceleration takes place, which would introduce additional boosts in the electron population energy reservoir. The simulations performed, together with previous results reported \citep[see][]{2021Mestre}, provide excellent prospects for detecting flares from the Crab Nebula with CTAO, especially for the LST subarray, featuring the best sensitivity at energies of a few tens of GeV.

\begin{figure}
   \centering
   \includegraphics[width=\columnwidth]{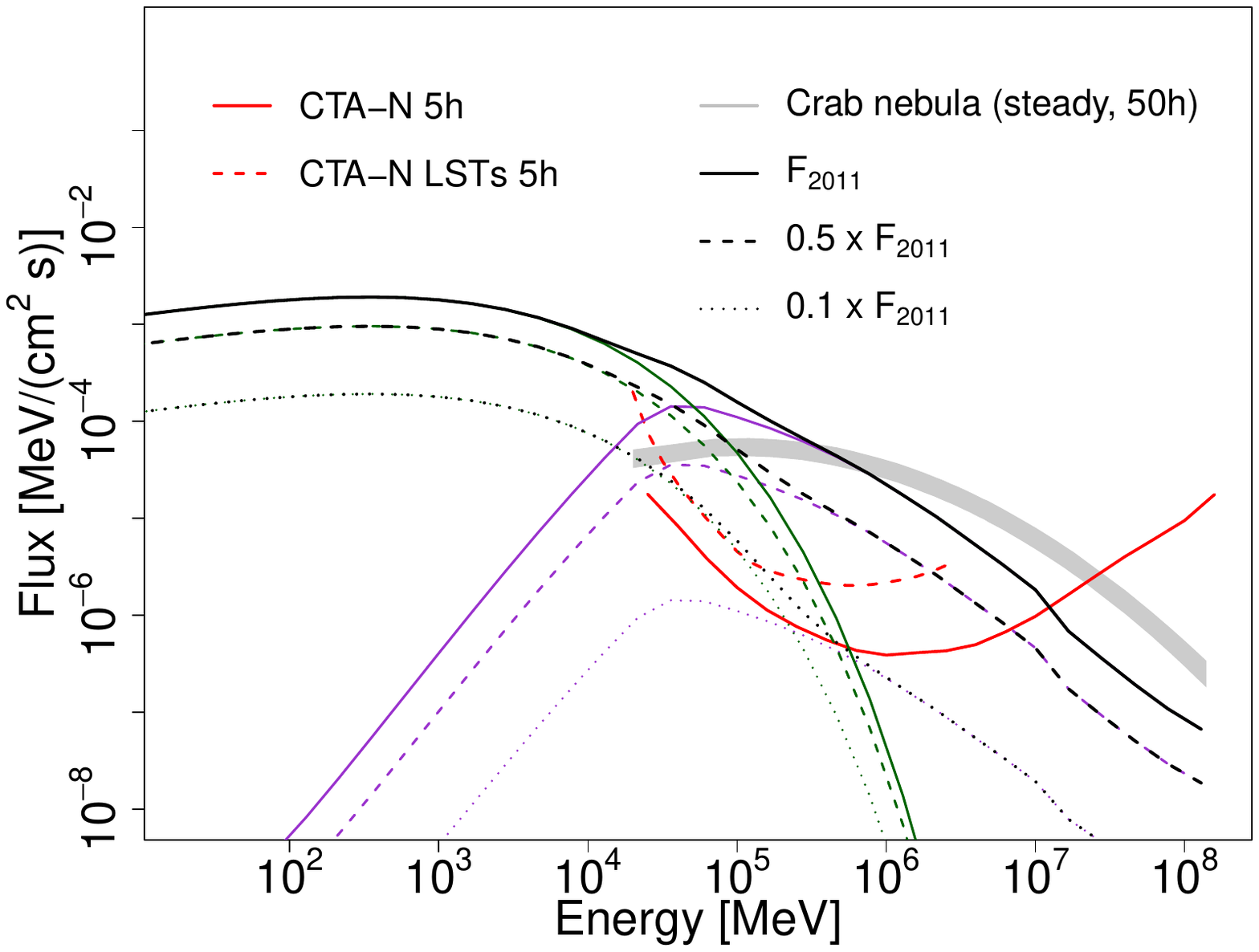}
    \caption{The synchrotron (green lines), IC (purple lines), and total (black lines) emission from the Crab Nebula for different flare models. The solid lines correspond to the model fitted to the \textit{Fermi}-LAT April 2011 flare data at energies above 80\,MeV for a particle index of $2.5$. The dashed and dotted lines correspond to the same model re-scaled by a factor of 0.5 and 0.1, respectively. All the models are computed for a magnetic field of 500\,$\mu$G. The red solid and dashed lines correspond to the sensitivities of the CTAO-N and, if considering only its four LSTs for 5\,h of integration time, respectively. The Crab Nebula steady spectrum simulations for 50\,h of observation time with CTAO-N are noted with the gray shaded area ($3\sigma$ region).}
    \label{FigCrabFlares}
 \end{figure}

\begin{table}
\caption{{In the second, third, and fourth columns, the observation time (in hours) necessary to detect different models of flares from 1.25\,TeV to 50\,TeV with CTAO-N {at $3\sigma$ ($5\sigma$) significance}. The first column indicates the magnetic field chosen for the acceleration region. The observation times in the second column are computed for flare models fitted to the \textit{Fermi}-LAT SED (at the moment of maximum flux level) of the April 2011 flare. For the third and fourth columns, the LAT SED (dubbed $F_{2011}$) was re-scaled prior to the fit by a factor of 0.5 and 0.1, respectively. The models with an asterisk imply an energy in electrons above 1\,TeV larger than $5 \times 10^{43}$ erg. We assume Crab is in flaring state during the entire observation time.}}
\centering
\footnotesize
\label{tab:Flaredetectionhi.}
\begin{tabular}{c|c|c|c}
\hline
Model B[$\mu$G] &  & Model SED & \\
\hline
 & $F_{2011}$ & $0.5 \times F_{2011}$ & $0.1 \times F_{2011}$ \\
\hline
$1000$ & $22\ (46)$ & $351^{*}\ (>500)$ & $> 500^{*}$  \\
$500$ & $0.8\ (1.6)$ & $14^{*}\ (31)$ & $> 500^{*}$ \\
$100$ & $0.03\ (0.05)$ & $0.13\ (0.24)$ & $68^{*}\ (133)$ \\
\hline
\end{tabular} 
\end{table}

\subsection{Novae}\label{CTAO_novae}


The only nova that has been detected at TeV energies so far is the symbiotic system RS Oph \citep{2022NatAs...6..689A,2022Sci...376...77H,2025A&A...695A.152A} which shows recurrent outburst every 15-20 years and harbours a white dwarf accreting from a late-type giant companion star. However, it could be argued that the detection is due to selection-effects based on the fact that RS Oph is relatively nearby (see below). By number, novae that involve a small, low-mass donor such as a main sequence star \citep[these are usually the \textit{classical novae}, see][]{chomiuk2021a} are by far the most common type of system.
The majority of novae have been observed in outburst only once in a human life timescale, and so far only a handful of novae are known to erupt with a recurrence time of ${\sim}$tens of years. It has been predicted from binary population synthesis studies \citep{2021MNRAS.504.6117K,kemp2022a} that most commonly, novae with evolved donors are more likely to contribute to the total current Galactic nova rate, even though by number these systems make up a smaller fraction of nova binaries.  

As pointed out recently by \citet{de2021a}, a large number of novae in optical bands might be being missed due to a number of sources residing behind and in the Galactic bulge. Taking obscuration by dust into account, \citet{de2021a} estimated a current Galactic nova rate of $43.7 \pm^{19.5}_{8.7}$ per year. This is notably much larger than the actual 
Galactic nova detection rate of $ \lesssim 10$ per year.

\citet{kemp2022a} estimated the Galactic nova rate to be $33$ per year. That study showed that the most common type of nova in our Galaxy today is expected to originate from a binary system involving a giant-like donor (see Fig. 11 in the aforementioned paper as a guide). 
As mentioned previously, currently the only system to have clearly been detected at VHE (and detected also at HE) is the symbiotic recurrent nova RS Oph. Though some groups have investigated detailed modelling of shock generation in nova systems \citep{hachisu2022a,metzger2016a}, which is believed to be mainly hadronic in nature, it is still not clear how many Galactic novae would be detectable by CTAO at and beyond the ${\sim}$TeV energy range. The majority of novae thus far detected at GeV energies with no clear evidence for a TeV component have been classical novae \citep[not symbiotic systems like RS Oph, see e.g.][]{zheng2022a}. Nonetheless, we anticipate several more novae could be observed with CTAO \citep[see][]{chomiuk2021a}, particularly if these novae are detected at other wavelengths early on enabling rapid triggering and follow-up. However, assuming we can expect of the order of $\thicksim$30 Galactic novae per year, even with adequate triggering, it is unlikely that all of these outbursts will be detectable by CTAO. If CTAO-N would have been operational since August 2008 to April 2023, it would have been able to perform observations of 7 novae detected in HE gamma-ray by \textit{Fermi}-LAT, in the 5 nights after their detection in optical. Assuming a similar rate of novae detected at HE gamma-ray in the future, it means that CTAO-N would observe $\sim0.5$ novae per year triggered by their HE gamma-ray emission.

CTAO observations will be important to put constraints on the maximum energies attainable in nova explosions and the physical mechanisms involved in the production of VHE gamma rays. We estimated the capability of CTAO to detect nova outbursts based on both theoretical modelling and empirical results. First, simple theoretical considerations based on the RS Oph detection are adopted to assess the gamma-ray emission at different outburst stages, following the approach in \cite{2022NatAs...6..689A,2022Sci...376...77H}. Second, a parametric study based on phenomenological parameters involved in the emission of gamma rays in nova outbursts is performed to estimate the parameter space we could constrain with CTAO observations. Finally, we considered dedicated numerical simulations of RS Oph to assess the expected detectability with CTAO.

\subsubsection{Modelling approach}

We explored the capability of CTAO to constrain the physical parameters of nova phenomena of different types, building up from basic arguments. The expected gamma-ray emission is obtained for different properties of a shock expanding with velocity ${\rm v_{sh}(t)}$, generated by ejected material of total mass $M_{\rm {ej}}(t)$ slamming into the companion star's wind and producing gamma rays through hadronic interactions. To accelerate protons to high energies via {diffusive} shock acceleration (DSA), the magnetic field has to be amplified in the shock. The maximum energy particles attain at a shock is limited ultimately by the Hillas condition \citep{1984ARA&A..22..425H}. However a more constraining limit is determined by either the time taken before radiative cooling dominates over acceleration, or by the necessary escape of the particles up-stream of the shock, in order to excite magnetic field fluctuations to a sufficient level ahead of the shock \citep{2004MNRAS.353..550Bell}. 
The maximum energy $E_{\rm {max}}$ in the particle spectrum, defined as as power-law function with an exponential cutoff, can be then described as: 
\begin{equation}
E_{\rm max} = 10 \left ( \frac{v_{\rm sh}}{5000~\mathrm{km/s}} \right )\left ( \frac{R_{\rm sh}}{1~\mathrm{AU}} \right )^{-1} \left ( \frac{B_{*}}{1~{\rm G}} \right )~~~{\rm TeV,}
\end{equation}

\noindent 
with $R_{\rm sh}$ the position of the shock with respect to the white dwarf, 
and $B_{*}$ the companion surface magnetic field (typically $\sim$1\,G for a red giant). {$R_{\rm sh}$ can be expressed in terms of} shock velocity, for which we assumed free expansion during the first few days, followed by radiative expansion when entering the Sedov-Taylor phase \citep{1985MNRAS.217..205Bode}. The particle flux per unit of time and energy is computed using the condition that a fixed fraction (50\%) of the kinetic energy of the protons (${\rm E_{kin} = \frac{1}{2} M_{ej}v_{sh}^2}$) is transferred to non-thermal particles \citep{2022NatAs...6..689A,2022Sci...376...77H}. Once the particle spectrum is defined as function of these three parameters ($v_{\rm sh}$, $B_{*}$, $M_{\rm ej}$) at different evolutionary stages of the shock expansion, we derive the gamma-ray emission originated by proton-proton interaction assuming a density of the ejecta which can be approximated 
following equation 4 in \cite{2022NatAs...6..689A}.
The particle spectrum and density was used to compute the non-thermal emission, using the \verb+Naima+ spectral model class \citep{2015ICRC...34..922Z} included in \verb+Gammapy+.

In Fig. \ref{fig:gammanova} we show the expected gamma-ray flux at different times from the nova explosion, considering several physical parameters in the shock. The upper panels show the expected emission for a explosion similar to RS Oph, with $B_{*}$, $M_{\rm ej}$ equal to 1~G and $10^{-6}M_\odot$, respectively, located at a distance of 2~kpc (left) and of 4~kpc (right). The effect of increasing the star surface magnetic field (10~G) and decreasing the ejecta mass ($10^{-7}M_\odot$) is shown in the bottom panels, for a fixed distance of 2~kpc. For reference, the isoflux line at $10^{-13}$ TeV\,cm$^{-2}$\,s$^{-1}$ is marked  when possible with a white line. The region below such line should be easily accessible to CTAO.

\begin{figure*}
    \begin{center}
        \includegraphics[width=0.38\textwidth]{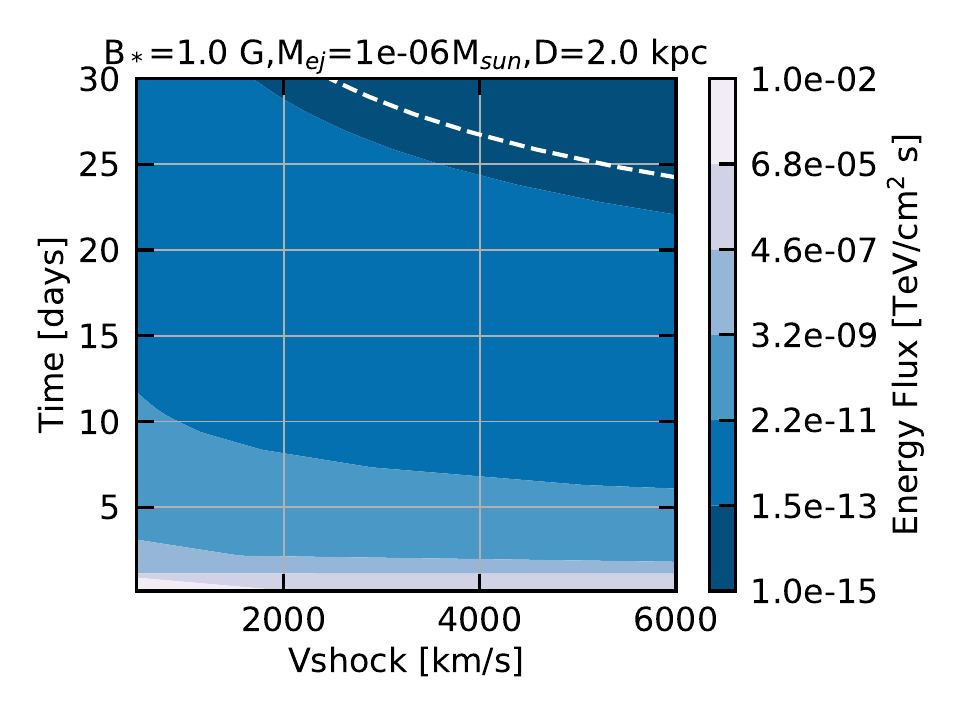}
        \includegraphics[width=0.38\textwidth]{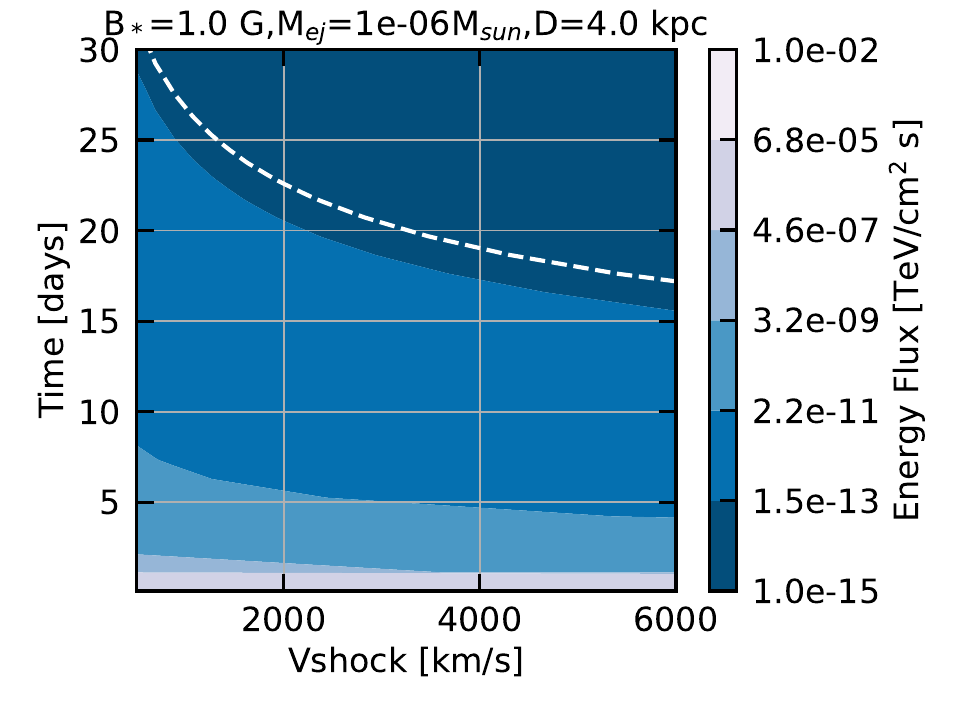}
        \includegraphics[width=0.38\textwidth]{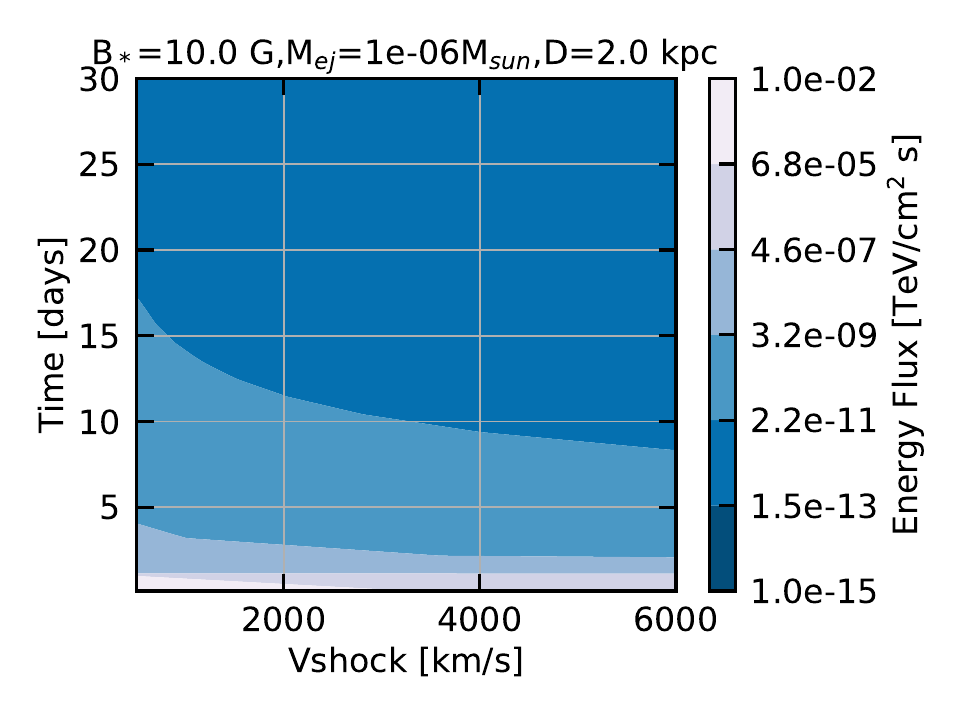}
        \includegraphics[width=0.38\textwidth]{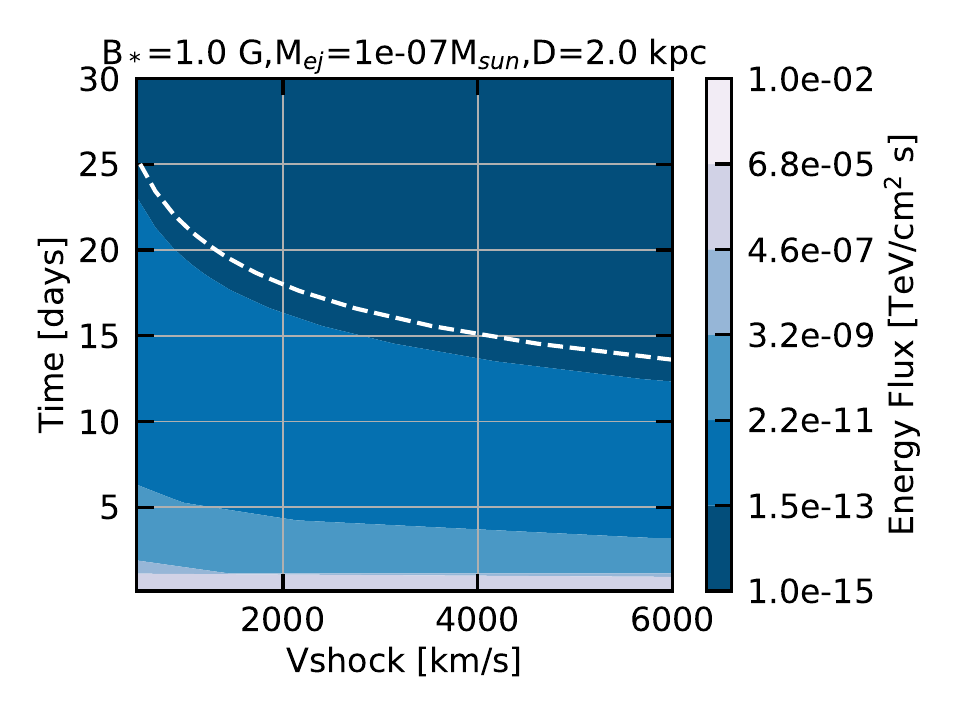}
        \caption{Expected energy flux in gamma rays with time after nova explosion as a function of shock velocity, integrated above 10\,GeV. Top left: for an RS Oph-like system at 2\,kpc. Top right: for the same physical properties yet at a {larger} distance of 4\,kpc. Bottom left: with increased magnetic field strength. Bottom right: with decreased mass-loss rate. In all plots the white dashed line indicates the CTAO {isoflux line at $10^{-13}$ TeV\,cm$^{-2}$\,s$^{-1}$}. Below this line the region should be easily accessible to CTAO.}
        \label{fig:gammanova}
    \end{center}
\end{figure*}

\subsubsection{Parametric space study}

We utilised a phenomenological approach to study the parameter space of gamma-ray emission from novae. The emission was assumed to be produced by hadronic processes from $\pi^{0}$ decay \citep{2014PhRvD..90l3014K}, as indicated by the gamma-ray emission of RS Oph (see Section~\ref{sect:RSOph}). The $\pi^{0}$ decay radiative model was parameterised using the target proton density ($n_{\rm{h}}$) and the relativistic proton energy distribution. For the latter, we considered a particle distribution function parameterised as a power-law model with an exponential cutoff. We described the parameter space under study as a 3D space, where we set the parameter space domain in the range of plausible values based on observed novae at gamma rays. A 2D grid was defined with different values for the prefactor ($A$) and the cutoff energy ($E_{\rm{cp}}$) of the proton energy distribution. The former in the range between 
$A=[10^{28},10^{32}]\,\mathrm{protons\, eV^{-1}}$ at a pivot energy of $100\,\rm{GeV}$ and the latter between $E_{\rm{cp}} =[10,1000]\,\rm{GeV}$. Two slices for the target proton density were used for the third axis, $n_{\rm{h}}=10^8\,\rm{cm^{-3}}$ and $n_{\rm{h}}=10^{11}\,\rm{cm^{-3}}$, which correspond to typical shock density values in novae \citep{metzger2016a}. The distance to the gamma-ray emitter was fixed to {$d=2\,\rm{kpc}$}. The spectral energy distribution for each model was obtained using the software package \verb+Naima+ \citep{2015ICRC...34..922Z}.

The emission detectability was assessed for both arrays of CTAO using the official IRFs from \verb+prod5-v0.1+ in the Alpha configuration (\verb+20deg-AverageAz+ for $5\,\rm{h}$ observation time) 
The results of the simulations for CTAO-N and CTAO-S are shown in {\textit{panels a} and \textit{b}} of Fig.~\ref{fig:param_space}, respectively. {The total proton energy above 100\,GeV ($W_{\rm p}$) multiplied by $\frac{n_{\rm h}}{d^2}$, hereafter ``effective proton energy reservoir'', was used as a function of $E_{\rm{cp}}$ to display} the ratio between the integral source flux and the CTAO sensitivity{. This ratio} was computed to obtain a qualitative estimation of the detectability of CTAO for each model in the parameter space. The higher the integral flux-to-CTAO-sensitivity ratio, the more feasible the detection. Moreover, the region where we would detect each model with CTAO in at least one energy bin is lower-delimited in Fig.~\ref{fig:param_space} by a dashed orange line to have a more precise boundary of the detection region. Therefore, the region between the dashed orange line and the white region (integral flux-to-CTAO-sensitivity close to 0) delimits the border of the parameter space where CTAO will likely begin to detect the gamma-ray emission of the models. Qualitatively, RS Oph would be located approximately in the top right corner of {Fig.~\ref{fig:param_space}}, while V959 Mon {(the first classical nova discovered by \textit{Fermi}-LAT; \citealp{2014Sci...345..554Ackermann})} would be in the lower left region of {the plots}. 

The integral gamma-ray emission and the integral flux-to-CTAO-sensitivity ratio in Fig.~\ref{fig:param_space} increases as {the effective proton energy reservoir and $E_{\rm{cp}}$} increase. {Both top regions of plots \textit{a} and \textit{b} in Fig.~\ref{fig:param_space} have positive values of integral flux-to-CTAO-sensitivity ratio (about 30\% of the total combinations), while the {bottom region} 
do not (about 70\% of the total combinations).
When comparing the results between CTAO-N and CTAO-S, the former extends the parameter space region with positive integral flux-to-CTAO-sensitivity towards models with $E_{\rm{cp}}<250\,\rm{GeV}$. On the other hand, the latter presents a wider detection region towards $E_{\rm{cp}}>250\,\rm{GeV}$ than CTAO-N. CTAO-N overperforms CTAO-S with about 10\% more detections.}
The better performance of CTAO-N at low energies is expected because the parameter space under study was restricted to produce most of the gamma-ray emission below $1\,\rm{TeV}$ {and it is also connected with the presence of four LSTs in the CTAO-N Alpha configuration}, as it is observed from current novae detected at gamma rays. Therefore, the lack of LSTs, which dominates the CTAO sensitivity at these energies \citep{2019sCTA.book.....C}, in the Southern array (Alpha configuration) will reduce the parameter space of detectability with CTAO-S.

\begin{figure*}
    \begin{center}
        \includegraphics[width=0.95\textwidth]{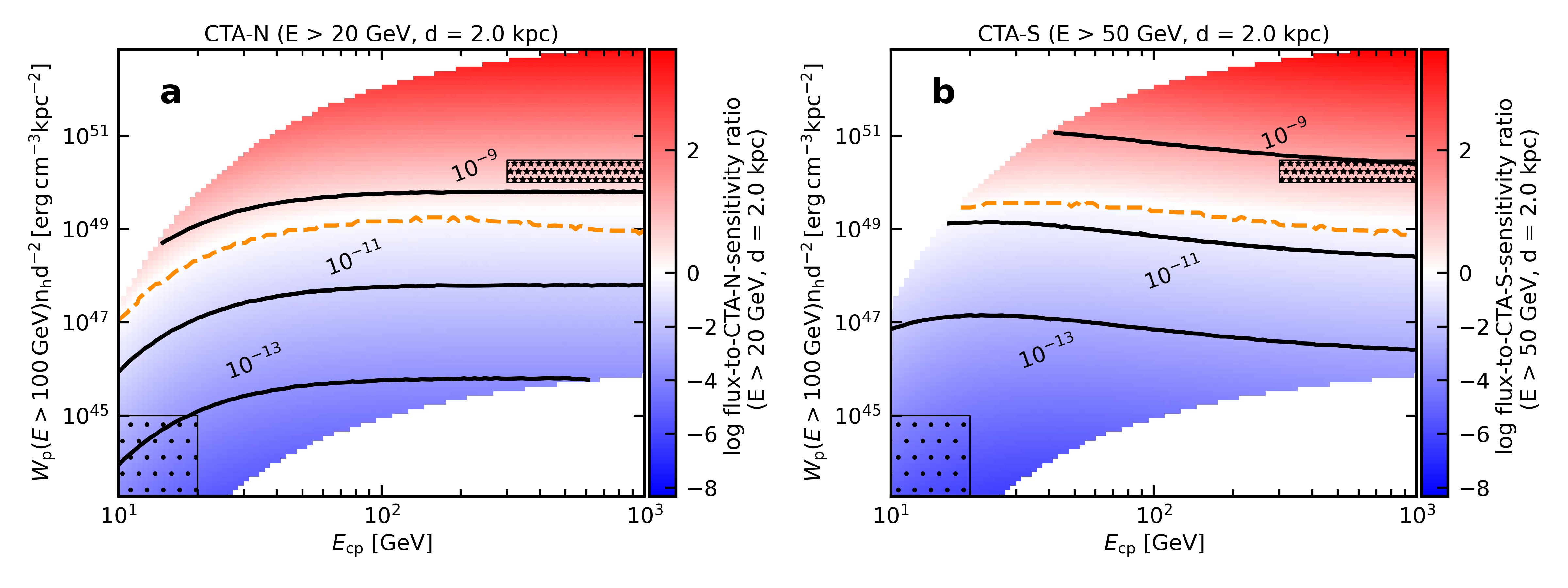}
        \caption{Logarithmic integral flux-to-CTAO-sensitivity ratio for CTAO-N {(\textit{panel a})}, integrated above $20\,\rm{GeV}$ and CTAO-S {(\textit{panel b})}, integrated above $50\,\rm{GeV}$ for different values {in the defined parameter space (see text)} of $n_{\rm{h}}$, cutoff energy ($E_{\rm{cp}}$) and prefactor ($A$) of the proton energy distribution function at a fixed distance of $\rm d =2.0\,\rm{kpc}$. The sensitivity was computed for a total observation time of 5 hours. The orange dashed line indicates the domain in the parameter space with detection in at least one energy bin for different values of $n_{\rm{h}}$, $E_{\rm{cp}}$ and $A$. Solid black lines are curves at constant integrated flux ($10^{-13}$, $10^{-11}$ and $10^{-9}\,\rm{cm^{-2}s^{-1}}$) above $20\,\rm{GeV}$ and $50\,\rm{GeV}$ for CTAO-N and CTAO-S panels, respectively. {The regions where V959 Mon and RS Oph would be approximately located in the parameter space are marked with black dots and stars, respectively.}}

        \label{fig:param_space}
    \end{center}
\end{figure*}

Thus, CTAO is expected to give strong constraints only to a sub-space of the whole parameter space under study. {For about 30\% of the area of the parameter space covered in Fig.~\ref{fig:param_space}}
could be likely detected with CTAO, in particular, where the relativistic protons have a high value of prefactor and cutoff energy. 
Assuming that the target proton number density is the number density of the main ejection of matter in the outburst, the results suggest that for denser ejecta, the detection region with CTAO will cover a wide range of parameter values of the relativistic proton energy distribution. CTAO-N should outperform CTAO-S for novae with $E_{\rm{cp}}<250\,\rm{GeV}$, while for $E_{\rm{cp}}>250\,\rm{GeV}$, CTAO-S should perform better than CTAO-N at high energies.

\subsubsection{RS Oph}
\label{sect:RSOph}
RS Oph is a symbiotic nova formed by a high-mass white dwarf ($1.2-1.4 \,M_{\odot}$) and a red giant star \citep[M0 III,][]{1999AA...344..177A}, which transfers material to the compact object. 
In the literature, its distance has been estimated ranging from $1\,\rm{kpc}$ to $5\,\rm{kpc}$ (\citealt{barry2008distance}, see also the discussion about the distance estimation in Section C.1 of the supplementary material of \citealt{2022NatAs...6..689A}), being the most recent value of about $2.68 \pm 0.16\,\rm{kpc}$ from \textit{Gaia} DR3 catalog \citep{2020yCat.1350....0G} 
RS Oph undergoes recurrent nova outbursts with a periodicity of about 15 years. Its last eruption occurred in August 2021. Covered widely at different wavelengths, the 2021 outburst was detected at VHE gamma rays, adding a new object class to the list of VHE emitters. The HE and VHE gamma-ray emission was consistent mainly with a hadronic origin (dominated by $\pi^{0}$ decay), likely originated by the interaction of the ejected material with the dense wind of the red giant  \citep{2022NatAs...6..689A,2022Sci...376...77H}. The gamma-ray spectrum showed hints of hardening with time produced by the migration of gamma rays to higher energies \citep{2022NatAs...6..689A}. The HE light curve presented a power-law decay after reaching the maximum emission phase. The index of the temporal decay at HE with \textit{Fermi}-LAT and the one obtained at VHE by the H.E.S.S. Collaboration were compatible within errors with values $1.35 \pm 0.07$ and $1.43 \pm 0.18$, respectively \citep{2022ApJ...935...44C, 2022Sci...376...77H}. 
It is expected that RS Oph will undergo another outburst when CTAO will be in operation. Hence, we carried out numerical simulations of RS Oph to estimate its detectability with CTAO along the temporal evolution of the outburst.

\subsubsection{RS Oph: CTAO simulations}
We performed the numerical simulations of RS Oph with CTAO using the official IRFs from \verb+prod5-v0.1+ for the CTAO northern and southern arrays 
In particular, the closest IRFs set to the culmination of RS Oph in the CTAO-N and CTAO-S site were used (\verb+North-40deg-SouthAz+, \verb+South-20deg-NorthAz+) for $0.5\,\rm{h}$ observation time. A total of 59 daily observations of 1 hour each were simulated starting one day after the beginning of the nova outburst {(batches of 100 simulations per day)}. We simulated this source based on the gamma-ray spectral and temporal profile reported by the MAGIC and H.E.S.S. Collaborations, respectively. The best daily-fit spectral log-parabola models from \cite{2022NatAs...6..689A} were considered to model the gamma-ray emission. Spectral variations were only contemplated for the simulations of the first four days, when spectral information in \cite{2022NatAs...6..689A} was available during the outburst. The spectral parameter values utilised in the different log-parabola models are shown in Table \ref{tab:model_RSOph}. After the fourth day, the spectral profile was fixed to the one from the last day with spectral information (fourth day), and the simulated gamma-ray emission was scaled to follow the power-law temporal decay reported by H.E.S.S. We set the index value of the power-law decay to {$\gamma$ =}$1.4$.

\begin{table}
\caption{Daily parameter values of the log-parabola spectral models used to simulate RS Oph. Adapted from \protect\cite{2022NatAs...6..689A}.}
\label{tab:model_RSOph}
\begin{center}
\begin{small}
	\begin{tabular}{|c|c|c|c|}
 \hline
    \multirow{2}{*}{Model day}  &  Prefactor at $130\,\rm{GeV}$  &  \multirow{2}{*}{$\alpha$}  &  \multirow{2}{*}{$\beta$}  \\
    & $[10^{-10}\,\rm{TeV^{-1}cm^{-2}s^{-1}}]$ &  & \\ \hline
    Day 1    &   5.40     & 3.86   &  0.194   \\ 
    Day 2    &   4.54     & 3.73   &  0.175   \\ 
    Day 3    &   5.37     & 3.64   &  0.173   \\ 
    Day $4-59$    &   5.00     & 3.44   &  0.147   \\ \hline
    \end{tabular}
\end{small}
\end{center}
\end{table}

The statistical detection significance as a function of time is shown in Fig.~\ref{fig:sig_vs_time}. The results confirm that RS Oph would be clearly detected with CTAO-N and CTAO-S for the first days, reaching a detection significance of about $60\sigma$ and $30\sigma$ in an hour with CTAO-N and CTAO-S, respectively. 
RS\,Oph is not only {detectable} with CTAO during the first days after the outburst, but CTAO would also daily detect RS Oph up to $20$ and $15$ days after the outburst with the northern and southern arrays, respectively. If we consider the combined data of 5 and 10 adjacent days with CTAO-N, the detection would be possible even up to $36-40$ and $46-55$ days, respectively. The $5\sigma$ detection would be limited down to $23-27$ and $38-47$ days with CTAO-S. The resulting SED for the first simulated observation (day 1 after the outburst) is shown in Fig.\,\ref{fig:SED_RSOph_day0} together with the observed spectrum obtained with MAGIC for the same observation time. The results suggest that CTAO will be able to probe the gamma-ray emission for several weeks after the outburst with a precise spectral coverage at least during the first days. Using 1-h observation, CTAO would be able to characterize the curvature of the VHE gamma-ray emission of RS Oph. For example, for the simulated observation of day 1, a log-parabola spectral model is preferred over a power-law model at 3.7$\sigma$.

Consequently, if one assumes that the next RS Oph outburst follows the same behaviour as the 2021 eruption, a plausible assumption based on the similarities observed at radio, optical and X-ray for the first weeks between 2006 and 2021 outbursts \citep{2022A&A...666L...6M,2022NatAs...6..689A,2022MNRAS.514.1557P}, CTAO observations can provide detailed coverage of the gamma-ray emission during the future RS Oph outburst. Also, we could probe the maximum energy of the accelerated particles and the nova physical conditions across different outburst stages.

\begin{figure}
\centering
\includegraphics[width=\columnwidth]{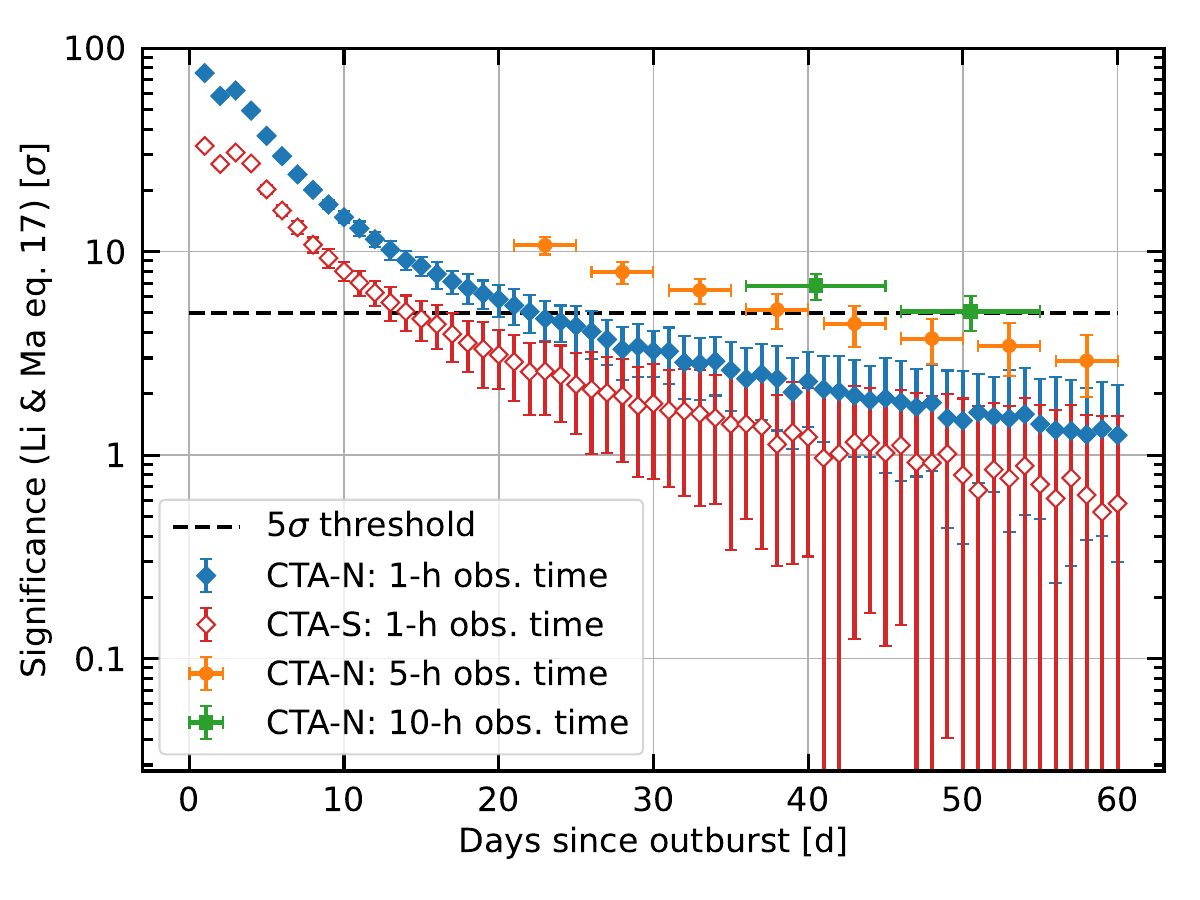} 
        \caption{Daily statistical detection significance \protect\citep[equation 17 of][]{1983ApJ...272..317L} from 1-hour simulated observation with CTAO-N and CTAO-S (blue filled diamonds and red empty diamonds, respectively) as a function of the number of days since the outburst of RS Oph. The 5-day (i.e. 5-h observation time, filled orange circles) and 10-day (i.e. 10-h observation time, filled green squares) combined significance for CTAO-N are computed when the daily and 5-day statistical detection significance reach a 5$\sigma$ detection (dashed black line), respectively. {Error bars correspond to the standard deviation of the statistical detection significance distribution for the 100 simulations per day.}}
        \label{fig:sig_vs_time}
\end{figure}

\begin{figure}
\centering
\includegraphics[width=\columnwidth]{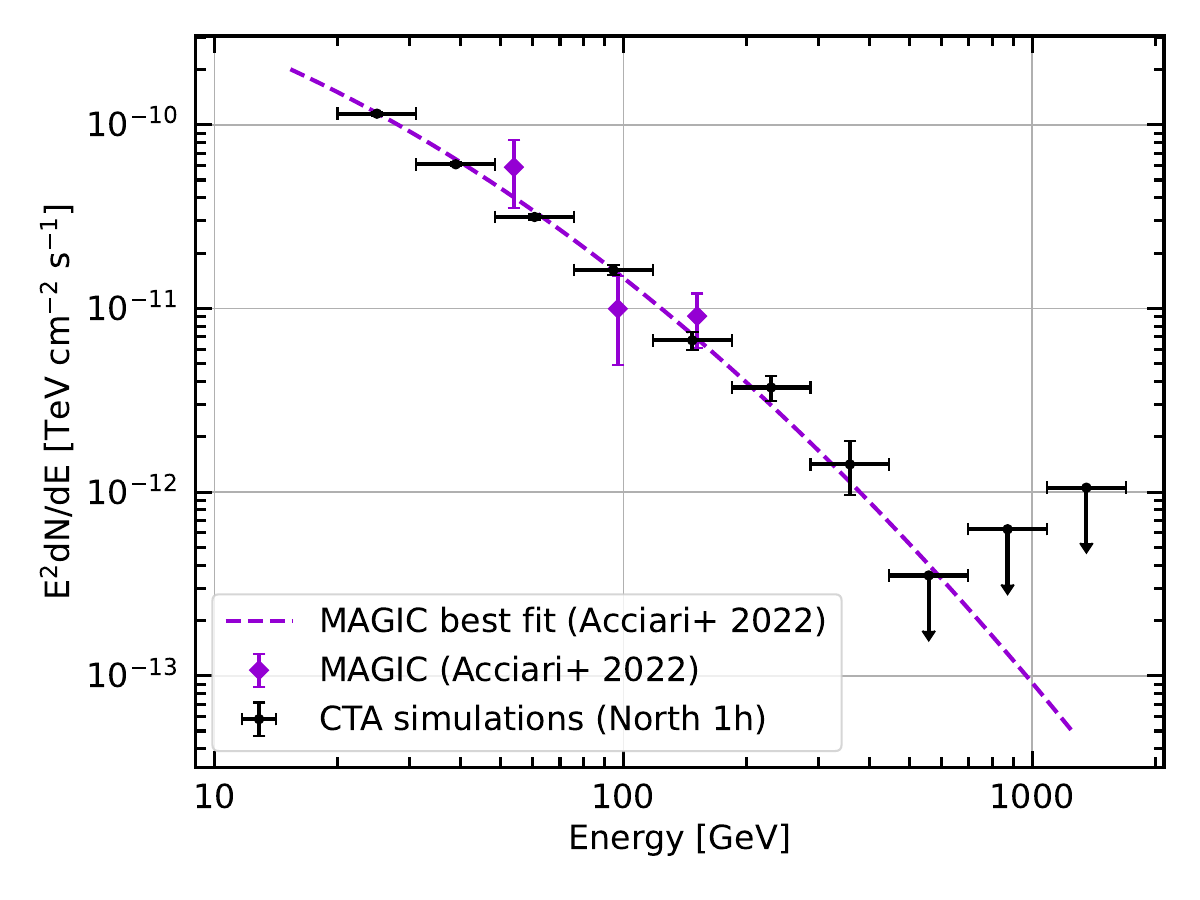} 
        \caption{VHE $\gamma$-ray SED of RS Oph after 1 day since the outburst. The best-fit model from \protect\cite{2022NatAs...6..689A} using MAGIC and \textit{Fermi}-LAT for the first night of observation is used (magenta dashed line) to simulate the source with CTAO. The CTAO-N flux points for a 1-hour observation time are shown as black points. Also, the MAGIC flux points (computed using 1-h observation time of data after cuts) from \protect\cite{2022NatAs...6..689A} are displayed (magenta points).}
        \label{fig:SED_RSOph_day0}
\end{figure}

\subsubsection{Other novae}

To date, RS Oph is the only recurrent nova system from which gamma-ray emission at TeV energies has been detected during outburst. Nevertheless, several other recurrent novae, in particular symbiotic binary systems with high mass-transfer rates and dense winds, are also promising potential gamma-ray emitters. T Coronae Borealis (T CrB) in particular is a nearby symbiotic binary system, located closer to Earth than RS Oph, from which two prior outbursts have been observed in optical wavelengths \citep{2023MNRAS.524.3146Schaefer_TCrB}. Models predict that the next outburst will occur in 2024.4 ± 0.3 years \citep{2023ATel16107....1S_TCrB} and, if a shock evolution comparable to RS Oph can be assumed, particle acceleration and detectable VHE gamma-ray emission is {highly expected.} 
{{Other novae that are expected to plausibly occur over the lifetime of CTAO are the recurrent novae V394 Coronae Australis, CI Aquilae or V3890 Sgr 
and possible superflares arising from V2487 Oph \citep{2019AAS...23412207S,2022MNRAS.512.1924S}.}}

{Recurrent novae - those from which more than a single outburst has been observed - tend to be associated with symbiotic binary systems due to their high mass transfer rate. This also renders them good candidates for particle acceleration to very high energies.Although MeV to ~GeV emission has been detected from classical novae by \textit{Fermi}-LAT, the extension of their spectral energy distributions into the energy range detectable by IACTs is not expected a priori. Only continued observations of a range of novae during outburst with different physical properties will provide further insights into particle acceleration occurring in these systems.  
}

\subsection{Magnetars: discussion} \label{CTAO_mag}

On April 15, 2020, the \textit{Fermi}-GBM and the \textit{Fermi}-LAT instruments detected MeV and GeV gamma-ray emission from a giant flare event of a magnetar located in the NGC 253 galaxy \citep{2021Natur.589..207R,2021NatAs.tmp...11F}. 
The first time detection of GeV emission from a giant magnetar flare is particularly interesting, with the detection of two photons with energies 1.3 GeV and 1.7 GeV. According to \cite{2021NatAs.tmp...11F}, these two GeV photons are produced via synchrotron emission considering the presence of a strong magnetic field which is generated in the shocks. 
It is proposed that these GeV photons are produced in the dissipation associated with the collision of the giant flare outflow and the external shell generated from swept-up material. This indicates that non-thermal processes accelerating particles at high-energies are at work. IC scattering can also occur in these events, and giant magnetar flares have been proposed as potential GeV-TeV emitters. TeV emission on millisecond timescales could be produced during giant flare events, which might be luminous enough to be detectable by IACTs \citep{Lyubarsky2014MNRAS.442L...9L,Murase2016MNRAS.461.1498M}, including CTAO. This TeV emission could be produced via synchrotron maser mechanism, triggered by strong magnetic disturbances from the magnetosphere and propagating outwards, until they dissipate by interacting with the ambient nebula. In the case of dissipation of disturbances within the magnetar wind, a non-thermal tail can plausibly arise, potentially leading to the production of VHE gamma-ray emission \citep{Metzger2020ApJ...902L..22M}.

Magnetars are also relevant for their possible connection with other transient sources, such as, e.g.,  GRBs, super luminous supernovae, and fast radio bursts (FRB). 

The association of a  burst from the Galactic magnetar SGR 1935$+$2153 with an extremely bright FRB-like radio pulse on April 28, 2020 led to the first unequivocal  association between FRBs and magnetars \citep{2020Natur.587...54C,2020Natur.587...59B,2020ApJ...898L..29M,Tavani2021NatAs...5..401T}. The radio flare showed a double-peak structure, pattern also detected at X rays. The X-ray burst was of intermediate energetics, significantly too faint to be  classified  as a giant flare. 
However, even if the X-ray emission was not particularly energetic, this  burst showed a harder spectra with respect to the typical bursts from SGR 1935$+$2154 and other magnetars. 

Comparing the  peak emission of the April 28, 2020 burst \citep{2020ApJ...898L..29M}, which reached a value of 50 ph cm$^{-2}$s$^{-1}$ (in the 15-50 keV band) to other energetic transient events, such as GRB 190114C \citep{2019Natur.575..455M}, we see that the X-ray energetics are compatible with those of the second X-ray peak. 
Current IACTs such as H.E.S.S. \citep{2021ApJ...919..106A} and MAGIC \citep{Lopez-Oramas:2021zd} have led campaigns to search for a VHE component in SGR 1935$+$2154. \cite{2021ApJ...919..106A} observed the source 2 hours prior the CHIME and STARE2 flare and then simultaneously to different X-ray flares. No VHE emission was detected and an UL at E$>$600 GeV of 2.4$\times$10$^{-12}$ erg cm$^{-2}$ s$^{-1}$ was established. 
Similarly, the MAGIC multi-wavelenght monitoring campaing \citep{Lopez-Oramas:2021zd} did not {find} any significant signal, even though some X-ray flares were present during the monitoring campaign. On October 8, CHIME detected three more millisecond events \citep{2020ATel14074....1G} from SGR 1935$+$2154 (with no X-ray counterpart reported), which were confirmed by FAST on October 9 \citep{FASToctober}. The radio fluence of these event was lower than that of April 2020. During simultaneous observations by Swift and MAGIC on October 9,  a forest of bursts was detected in the hard X-ray range, but no TeV emission was revealed \citep{Lopez-Oramas:2021zd}. The H.E.S.S. and MAGIC observations set constraints to the persistent emission in SGR1935+2154 and to the bursting emission. However, depending on the emission region in the magnetar and the interaction with the surrounding nebula (existing in the case of SGR 1935$+$2154 ), future detection of VHE bursts is still plausible. 


The new radio facilities that will operate at the time of CTAO will provide the detection of up to hundreds of FRB per day. Many of these will have good localizations and will be inside the CTAO field of view, making it possible to search for prompt and/or delayed very high-energy emission. Although no magnetar outburst has been yet detected at TeV, the existence of MeV and GeV emission maintains the expectations of a possible TeV component, making magnetars good source candidates for CTAO. The CTAO Observatory {should} aim at observing magnetar flares as soon as possible, triggering on external alerts. Automatic re-pointing of the telescopes can take place whenever certain observational criteria (such as i.e. flare type, brightness, multi-wavelength counterparts or distance) are fulfilled. The high sensitivity of CTAO at short timescales while provide new insights onto the physics of magnetars at VHE.

\section{Synergies with large astronomical facilities} \label{sec:synergies}
Simultaneous coordinated observations with telescopes and facilities at different wavelengths are crucial for understanding the processes and mechanisms at work in the sources of our interest (see \citealp{2019sCTA.book.....C} for a detailed review). Coordination, response to real-time alerts and target-of-opportunity (ToOs) are key in time domain astrophysics when observing transient events. In the case of Galactic transients, external alerts can trigger ToO observations by CTAO of new Galactic events or renewed activity of known sources. The trigger criteria is dependent on the type of source observed, with different time urgency and duration varying in terms of the evolution of the specific phenomena. {Regarding the observational strategy, a first description on the triggering criteria, expected trigger rate and target priority is already provided in \cite{2019sCTA.book.....C}. Updated strategies for multi-wavelength follow-up and needs are already being collected and discussed in a dedicated operations requirement document\footnote{In preparation}, where different studies for the planning and optimization of the CTAO follow-up programme are being conducted.}

{For a prompt reaction and fast reposition, CTAO will need to manage external and internal scientific alerts, for which a \textit{transient handler} subsystem is under development \citep{Egberts2022SPIE12186E..0LE}. The LST-1 transient handler is already operational and has allowed the follow-up of different transient alerts \citep{Carosi2021arXiv210804309C}. Furthermore, a real-time analysis will be performed by CTAO via the \textit{Science Alert Generation} system with the goal of issuing science alerts of transient phenomena to the community with a latency of about 20 seconds \citep{2024SPIE13101E..29B}, increasing the synergies of CTAO with the astronomical community.}



The radio band is of general interest for providing information of the non-thermal processes, and radio facilities are key for locating acceleration sites and shocks {and to provide feedback on fast events such as magnetar flares or stellar flares.} 
For that purpose, {CTAO} will need an external trigger from a radio observatory sensitive to (millisecond) bursts and capable of issuing prompt alerts. Current generation of telescopes such as CHIME, with a large field of view and collecting area, have proven {their efficiency for detecting these alerts \citep{2020Natur.587...54C}. An example of clear synergies are those with the MeerTRAP project to follow-up fast radio signals \citep{Meertrap2020SPIE11447E..0JR,Meertrap2022MNRAS.512.1483B}}

In the optical and infrared domain prompt reaction would be facilitated by coordinating with external observatories within time zones close to those of CTAO sites. In the case of CTAO North, {it} should be pursuing synergies mainly with La Palma, Tenerife (and CAHA) telescopes in Spain, while in the CTAO South the most appropriate choice would be the different ESO telescopes and other large facilities in Chile, such as the Rubin Observatory Legacy Survey of Space and Time (LSST), which will be key for time-domain astronomy \citep[see][]{2022arXiv220804499H}. 
The availability of imaging and spectroscopic instrumentation in both hemispheres {does} largely fulfill the observational needs required for coordinated campaigns. Specific target of opportunity proposals could be eventually placed by the CTAO community. In parallel, small optical telescopes either on-site or operated by nearby CTAO institutions could also play a key role in this effort if the targets are bright enough as in the case of novae or gamma-ray binaries. This could serve to mitigate the often high-time pressure on large telescopes. Additionally, support optical telescopes would ensure a fast follow-up on transients and would provide sufficient coverage in the case of bright sources.

Space missions will allow to study the X-ray domain with improved capabilities. The Space Variable Objects Monitor (SVOM) [4 - 150 keV range] 
has been recently launched and although its core science is focused in GRB detection, it has the capabilities for detecting i.e. thermonuclear explosions and outbursts of Galactic origin due to the sensitivity of the ECLAIRs (50mCrab per orbit ) telescope \citep{SVOM2016arXiv161006892W}. {Synergies with future instruments such as} the Advanced Telescope for High-ENergy Astrophysics (Athena) [0.2–12 keV] \citep{Athena2017AN....338..153B}, scheduled for launch in {mid} 2030s {contemporaneous to the CTAO era, are also expected}. 

The high-energy band (E$>$100 MeV) is currently explored by the \textit{Fermi}-LAT satellite \citep{2009ApJ...697.1071A}. Triggers from \textit{Fermi}-LAT have been important for the detection of {Galactic transients such as} novae in the VHE domain. {Simultaneous} MeV-GeV information is also important to disentangle between hadronic and leptonic processes. If \textit{Fermi} lifetime extends through (at least partially) CTAO lifetime, it will provide the needed coverage and alerts to certain observations/triggers. {Just recently, the AGILE satellite, that also operated in this energy range, ceased its operations and re-entered the atmosphere \citep{AGILE2024}.}  For the moment, there is no obvious successor to Fermi {or AGILE} in the same energy domain, although some missions could take over. {A successor will be crucial to fully exploit the full potential of gamma-ray astronomy.} The High Energy cosmic-Radiation Detection facility (HERD) is expected to be operational from 2026 on board of the Chinese Space Station, detecting gamma rays above 0.5 GeV {and with potential for transient detection \citep{HERD2024NIMPA106869788C}}.

Finally, in a multi-messenger context, CTAO will be able to search for the electromagnetic counterpart of gravitational waves (GWs) {and neutrinos} in the Milky Way. The Laser Interferometer Space Antenna (LISA) mission will be the first GW observatory on space (scheduled launch on 2030) and will key to study white dwarf mergers \citep{2019MNRAS.490.5888L,2023MNRAS.519.2552G}. 
{Neutrino alerts from the next generation of detectors such as IceCube-Gen2 will also be crucial in the case of a Galactic core-collapse supernova \citep{2021JPhG...48f0501A}.}


\section{Summary and conclusions} \label{sec:summary}

We have studied the capabilities of the forthcoming CTAO to detect transient and variable emission from Galactic sources of different nature. {We additionally tested longer exposure times in specific cases of high interest for the observatory}. CTAO will be able to discover new transients with {not significant} degradation in the sensitivity, with a maximum of 15\% in the crowded inner regions of the Galaxy when overlapping with strong emitters. Similarly, in order to detect variability from dim systems, our simulations have shown that sources with a photon flux $<1\times10^{-13}$ ph cm$^{-2}$ s$^{-1}$ will require $>10$ h of observations to detect this variability. For sources with fluxes above this threshold, only 
$5-10$ h are required. In the case of strong sources ($\ge3\times10^{-12}$ ph cm$^{-2}$ s$^{-1}$), short exposures $0.5-1.0$ h are {required}, implying that low variations in the flux can be detectable from  bright sources. For the case of generic transient sources  with the fluxes $<10^{-13}$ ph cm$^{-2}$ s$^{-1}$ TeV$^{-1}$ CTAO will not detect any source in about one hour observation time, while for those with fluxes $<10^{-9}$ ph cm$^{-2}$ s$^{-1}$ TeV$^{-1}$ in an uncrowded region, CTAO {would} be able to detect $\ge65\%$ of sources within just 1 hour of observation for both arrays for the different zenith angles and configurations of the geomagnetic field used in this study {if all sources are visible. Any visibility constraints will significantly lower this number.}

The unique sensitivity at short timescales together with the fast slewing capabilities of the LSTs and the aforementioned capacities of CTAO will allow the detection and discovery of a variety of sources of different nature, according to our simulations:

   \begin{enumerate}
      \item CTAO will detect VHE from microquasars and from the interaction between their jets and the surrounding environment. Our simulations show that CTAO will {likely} detect both transient and persistent emission from the massive microquasars Cyg X-1 and Cyg X-3. CTAO will also significantly detect SS 433 
      {including possible flux variability}. In the case of LMXBs, CTAO will detect outbursts within few tens of minutes from a nearby source ($<$4 kpc) with relatively small inclination angle ($<$30$^{\circ}$). Longer exposure times are required for LMXBs with larger angles. 
      
      \item We tested the case of tMSPs, concluding that CTAO will need long integration times ($>$50 h) to be able to detect the possible emission of tMSPs when they are in the LMXB state. These systems could be detected during a transition from RSMP to LMXB if an additional VHE component is present, which could provide crucial information on particle interaction.
         
      \item Flaring emission from the Crab Nebula will be best detected by CTAO (or LST sub-array) at low energies (E$<$200 GeV) in less than 1 h. In the TeV regime, integration times of $<$10 h will be needed, specially for the detection of dimmer flares. 
      
      \item In the case of novae, CTAO will be able to detect close-by novae. 
      As an example of the only VHE novae known to date, our simulations reveal that CTAO will detect the symbiotic recurrent nova RS Oph with high significance in only 30 min, allowing for a detailed measurement, and therefore detailed modeling, of its SED from energies as low as 20 GeV.
    Combined with multi-wavelength observations, the temporal and spectral analyses of CTAO observation would improve our understanding of the acceleration processes in novae.

      
   \end{enumerate}
   
Regarding sources of different nature from the aforementioned, we {could} expect CTAO to detect emission from magnetars during a giant flare and even likely during intermediate flares associated with an FRB. Other possible transient events are flares from SFXTs, for which a detection would definitely identify SFXTs as VHE emitters. Other variable VHE candidates are runaway stars and  young stellar objects. Serendipitous discoveries are also {possible} while performing, {e.g.}, surveys. Simultaneous multi-wavelength and multi-messenger observations will be crucial to maximize the scientific output of the CTAO observatory.




\section*{Acknowledgements}


We gratefully acknowledge financial support from the following agencies and organizations:

\bigskip

State Committee of Science of Armenia, Armenia;
The Australian Research Council, Astronomy Australia Ltd, The University of Adelaide, Australian National University, Monash University, The University of New South Wales, The University of Sydney, Western Sydney University, Australia; Federal Ministry of Education, Science and Research, and Innsbruck University, Austria;
Conselho Nacional de Desenvolvimento Cient\'{\i}fico e Tecnol\'{o}gico (CNPq), Funda\c{c}\~{a}o de Amparo \`{a} Pesquisa do Estado do Rio de Janeiro (FAPERJ), Funda\c{c}\~{a}o de Amparo \`{a} Pesquisa do Estado de S\~{a}o Paulo (FAPESP), Funda\c{c}\~{a}o de Apoio \`{a} Ci\^encia, Tecnologia e Inova\c{c}\~{a}o do Paran\'a - Funda\c{c}\~{a}o Arauc\'aria, Ministry of Science, Technology, Innovations and Communications (MCTIC), Brasil;
Ministry of Education and Science, National RI Roadmap Project DO1-153/28.08.2018, Bulgaria; 
The Natural Sciences and Engineering Research Council of Canada and the Canadian Space Agency, Canada; 
ANID PIA/APOYO AFB230003, ANID-Chile Basal grant FB 210003, N\'ucleo Milenio TITANs (NCN19-058), FONDECYT-Chile grants 1201582, 1210131, 1230345, and 1240904; 
Croatian Science Foundation, Rudjer Boskovic Institute, University of Osijek, University of Rijeka, University of Split, Faculty of Electrical Engineering, Mechanical Engineering and Naval Architecture, University of Zagreb, Faculty of Electrical Engineering and Computing, Croatia;
Ministry of Education, Youth and Sports, MEYS  LM2018105, LM2023047, EU/MEYS CZ.02.1.01/0.0/0.0/16\_013/0001403, CZ.02.1.01/0.0/0.0/18\_046/0016007,  CZ.02.1.01/0.0/0.0/16\_019/0000754 and CZ.02.01.01/00/22\_008/0004632, Czech Republic; 
Academy of Finland (grant nr.317636 and 320045), Finland;
Ministry of Higher Education and Research, CNRS-INSU and CNRS-IN2P3, CEA-Irfu, ANR, Regional Council Ile de France, Labex ENIGMASS, OCEVU, OSUG2020 and P2IO, France; 
The German Ministry for Education and Research (BMBF), the Max Planck Society, the German Research Foundation (DFG, with Collaborative Research Centres 876 \& 1491), and the Helmholtz Association, Germany; 
Department of Atomic Energy, Department of Science and Technology, India; 
Istituto Nazionale di Astrofisica (INAF), Istituto Nazionale di Fisica Nucleare (INFN), MIUR, Istituto Nazionale di Astrofisica (INAF-OABRERA) Grant Fondazione Cariplo/Regione Lombardia ID 2014-1980/RST\_ERC, Italy; 
ICRR, University of Tokyo, JSPS, MEXT, Japan; 
Netherlands Research School for Astronomy (NOVA), Netherlands Organization for Scientific Research (NWO), Netherlands; 
University of Oslo, Norway; 
Ministry of Science and Higher Education, DIR/WK/2017/12, the National Centre for Research and Development and the National Science Centre, UMO-2016/22/M/ST9/00583, Poland; 
Slovenian Research Agency, grants P1-0031, P1-0385, I0-0033, J1-9146, J1-1700, N1-0111, and the Young Researcher program, Slovenia; 
South African Department of Science and Technology and National Research Foundation through the South African Gamma-Ray Astronomy Programme, South Africa; 
The Spanish groups acknowledge funds from "ERDF A way of making Europe" and the Spanish Ministry of Science and Innovation and the Spanish Research State Agency (AEI) via MCIN/AEI/10.13039/501100011033 through government budget lines PGE2021/28.06.000X.411.01, PGE2022/28.06.000X.411.01, PGE2022/28.06.000X.711.04, and grants PID2022-137810NB-C22, PID2022-136828NB-C42, PID2022-139117NB-C42, PID2022-139117NB-C41, PID2022-136828NB-C41, PID2022-138172NB-C43, PID2022-138172NB-C42, PID2022-139117NB-C44, PID2021-124581OB-I00, PID2021-125331NB-I00, PID2019-104114RB-C31,  PID2019-107847RB-C44, PID2019-104114RB-C32, PID2019-105510GB-C31, PID2019-104114RB-C33, PID2019-107847RB-C41, PID2019-107847RB-C43, PID2019-107847RB-C42; the "Centro de Excelencia Severo Ochoa" program through grants no. CEX2019-000920-S, CEX2020-001007-S, CEX2021-001131-S; the "Unidad de Excelencia Mar\'ia de Maeztu" program through grants no. CEX2019-000918-M, CEX2020-001058-M; the "Ram\'on y Cajal" program through grants RYC2021-032552-I, RYC2021-032991-I, RYC2020-028639-I and RYC-2017-22665; and the "Juan de la Cierva" program through grants no. IJC2019-040315-I and JDC2022-049705-I. La Caixa Banking Foundation is also acknowledged, grant no. LCF/BQ/PI21/11830030. They also acknowledge the project "Tecnolog\'ias avanzadas para la exploraci\'on del universo y sus componentes" (PR47/21 TAU), funded by Comunidad de Madrid regional government. Funds were also granted by the Junta de Andaluc\'ia regional government under the "Plan Complementario de I+D+I" (Ref. AST22\_00001) and "Plan Andaluz de Investigaci\'on, Desarrollo e Innovaci\'on" (Ref. FQM-322); by the "Programa Operativo de Crecimiento Inteligente" FEDER 2014-2020 (Ref.~ESFRI-2017-IAC-12) and Spanish Ministry of Science and Innovation, 15\% co-financed by "Consejer\'ia de Econom\'ia, Industria, Comercio y Conocimiento" of the Gobierno de Canarias regional government. The Generalitat de Catalunya regional government is also gratefully acknowledged via its "CERCA'' program and grants 2021SGR00426 and 2021SGR00679. Spanish groups were also kindly supported by European Union funds via the "Horizon 2020" program, grant no. GA:824064, and NextGenerationEU, grants no. PRTR-C17.I1, CT19/23-INVM-109, and "Mar\'ia Zambrano" program, BDNS: 572725. This research used computing and storage resources provided by the Port d'Informaci\'o Cient\'ifica (PIC) data center; 
Swedish Research Council, Royal Physiographic Society of Lund, Royal Swedish Academy of Sciences, The Swedish National Infrastructure for Computing (SNIC) at Lunarc (Lund), Sweden; 
State Secretariat for Education, Research and Innovation (SERI) and Swiss National Science Foundation (SNSF), Switzerland; Durham University, Leverhulme Trust, Liverpool University, University of Leicester, University of Oxford, Royal Society, Science and Technology Facilities Council, UK; 
U.S. National Science Foundation, U.S. Department of Energy, Argonne National Laboratory, Barnard College, University of California, University of Chicago, Columbia University, Georgia Institute of Technology, Institute for Nuclear and Particle Astrophysics (INPAC-MRPI program), Iowa State University, the Smithsonian Institution, V.V.D. is funded by NSF grant AST-1911061, Washington University McDonnell Center for the Space Sciences, The University of Wisconsin and the Wisconsin Alumni Research Foundation, USA.

\bigskip

The research leading to these results has received funding from the European Union's Seventh Framework Programme (FP7/2007-2013) under grant agreements No~262053 and No~317446.
This project is receiving funding from the European Union's Horizon 2020 research and innovation programs under agreement No~676134.

\bigskip

ALO acknowledges the programa Ram\'on y Cajal for which this publication is part through the Project RYC2021-032991-I, funded by MICIU/AEI/10.13039/501100011033, and the European Union “NextGenerationEU”/PRTR. ALO also acknowledges past support from the JSPS Fellowship for Overseas Researchers of the Japan Society for the Promotion of Science. AJR acknowledges support from an ARC fellowship through award number FT170100243. Part of this research was undertaken with the assistance of resources and services from the National Computational Infrastructure (NCI), which is supported by the Australian Government, through the UNSW HPC Resource Allocation Scheme. DdM acknowledges financial support from INAF AstroFund2022 FANS project. AP acknowledges financial support from the National Institute for Astrophysics (INAF) Research Grant "Uncovering the optical beat of the fastest magnetised neutron stars (FANS)" and from the Italian Ministry of University and Research (MUR), PRIN 2020 (prot. 2020BRP57Z) "Gravitational and Electromagnetic-wave Sources in the Universe with current and next-generation detectors (GEMS)".

This research made use of \verb+ctools+, a community-developed gamma-ray astronomy science analysis software. \verb+ctools+ is based on \verb+GammaLib+, a community-developed toolbox for the scientific analysis of astronomical gamma-ray data. This research made use of \verb+gammapy+,\footnote{\href{https://www.gammapy.org}{https://www.gammapy.org}} a community-developed core Python package for TeV gamma-ray astronomy.

\section*{Data Availability}

The official CTAO observatory IRFs \citep{maier_g_2023_8050921}  used in this manuscript are available in: \verb+prod3b-v2+ \url{https://zenodo.org/record/5163273} and \verb+prod5-v0.1+ \url{https://zenodo.org/record/5499840}. The \verb+ctools+ (\verb+ascl+:1601.005) DOI is: \url{10.5281/zenodo.4727876}.  The \verb+gammapy+ 
DOI is 10.5281/zenodo.4701488. The simulations of the Galactic Plane Survey used in Section \ref{sec:sensitivity} were retrieved from  \url{https://zenodo.org/doi/10.5281/zenodo.8402519}.\\




\section*{Author contributions}

The main individual authors that contributed to this manuscript are, in alphabetical order: 

A. Aguasca-Cabot, simulations and drafting of the novae subsection;
M. Chernyakova, C.~Dignam, simulations and drafting of the SS 433 and microquasars subsections; 
D. Kantzas, simulations and drafting of the LMXBs subsection and modeling of Cygnus X-1 microquasar; 
A. L\'opez-Oramas, coordinator and editor of the manuscript, general drafting;
P. L. Luque-Escamilla, drafting of optical and infrared synergies;
S. Markoff, simulations and drafting of the LMXBs subsection and modeling of Cygnus X-1 microquasar; 
J. Mart\'i, drafting of optical and infrared synergies;
D. de Martino, internal review of the manuscript;
S. McKeague,  simulations and drafting of the SS 433 and microquasars subsections;
S. Mereghetti, drafting of the magnetar subsection;
E. Mestre, simulations and drafting of the Crab Nebula flares subsection;
A. Mitchell, simulations and drafting of the novae subsection;
E. de O\~na-Wilhelmi, simulations and drafting of the Crab Nebula flares and the novae subsections; 
G. Piano, simulations and drafting of the Cygnus X-3, Cygnus X-1, V404 Cyg and microquasars subsections;
P. Romano, early perspectives with CTAO, SFXT discussion and internal review of the manuscript; 
A. J. Ruiter, simulations and drafting of the nova subsection;
I. Sadeh, simulations and drafting of the capabilities of CTAO for transient detection section;
O. Sergijenko, simulations and drafting of the detectability of transients of unknown origin section;
L. Sidoli, SFXTs and other sources;
A. Spolon, simulations and drafting of the tMSPs subsection; 
L. Zampieri, simulations of the tMSPs subsection.

\section*{Affiliations}
\begin{enumerate}[label=$^{\arabic*}$,ref=\arabic*,leftmargin=1.5em,labelsep=0.25em,labelwidth=1.25em]
\item Department of Physics, Tokai University, 4-1-1, Kita-Kaname, Hiratsuka, Kanagawa 259-1292, Japan\label{AFFIL::JapanUTokai}
\item Institute for Cosmic Ray Research, University of Tokyo, 5-1-5, Kashiwa-no-ha, Kashiwa, Chiba 277-8582, Japan\label{AFFIL::JapanUTokyoICRR}
\item ETH Z\"urich, Institute for Particle Physics and Astrophysics, Otto-Stern-Weg 5, 8093 Z\"urich, Switzerland\label{AFFIL::SwitzerlandETHZurich}
\item INFN and Universit\`a degli Studi di Siena, Dipartimento di Scienze Fisiche, della Terra e dell'Ambiente (DSFTA), Sezione di Fisica, Via Roma 56, 53100 Siena, Italy\label{AFFIL::ItalyUSienaandINFN}
\item Universit\'e Paris-Saclay, Universit\'e Paris Cit\'e, CEA, CNRS, AIM, F-91191 Gif-sur-Yvette Cedex, France\label{AFFIL::FranceCEAIRFUDAp}
\item FSLAC IRL 2009, CNRS/IAC, La Laguna, Tenerife, Spain\label{AFFIL::SpainFSLACIRLCNRSIAC}
\item University of Alabama, Tuscaloosa, Department of Physics and Astronomy, Gallalee Hall, Box 870324 Tuscaloosa, AL 35487-0324, USA\label{AFFIL::USAUAlabamaTuscaloosa}
\item Universit\'e C\^ote d'Azur, Observatoire de la C\^ote d'Azur, CNRS, Laboratoire Lagrange, France\label{AFFIL::FranceOCotedAzur}
\item Laboratoire Leprince-Ringuet, CNRS/IN2P3, \'Ecole polytechnique, Institut Polytechnique de Paris, 91120 Palaiseau, France\label{AFFIL::FranceLLREcolePolytechnique}
\item Departament de F{\'\i}sica Qu\`antica i Astrof{\'\i}sica, Institut de Ci\`encies del Cosmos, Universitat de Barcelona, IEEC-UB, Mart{\'\i} i Franqu\`es, 1, 08028, Barcelona, Spain\label{AFFIL::SpainICCUB}
\item Instituto de Astrof{\'\i}sica de Andaluc{\'\i}a-CSIC, Glorieta de la Astronom{\'\i}a s/n, 18008, Granada, Spain\label{AFFIL::SpainIAACSIC}
\item Institute for Computational Cosmology and Department of Physics, Durham University, South Road, Durham DH1 3LE, United Kingdom\label{AFFIL::UnitedKingdomICCUDurham}
\item Pontificia Universidad Cat\'olica de Chile, Av. Libertador Bernardo O'Higgins 340, Santiago, Chile\label{AFFIL::ChileUPontificiaCatolicadeChile}
\item Universidad Nacional Aut\'onoma de M\'exico, Delegaci\'on Coyoac\'an, 04510 Ciudad de M\'exico, Mexico\label{AFFIL::MexicoUNAMMexico}
\item IPARCOS-UCM, Instituto de F{\'\i}sica de Part{\'\i}culas y del Cosmos, and EMFTEL Department, Universidad Complutense de Madrid, E-28040 Madrid, Spain\label{AFFIL::SpainUCMAltasEnergias}
\item Instituto de F{\'\i}sica Te\'orica UAM/CSIC and Departamento de F{\'\i}sica Te\'orica, Universidad Aut\'onoma de Madrid, c/ Nicol\'as Cabrera 13-15, Campus de Cantoblanco UAM, 28049 Madrid, Spain\label{AFFIL::SpainIFTUAMCSIC}
\item LUTH, GEPI and LERMA, Observatoire de Paris, Universit\'e PSL, Universit\'e Paris Cit\'e, CNRS, 5 place Jules Janssen, 92190, Meudon, France\label{AFFIL::FranceObservatoiredeParis}
\item INAF - Osservatorio Astrofisico di Arcetri, Largo E. Fermi, 5 - 50125 Firenze, Italy\label{AFFIL::ItalyOArcetri}
\item INFN Sezione di Perugia and Universit\`a degli Studi di Perugia, Via A. Pascoli, 06123 Perugia, Italy\label{AFFIL::ItalyUPerugiaandINFN}
\item INAF - Osservatorio Astronomico di Roma, Via di Frascati 33, 00078, Monteporzio Catone, Italy\label{AFFIL::ItalyORoma}
\item T\"UB\.ITAK Research Institute for Fundamental Sciences, 41470 Gebze, Kocaeli, Turkey\label{AFFIL::TurkeyTubitak}
\item INFN Sezione di Napoli, Via Cintia, ed. G, 80126 Napoli, Italy\label{AFFIL::ItalyINFNNapoli}
\item INFN Sezione di Padova, Via Marzolo 8, 35131 Padova, Italy\label{AFFIL::ItalyINFNPadova}
\item Instituto de Astrof{\'\i}sica de Canarias and Departamento de Astrof{\'\i}sica, Universidad de La Laguna, La Laguna, Tenerife, Spain\label{AFFIL::SpainIAC}
\item Kapteyn Astronomical Institute, University of Groningen, Landleven 12, 9747 AD, Groningen, The Netherlands\label{AFFIL::NetherlandsUGroningen}
\item Instituto de F{\'\i}sica de S\~ao Carlos, Universidade de S\~ao Paulo, Av. Trabalhador S\~ao-carlense, 400 - CEP 13566-590, S\~ao Carlos, SP, Brazil\label{AFFIL::BrazilIFSCUSaoPaulo}
\item Astroparticle Physics, Department of Physics, TU Dortmund University, Otto-Hahn-Str. 4a, 44227 Dortmund, Germany\label{AFFIL::GermanyUDortmundTU}
\item Department of Physics, Chemistry \& Material Science, University of Namibia, Private Bag 13301, Windhoek, Namibia\label{AFFIL::NamibiaUNamibia}
\item Centre for Space Research, North-West University, Potchefstroom, 2520, South Africa\label{AFFIL::SouthAfricaNWU}
\item School of Physics and Astronomy, Monash University, Melbourne, Victoria 3800, Australia\label{AFFIL::AustraliaUMonash}
\item D\'epartement de physique nucl\'eaire et corpusculaire, University de Gen\`eve,  Facult\'e de Sciences, 1205 Gen\`eve, Switzerland\label{AFFIL::SwitzerlandUGenevaDPNC}
\item Faculty of Science and Technology, Universidad del Azuay, Cuenca, Ecuador.\label{AFFIL::EcuadorUAzuay}
\item Deutsches Elektronen-Synchrotron, Platanenallee 6, 15738 Zeuthen, Germany\label{AFFIL::GermanyDESY}
\item Centro Brasileiro de Pesquisas F{\'\i}sicas, Rua Xavier Sigaud 150, RJ 22290-180, Rio de Janeiro, Brazil\label{AFFIL::BrazilCBPF}
\item Instituto de Astronomia, Geof{\'\i}sica e Ci\^encias Atmosf\'ericas - Universidade de S\~ao Paulo, Cidade Universit\'aria, R. do Mat\~ao, 1226, CEP 05508-090, S\~ao Paulo, SP, Brazil\label{AFFIL::BrazilIAGUSaoPaulo}
\item INFN Sezione di Padova and Universit\`a degli Studi di Padova, Via Marzolo 8, 35131 Padova, Italy\label{AFFIL::ItalyUPadovaandINFN}
\item Institut f\"ur Physik \& Astronomie, Universit\"at Potsdam, Karl-Liebknecht-Strasse 24/25, 14476 Potsdam, Germany\label{AFFIL::GermanyUPotsdam}
\item Institut f\"ur Theoretische Physik, Lehrstuhl IV: Plasma-Astroteilchenphysik, Ruhr-Universit\"at Bochum, Universit\"atsstra{\ss}e 150, 44801 Bochum, Germany\label{AFFIL::GermanyUBochum}
\item Universit\'e Paris Cit\'e, Universit\'e Paris-Saclay, CEA, CNRS, AIM, F-91191 Gif-sur-Yvette, France\label{AFFIL::FranceCEAIRFUDApUParisCiteaffiliatedpersonnel}
\item Center for Astrophysics | Harvard \& Smithsonian, 60 Garden St, Cambridge, MA 02138, USA\label{AFFIL::USACfAHarvardSmithsonian}
\item CIEMAT, Avda. Complutense 40, 28040 Madrid, Spain\label{AFFIL::SpainCIEMAT}
\item Max-Planck-Institut f\"ur Kernphysik, Saupfercheckweg 1, 69117 Heidelberg, Germany\label{AFFIL::GermanyMPIK}
\item Max-Planck-Institut f\"ur Physik, Boltzmannstr. 8, 85748 Garching, Germany\label{AFFIL::GermanyMPP}
\item Pidstryhach Institute for Applied Problems in Mechanics and Mathematics NASU, 3B Naukova Street, Lviv, 79060, Ukraine\label{AFFIL::UkraineIAPMMLviv}
\item Univ. Savoie Mont Blanc, CNRS, Laboratoire d'Annecy de Physique des Particules - IN2P3, 74000 Annecy, France\label{AFFIL::FranceLAPPUSavoieMontBlanc}
\item Center for Astrophysics and Cosmology (CAC), University of Nova Gorica, Nova Gorica, Slovenia\label{AFFIL::SloveniaUNovaGoricaCAC}
\item Politecnico di Bari, via Orabona 4, 70124 Bari, Italy\label{AFFIL::ItalyPolitecnicoBari}
\item INFN Sezione di Bari, via Orabona 4, 70126 Bari, Italy\label{AFFIL::ItalyINFNBari}
\item Institut de Fisica d'Altes Energies (IFAE), The Barcelona Institute of Science and Technology, Campus UAB, 08193 Bellaterra (Barcelona), Spain\label{AFFIL::SpainIFAEBIST}
\item FZU - Institute of Physics of the Czech Academy of Sciences, Na Slovance 1999/2, 182 00 Praha 8, Czech Republic\label{AFFIL::CzechRepublicFZU}
\item INAF - Osservatorio Astronomico di Palermo {\textquotedblleft}G.S. Vaiana{\textquotedblright}, Piazza del Parlamento 1, 90134 Palermo, Italy\label{AFFIL::ItalyOPalermo}
\item Sorbonne Universit\'e, CNRS/IN2P3, Laboratoire de Physique Nucl\'eaire et de Hautes Energies, LPNHE, 4 place Jussieu, 75005 Paris, France\label{AFFIL::FranceLPNHEUSorbonne}
\item INAF - Osservatorio Astronomico di Brera, Via Brera 28, 20121 Milano, Italy\label{AFFIL::ItalyOBrera}
\item INFN Sezione di Pisa, Edificio C {\textendash} Polo Fibonacci, Largo Bruno Pontecorvo 3, 56127 Pisa\label{AFFIL::ItalyINFNPisa}
\item University School for Advanced Studies IUSS Pavia, Palazzo del Broletto, Piazza della Vittoria 15, 27100 Pavia, Italy\label{AFFIL::ItalyIUSSPaviaINAF}
\item Universit\`a degli Studi di Trento, Via Calepina, 14, 38122 Trento, Italy\label{AFFIL::ItalyUTrento}
\item University of Zagreb, Faculty of electrical engineering and computing, Unska 3, 10000 Zagreb, Croatia\label{AFFIL::CroatiaUZagreb}
\item IRFU, CEA, Universit\'e Paris-Saclay, B\^at 141, 91191 Gif-sur-Yvette, France\label{AFFIL::FranceCEAIRFUDPhP}
\item INAF - Osservatorio di Astrofisica e Scienza dello spazio di Bologna, Via Piero Gobetti 93/3, 40129  Bologna, Italy\label{AFFIL::ItalyOASBologna}
\item Centre for Advanced Instrumentation, Department of Physics, Durham University, South Road, Durham, DH1 3LE, United Kingdom\label{AFFIL::UnitedKingdomUDurham}
\item INFN Sezione di Trieste and Universit\`a degli Studi di Udine, Via delle Scienze 208, 33100 Udine, Italy\label{AFFIL::ItalyUUdineandINFNTrieste}
\item Dublin Institute for Advanced Studies, 31 Fitzwilliam Place, Dublin 2, Ireland\label{AFFIL::IrelandDIAS}
\item Armagh Observatory and Planetarium, College Hill, Armagh BT61 9DB, United Kingdom\label{AFFIL::UnitedKingdomArmaghObservatoryandPlanetarium}
\item School of Physics, University of New South Wales, Sydney NSW 2052, Australia\label{AFFIL::AustraliaUNewSouthWales}
\item INFN Sezione di Catania, Via S. Sofia 64, 95123 Catania, Italy\label{AFFIL::ItalyINFNCatania}
\item Unitat de F{\'\i}sica de les Radiacions, Departament de F{\'\i}sica, and CERES-IEEC, Universitat Aut\`onoma de Barcelona, Edifici C3, Campus UAB, 08193 Bellaterra, Spain\label{AFFIL::SpainUABandCERESIEEC}
\item Department of Physics, Faculty of Science, Kasetsart University, 50 Ngam Wong Wan Rd., Lat Yao, Chatuchak, Bangkok, 10900, Thailand\label{AFFIL::ThailandUKasetsart}
\item National Astronomical Research Institute of Thailand, 191 Huay Kaew Rd., Suthep, Muang, Chiang Mai, 50200, Thailand\label{AFFIL::ThailandNARIT}
\item INAF - Osservatorio Astronomico di Capodimonte, Via Salita Moiariello 16, 80131 Napoli, Italy\label{AFFIL::ItalyOCapodimonte}
\item Universidade Cidade de S\~ao Paulo, N\'ucleo de Astrof{\'\i}sica, R. Galv\~ao Bueno 868, Liberdade, S\~ao Paulo, SP, 01506-000, Brazil\label{AFFIL::BrazilUCidadeSPaulo}
\item Dep. of Physics, Sapienza, University of Roma, Piazzale A. Moro 5, 00185, Roma, Italy \label{AFFIL::ItalyURomaSapienza}
\item INAF - Istituto di Astrofisica Spaziale e Fisica Cosmica di Milano, Via A. Corti 12, 20133 Milano, Italy\label{AFFIL::ItalyIASFMilano}
\item CCTVal, Universidad T\'ecnica Federico Santa Mar{\'\i}a, Avenida Espa\~na 1680, Valpara{\'\i}so, Chile\label{AFFIL::ChileUTecnicaFedericoSantaMaria}
\item Aix Marseille Univ, CNRS/IN2P3, CPPM, Marseille, France\label{AFFIL::FranceCPPMUAixMarseille}
\item Universidad de Alcal\'a - Space \& Astroparticle group, Facultad de Ciencias, Campus Universitario Ctra. Madrid-Barcelona, Km. 33.600 28871 Alcal\'a de Henares (Madrid), Spain\label{AFFIL::SpainUAlcala}
\item INFN Sezione di Bari and Universit\`a degli Studi di Bari, via Orabona 4, 70124 Bari, Italy\label{AFFIL::ItalyUandINFNBari}
\item Universit\'e Paris Cit\'e, CNRS, Astroparticule et Cosmologie, F-75013 Paris, France\label{AFFIL::FranceAPCUParisCite}
\item University of the Witwatersrand, 1 Jan Smuts Avenue, Braamfontein, 2000 Johannesburg, South Africa\label{AFFIL::SouthAfricaUWitwatersrand}
\item Dublin City University, Glasnevin, Dublin 9, Ireland\label{AFFIL::IrelandDCU}
\item INFN Sezione di Torino, Via P. Giuria 1, 10125 Torino, Italy\label{AFFIL::ItalyINFNTorino}
\item Dipartimento di Fisica - Universit\`a degli Studi di Torino, Via Pietro Giuria 1 - 10125 Torino, Italy\label{AFFIL::ItalyUTorino}
\item Dipartimento di Fisica e Chimica {\textquotedblleft}E. Segr\`e{\textquotedblright}, Universit\`a degli Studi di Palermo, Via Archirafi 36, 90123, Palermo, Italy\label{AFFIL::ItalyUPalermo}
\item Universidade Federal Do Paran\'a - Setor Palotina, Departamento de Engenharias e Exatas, Rua Pioneiro, 2153, Jardim Dallas, CEP: 85950-000 Palotina, Paran\'a, Brazil\label{AFFIL::BrazilUFPR}
\item INAF - Osservatorio Astrofisico di Catania, Via S. Sofia, 78, 95123 Catania, Italy\label{AFFIL::ItalyOCatania}
\item University of Oxford, Department of Physics, Clarendon Laboratory, Parks Road, Oxford, OX1 3PU, United Kingdom\label{AFFIL::UnitedKingdomUOxford}
\item Universidad de Valpara{\'\i}so, Blanco 951, Valparaiso, Chile\label{AFFIL::ChileUdeValparaiso}
\item University of Wisconsin, Madison, 500 Lincoln Drive, Madison, WI, 53706, USA\label{AFFIL::USAUWisconsin}
\item INAF - Istituto di Astrofisica Spaziale e Fisica Cosmica di Palermo, Via U. La Malfa 153, 90146 Palermo, Italy\label{AFFIL::ItalyIASFPalermo}
\item Department of Physics and Technology, University of Bergen, Museplass 1, 5007 Bergen, Norway\label{AFFIL::NorwayUBergen}
\item INAF - Istituto di Radioastronomia, Via Gobetti 101, 40129 Bologna, Italy\label{AFFIL::ItalyRadioastronomiaINAF}
\item Western Sydney University, Locked Bag 1797, Penrith, NSW 2751, Australia\label{AFFIL::AustraliaUWesternSydney}
\item INAF - Istituto Nazionale di Astrofisica, Viale del Parco Mellini 84, 00136 Rome, Italy\label{AFFIL::ItalyINAF}
\item Laboratoire Univers et Particules de Montpellier, Universit\'e de Montpellier, CNRS/IN2P3, CC 72, Place Eug\`ene Bataillon, F-34095 Montpellier Cedex 5, France\label{AFFIL::FranceLUPMUMontpellier}
\item Universit\`a degli Studi di Napoli {\textquotedblleft}Federico II{\textquotedblright} - Dipartimento di Fisica {\textquotedblleft}E. Pancini{\textquotedblright}, Complesso Universitario di Monte Sant'Angelo, Via Cintia - 80126 Napoli, Italy\label{AFFIL::ItalyUNapoli}
\item Universit\`a degli Studi di Modena e Reggio Emilia, Dipartimento di Ingegneria ''Enzo Ferrari'', via Pietro Vivarelli 10, 41125, Modena, Italy\label{AFFIL::ItalyUModena}
\item Institut f\"ur Astronomie und Astrophysik, Universit\"at T\"ubingen, Sand 1, 72076 T\"ubingen, Germany\label{AFFIL::GermanyIAAT}
\item University of Rijeka, Faculty of Physics, Radmile Matejcic 2, 51000 Rijeka, Croatia\label{AFFIL::CroatiaURijeka}
\item Institute for Theoretical Physics and Astrophysics, Universit\"at W\"urzburg, Campus Hubland Nord, Emil-Fischer-Str. 31, 97074 W\"urzburg, Germany\label{AFFIL::GermanyUWurzburg}
\item Universit\'e Paris-Saclay, CNRS/IN2P3, IJCLab, 91405 Orsay, France\label{AFFIL::FranceIJCLab}
\item Department of Astronomy and Astrophysics, University of Chicago, 5640 S Ellis Ave, Chicago, Illinois, 60637, USA\label{AFFIL::USAUChicagoDAA}
\item LAPTh, CNRS, USMB, F-74940 Annecy, France\label{AFFIL::FranceLAPTh}
\item School of Physics, Chemistry and Earth Sciences, University of Adelaide, Adelaide SA 5005, Australia\label{AFFIL::AustraliaUAdelaide}
\item Department of Physics, Washington University, St. Louis, MO 63130, USA\label{AFFIL::USAWashingtonU}
\item Escuela de Ingenier{\'\i}a El\'ectrica, Facultad de Ingenier{\'\i}a, Pontificia Universidad Cat\'olica de Valpara{\'\i}so, Avenida Brasil 2147, Valpara{\'\i}so, Chile\label{AFFIL::ChileEscIngElec}
\item Santa Cruz Institute for Particle Physics and Department of Physics, University of California, Santa Cruz, 1156 High Street, Santa Cruz, CA 95064, USA\label{AFFIL::USASCIPP}
\item Escola de Artes, Ci\^encias e Humanidades, Universidade de S\~ao Paulo, Rua Arlindo Bettio, CEP 03828-000, 1000 S\~ao Paulo, Brazil\label{AFFIL::BrazilEACHUSaoPaulo}
\item Department of Physics and Astronomy, University of Utah, Salt Lake City, UT 84112-0830, USA\label{AFFIL::USAUUtah}
\item The University of Manitoba, Dept of Physics and Astronomy, Winnipeg, Manitoba R3T 2N2, Canada\label{AFFIL::CanadaUManitoba}
\item RIKEN, Institute of Physical and Chemical Research, 2-1 Hirosawa, Wako, Saitama, 351-0198, Japan\label{AFFIL::JapanRIKEN}
\item INFN Sezione di Roma La Sapienza, P.le Aldo Moro, 2 - 00185 Roma, Italy\label{AFFIL::ItalyINFNRomaLaSapienza}
\item INAF - Osservatorio Astronomico di Padova, Vicolo dell'Osservatorio 5, 35122 Padova, Italy\label{AFFIL::ItalyOPadova}
\item INAF - Istituto di Astrofisica e Planetologia Spaziali (IAPS), Via del Fosso del Cavaliere 100, 00133 Roma, Italy\label{AFFIL::ItalyIAPS}
\item Ruhr University Bochum, Faculty of Physics and Astronomy, Astronomical Institute (AIRUB), Universit\"atsstra{\ss}e 150, 44801 Bochum, Germany\label{AFFIL::GermanyUBochumPhysAst}
\item Physics Program, Graduate School of Advanced Science and Engineering, Hiroshima University, 739-8526 Hiroshima, Japan\label{AFFIL::JapanUHiroshima}
\item Department of Physics, Nagoya University, Chikusa-ku, Nagoya, 464-8602, Japan\label{AFFIL::JapanUNagoya}
\item Friedrich-Alexander-Universit\"at Erlangen-N\"urnberg, Erlangen Centre for Astroparticle Physics, Nikolaus-Fiebiger-Str. 2, 91058 Erlangen, Germany\label{AFFIL::GermanyUErlangenECAP}
\item Department of Information Technology, Escuela Polit\'ecnica Superior, Universidad San Pablo-CEU, CEU Universities, Campus Montepr{\'\i}ncipe, Boadilla del Monte, Madrid 28668, Spain\label{AFFIL::SpainUniversidadSanPabloCEU}
\item INFN Sezione di Roma Tor Vergata, Via della Ricerca Scientifica 1, 00133 Rome, Italy\label{AFFIL::ItalyINFNRomaTorVergata}
\item Alikhanyan National Science Laboratory, Yerevan Physics Institute, 2 Alikhanyan Brothers St., 0036, Yerevan, Armenia\label{AFFIL::ArmeniaNSLAlikhanyan}
\item Universit\'e Paris Cit\'e, CNRS, CEA, Astroparticule et Cosmologie, F-75013 Paris, France\label{AFFIL::FranceAPCUParisCiteCEAaffiliatedpersonnel}
\item Universidad Andr\'es Bello, Av. Fern\'andez Concha 700, Las Condes, Santiago, Chile\label{AFFIL::ChileUAndresBello}
\item N\'ucleo de Astrof{\'\i}sica e Cosmologia (Cosmo-ufes) \& Departamento de F{\'\i}sica, Universidade Federal do Esp{\'\i}rito Santo (UFES), Av. Fernando Ferrari, 514. 29065-910. Vit\'oria-ES, Brazil\label{AFFIL::BrazilUFES}
\item Astrophysics Research Center of the Open University (ARCO), The Open University of Israel, P.O. Box 808, Ra{\textquoteright}anana 4353701, Israel\label{AFFIL::IsraelOpenUniversityofIsrael}
\item Department of Physics, The George Washington University, Washington, DC 20052, USA\label{AFFIL::USAGWUWashingtonDC}
\item University of Liverpool, Oliver Lodge Laboratory, Liverpool L69 7ZE, United Kingdom\label{AFFIL::UnitedKingdomULiverpool}
\item King's College London, Strand, London, WC2R 2LS, United Kingdom\label{AFFIL::UnitedKingdomKingsCollege}
\item Cherenkov Telescope Array Observatory gGmbH, Via Gobetti, Bologna, Italy\label{AFFIL::ItalyCTAOBologna}
\item General Education Center, Yamanashi-Gakuin University, Kofu, Yamanashi 400-8575, Japan\label{AFFIL::JapanUYamanashiGakuin}
\item Sendai College, National Institute of Technology, Natori, Miyagi 981-1239, Japan\label{AFFIL::JapanNITSendaiNatori}
\item Universit\"at Innsbruck, Institut f\"ur Astro- und Teilchenphysik, Technikerstr. 25/8, 6020 Innsbruck, Austria\label{AFFIL::AustriaUInnsbruck}
\item Department of Physics, Faculty of Engineering Science, Yokohama National University, Yokohama 240{\textendash}8501, Japan\label{AFFIL::JapanUYokohamaNational}
\item Astronomical Observatory of Taras Shevchenko National University of Kyiv, 3 Observatorna Street, Kyiv, 04053, Ukraine\label{AFFIL::UkraineAstObsofUKyiv}
\item Palack\'y University Olomouc, Faculty of Science, Joint Laboratory of Optics of Palack\'y University and Institute of Physics of the Czech Academy of Sciences, 17. listopadu 1192/12, 779 00 Olomouc, Czech Republic\label{AFFIL::CzechRepublicUOlomouc}
\item Finnish Centre for Astronomy with ESO, University of Turku, Finland, FI-20014 University of Turku, Finland\label{AFFIL::FinlandFINCA}
\item Aalto University, Mets\"ahovi Radio Observatory, Mets\"ahovintie 114, FI-02540 Kylm\"al\"a, Finland\label{AFFIL::FinlandUAalto}
\item Josip Juraj Strossmayer University of Osijek, Trg Ljudevita Gaja 6, 31000 Osijek, Croatia\label{AFFIL::CroatiaUOsijek}
\item CETEMPS Dipartimento di Scienze Fisiche e Chimiche, Universit\`a degli Studi dell{\textquoteright}Aquila and GSGC-LNGS-INFN, Via Vetoio 1, L{\textquoteright}Aquila, 67100, Italy\label{AFFIL::ItalyCETEMPSandUandINFNAquila}
\item Chiba University, 1-33, Yayoicho, Inage-ku, Chiba-shi, Chiba, 263-8522 Japan\label{AFFIL::JapanUChiba}
\item Department of Earth and Space Science, Graduate School of Science, Osaka University, Toyonaka 560-0043, Japan\label{AFFIL::JapanUOsaka}
\item Astronomical Observatory, Jagiellonian University, ul. Orla 171, 30-244 Cracow, Poland\label{AFFIL::PolandUJagiellonian}
\item Landessternwarte, Zentrum f\"ur Astronomie  der Universit\"at Heidelberg, K\"onigstuhl 12, 69117 Heidelberg, Germany\label{AFFIL::GermanyLSW}
\item IRAP, Universit\'e de Toulouse, CNRS, CNES, UPS, 9 avenue Colonel Roche, 31028 Toulouse, Cedex 4, France\label{AFFIL::FranceIRAPUToulouse}
\item Department of Physics and Astronomy, University of California, Los Angeles, CA 90095, USA\label{AFFIL::USAUCLA}
\item Astronomical Institute of the Czech Academy of Sciences, Bocni II 1401 - 14100 Prague, Czech Republic\label{AFFIL::CzechRepublicASU}
\item Faculty of Science, Ibaraki University, Mito, Ibaraki, 310-8512, Japan\label{AFFIL::JapanUIbaraki}
\item Faculty of Science and Engineering, Waseda University, Shinjuku, Tokyo 169-8555, Japan\label{AFFIL::JapanUWaseda}
\item University of Oslo, Department of Physics, Sem Saelandsvei 24 - PO Box 1048 Blindern, N-0316 Oslo, Norway\label{AFFIL::NorwayUOslo}
\item Nicolaus Copernicus Astronomical Center, Polish Academy of Sciences, ul. Bartycka 18, 00-716 Warsaw, Poland\label{AFFIL::PolandNicolausCopernicusAstronomicalCenter}
\item National Astronomical Observatory of Japan (NAOJ), Division of Science, 2-21-1, Osawa, Mitaka, Tokyo 181-8588, Japan\label{AFFIL::JapanNAOJ}
\item Institute of Particle and Nuclear Studies,  KEK (High Energy Accelerator Research Organization), 1-1 Oho, Tsukuba, 305-0801, Japan\label{AFFIL::JapanKEK}
\item School of Physics and Astronomy, University of Leicester, Leicester, LE1 7RH, United Kingdom\label{AFFIL::UnitedKingdomULeicester}
\item Universit\'e Bordeaux, CNRS, LP2I Bordeaux, UMR 5797, 19 Chemin du Solarium, F-33170 Gradignan, France\label{AFFIL::FranceLP2IUBordeaux}
\item Universit\`a degli studi di Catania, Dipartimento di Fisica e Astronomia {\textquotedblleft}Ettore Majorana{\textquotedblright}, Via S. Sofia 64, 95123 Catania, Italy\label{AFFIL::ItalyUCatania}
\item Department of Physics and Astronomy, University of Turku, Finland, FI-20014 University of Turku, Finland\label{AFFIL::FinlandUTurku}
\item ASI - Space Science Data Center, Via del Politecnico s.n.c., 00133, Rome, Italy\label{AFFIL::ASISpaceScienceDataCenter}
\item INFN Sezione di Trieste and Universit\`a degli Studi di Trieste, Via Valerio 2 I, 34127 Trieste, Italy\label{AFFIL::ItalyUandINFNTrieste}
\item Escuela Polit\'ecnica Superior de Ja\'en, Universidad de Ja\'en, Campus Las Lagunillas s/n, Edif. A3, 23071 Ja\'en, Spain\label{AFFIL::SpainUJaen}
\item Department of Astronomy, University of Geneva, Chemin d'Ecogia 16, CH-1290 Versoix, Switzerland\label{AFFIL::SwitzerlandUGenevaISDC}
\item Anton Pannekoek Institute/GRAPPA, University of Amsterdam, Science Park 904 1098 XH Amsterdam, The Netherlands\label{AFFIL::NetherlandsUAmsterdam}
\item Saha Institute of Nuclear Physics, A CI of Homi Bhabha National Institute, Kolkata 700064, West Bengal, India\label{AFFIL::IndiaSahaInstitute}
\item Institute for Nuclear Research and Nuclear Energy, Bulgarian Academy of Sciences, 72 boul. Tsarigradsko chaussee, 1784 Sofia, Bulgaria\label{AFFIL::BulgariaINRNEBAS}
\item Kavli Institute for Particle Astrophysics and Cosmology, Stanford University, Stanford, CA 94305, USA\label{AFFIL::USAStanford}
\item UCM-ELEC group, EMFTEL Department, University Complutense of Madrid, 28040 Madrid, Spain\label{AFFIL::SpainUCMElectronica}
\item Departamento de Ingenier{\'\i}a El\'ectrica, Universidad Pontificia Comillas - ICAI, 28015 Madrid\label{AFFIL::SpainUPCMadrid}
\item Institute of Space Sciences (ICE, CSIC), and Institut d'Estudis Espacials de Catalunya (IEEC), and Instituci\'o Catalana de Recerca I Estudis Avan\c{c}ats (ICREA), Campus UAB, Carrer de Can Magrans, s/n 08193 Cerdanyola del Vall\'es, Spain\label{AFFIL::SpainICECSIC}
\item The Henryk Niewodnicza\'nski Institute of Nuclear Physics, Polish Academy of Sciences, ul. Radzikowskiego 152, 31-342 Cracow, Poland\label{AFFIL::PolandIFJ}
\item IPARCOS Institute, Faculty of Physics (UCM), 28040 Madrid, Spain\label{AFFIL::SpainIPARCOSInstitute}
\item Department of Physics, Konan University, Kobe, Hyogo, 658-8501, Japan\label{AFFIL::JapanUKonan}
\item Hiroshima Astrophysical Science Center, Hiroshima University, Higashi-Hiroshima, Hiroshima 739-8526, Japan\label{AFFIL::JapanHASC}
\item Joseph-von-Fraunhofer-Str. 25, 44227 Dortmund, Germany\label{AFFIL::LamarrInstituteGermany}
\item School of Allied Health Sciences, Kitasato University, Sagamihara, Kanagawa 228-8555, Japan\label{AFFIL::JapanUKitasato}
\item Department of Physics, Yamagata University, Yamagata, Yamagata 990-8560, Japan\label{AFFIL::JapanUYamagata}
\item Departamento de F{\'\i}sica, Universidade Federal do Rio Grande do Norte, 59078-970, Natal, RN, Brasil\label{AFFIL::BrazilURioGrandedoNortePhys}
\item International Institute of Physics, Universidade Federal do Rio Grande do Norte, 59078-970, Natal, RN, Brasil\label{AFFIL::BrazilURioGrandedoNorteIIP}
\item University of Bia{\l}ystok, Faculty of Physics, ul. K. Cio{\l}kowskiego 1L, 15-245 Bia{\l}ystok, Poland\label{AFFIL::PolandUBiaystok}
\item Charles University, Institute of Particle \& Nuclear Physics, V Hole\v{s}ovi\v{c}k\'ach 2, 180 00 Prague 8, Czech Republic\label{AFFIL::CzechRepublicUPrague}
\item Institute for Space{\textemdash}Earth Environmental Research, Nagoya University, Furo-cho, Chikusa-ku, Nagoya 464-8601, Japan\label{AFFIL::JapanUNagoyaISEE}
\item Kobayashi{\textemdash}Maskawa Institute for the Origin of Particles and the Universe, Nagoya University, Furo-cho, Chikusa-ku, Nagoya 464-8602, Japan\label{AFFIL::JapanUNagoyaKMI}
\item Graduate School of Technology, Industrial and Social Sciences, Tokushima University, Tokushima 770-8506, Japan\label{AFFIL::JapanUTokushima}
\item Cherenkov Telescope Array Observatory, Saupfercheckweg 1, 69117 Heidelberg, Germany\label{AFFIL::GermanyCTAOHeidelberg}
\item University of Pisa, Largo B. Pontecorvo 3, 56127 Pisa, Italy \label{AFFIL::ItalyUPisa}
\item Rudjer Boskovic Institute, Bijenicka 54, 10 000 Zagreb, Croatia\label{AFFIL::CroatiaIRB}
\item INAF - Osservatorio Astronomico di Padova and INFN Sezione di Trieste, gr. coll. Udine, Via delle Scienze 208 I-33100 Udine, Italy\label{AFFIL::ItalyOandINFNTrieste}
\item Univ. Grenoble Alpes, CNRS, IPAG, 38000 Grenoble, France\label{AFFIL::FranceIPAGUGrenobleAlpes}
\item Dipartimento di Scienze Fisiche e Chimiche, Universit\`a degli Studi dell'Aquila and GSGC-LNGS-INFN, Via Vetoio 1, L'Aquila, 67100, Italy\label{AFFIL::ItalyUandINFNAquila}
\item Centre for Astro-Particle Physics (CAPP) and Department of Physics, University of Johannesburg, PO Box 524, Auckland Park 2006, South Africa\label{AFFIL::SouthAfricaUJohannesburg}
\item Departamento de F{\'\i}sica, Facultad de Ciencias B\'asicas, Universidad Metropolitana de Ciencias de la Educaci\'on, Avenida Jos\'e Pedro Alessandri 774, \~Nu\~noa, Santiago, Chile\label{AFFIL::ChileUMCE}
\item School of Physics and Astronomy, University of Minnesota, 116 Church Street S.E. Minneapolis, Minnesota 55455-0112, USA\label{AFFIL::USAUMinnesota}
\item Instituto de Estudios Astrof{\'\i}sicos, Facultad de Ingenier{\'\i}a y Ciencias, Universidad Diego Portales, Av. Ej\'ercito Libertador 441, 8370191 Santiago, Chile\label{AFFIL::ChileUniversidadDiegoPortales}
\item Departamento de Astronom{\'\i}a, Universidad de Concepci\'on, Barrio Universitario S/N, Concepci\'on, Chile\label{AFFIL::ChileUdeConcepcion}
\item Instituto de F{\'\i}sica - Universidade de S\~ao Paulo, Rua do Mat\~ao Travessa R Nr.187 CEP 05508-090  Cidade Universit\'aria, S\~ao Paulo, Brazil\label{AFFIL::BrazilIFUSaoPaulo}
\item University of New South Wales, School of Science, Australian Defence Force Academy, Canberra, ACT 2600, Australia \label{AFFIL::AustraliaUNewSouthWalesCanberra}
\item Gifu University, Faculty of Engineering, 1-1 Yanagido, Gifu 501-1193, Japan \label{AFFIL::JapanUGifu}
\item University of Split  - FESB, R. Boskovica 32, 21 000 Split, Croatia\label{AFFIL::CroatiaFESB}
\item Departamento de F{\'\i}sica, Universidad de Santiago de Chile (USACH), Av. Victor Jara 3493, Estaci\'on Central, Santiago, Chile\label{AFFIL::ChileUniversidaddeSantiagodeChile}
\item Main Astronomical Observatory of the National Academy of Sciences of Ukraine, Zabolotnoho str., 27, 03143, Kyiv, Ukraine\label{AFFIL::UkraineObsNASUkraine}
\item Space Technology Centre, AGH University of Krakow, Aleja Mickiewicza 30, Krak\'ow 30-059, Poland\label{AFFIL::PolandAGHCracowSTC}
\item Academic Computer Centre CYFRONET AGH, ul. Nawojki 11, 30-950, Krak\'ow, Poland\label{AFFIL::PolandCYFRONETAGH}
\item Institute of Astronomy, Faculty of Physics, Astronomy and Informatics, Nicolaus Copernicus University in Toru\'n, ul. Grudzi\k{a}dzka 5, 87-100 Toru\'n, Poland\label{AFFIL::PolandTorunInstituteofAstronomy}
\item Warsaw University of Technology, Faculty of Electronics and Information Technology, Institute of Electronic Systems, Nowowiejska 15/19, 00-665 Warsaw, Poland\label{AFFIL::PolandWUTElectronics}
\item Department of Physical Sciences, Aoyama Gakuin University, Fuchinobe, Sagamihara, Kanagawa, 252-5258, Japan\label{AFFIL::JapanUAoyamaGakuin}
\item Division of Physics and Astronomy, Graduate School of Science, Kyoto University, Sakyo-ku, Kyoto, 606-8502, Japan\label{AFFIL::JapanUKyotoPhysicsandAstronomy}
\item INAF - Osservatorio Astronomico di Cagliari, Via della Scienza 5, I-09047 Selargius (CA), Italy\label{AFFIL::ItalyINAFCagliari}
\item INAF - Osservatorio Astrofisico di Torino, Strada Osservatorio 20, 10025  Pino Torinese (TO), Italy\label{AFFIL::ItalyOTorino}
\item Departamento de F{\'\i}sica, Universidad T\'ecnica Federico Santa Mar{\'\i}a, Avenida Espa\~na, 1680 Valpara{\'\i}so, Chile\label{AFFIL::ChileDepFisUTecnicaFedericoSantaMaria}
\item School of Physics and Astronomy, Sun Yat-sen University, Zhuhai, China\label{AFFIL::ChinaUSunYatsen}
\end{enumerate}



\bibliographystyle{mnras}
\bibliography{references} 








\bsp	
\label{lastpage}
\end{document}